\documentclass[twocolumn,superscriptaddress]{revtex4}

\usepackage[latin1]{inputenc}
\usepackage[english]{babel}
\usepackage{amsmath}
\usepackage{amsfonts}
\usepackage{amssymb}
\usepackage{graphicx}
\usepackage[normalem]{ulem}
\usepackage{color}
\usepackage{filecontents}

\makeatletter
\newcommand{\settitle}{\@maketitle}
\makeatother

\setcitestyle{super}

\begin{document}

\title{Optomechanical Collective Effects in Surface-Enhanced Raman Scattering from Many Molecules}

\author{Yuan Zhang}
\email{yzhuaudipc@zzu.edu.cn}
\affiliation{Donostia International Physics Center, Paseo Manuel de Lardizabal 4, 20018 Donostia-San Sebastian, Spain}
\affiliation{School of Physics and Microelectronics, Zhengzhou University, Daxue Road 75, Zhengzhou 450052 China}

\author{Javier Aizpurua}
\email{aizpurua@ehu.eus}
\affiliation{Donostia International Physics Center, Paseo Manuel de Lardizabal 4, 20018 Donostia-San Sebastian, Spain}
\affiliation{Center for Material Physics (CSIC - UPV/EHU), Paseo Manuel de Lardizabal 5, 20018 Donostia-San Sebastian, Spain}

\author{Ruben Esteban}
\email{ruben_esteban@ehu.eus}
\affiliation{Donostia International Physics Center, Paseo Manuel de Lardizabal 4, 20018 Donostia-San Sebastian, Spain}
\affiliation{IKERBASQUE, Basque Foundation for Science, Maria Diaz de Haro 3, 48013 Bilbao, Spain}


\begin{abstract}
The interaction between molecules is commonly ignored in surface-enhanced Raman scattering (SERS). Under this assumption, the total SERS signal is  described as the sum of the individual contributions of each molecule treated independently. We adopt here an optomechanical description of SERS within a cavity quantum electrodynamics framework to study how collective effects emerge from the quantum correlations of distinct molecules. We derive analytical expressions for identical molecules and implement numerical simulations to analyze two types of collective phenomena: (i) a decrease of the laser intensity threshold to observe strong non-linearities as the number of molecules increases, within intense illumination, and (ii) identification of superradiance in the SERS signal, namely a quadratic scaling with the number of molecules.  The laser intensity required to observe the latter in the anti-Stokes scattering is relatively moderate, which makes it particularly accessible to experiments. Our results also show that collective phenomena can survive in the presence of moderate homogeneous and inhomogeneous broadening.
\end{abstract}

\maketitle 

\section{Introduction}
The interaction between molecular vibrations and photons of an external laser as measured in surface-enhanced Raman scattering (SERS)\citep{MMoskovits,EricLR}  is strongly enhanced by the presence of nearby metallic nanostructures acting as effective optical nanoantennas \citep{MuhP,BhaP}, such as nanorods \citep{STSivapalan,TamTH,MuskOL,RogL}, nanostars \citep{WNiu,KumPS,HaoF}, nanoparticle dimers \citep{WZhu,AizpJ,RomI, EsteR,NordP}, nanoparticle-on-a-mirror configurations \citep{ALombardi-0,BaumbergJJ} and atomic force microscope or scanning tunnelling microscope (STM) tips \citep{RZhang,LiuS,QiuXH,ImadaH,PettB,StoRM,KazE,DopB}. This enhancement is  partially attributed to the chemical interaction between the molecules and the metal \citep{JRLombardi}, but it is mainly boosted by the strong increase of the electromagnetic field strength near the nanostructures \citep{MMoskovits} due to the collective excitation of electrons in the metal, i.e. localized surface plasmon polaritons. Because the characteristic narrow Raman peaks can be associated with unique vibrational frequencies of molecules, SERS is standardly applied to detect particular molecular fingerprints and to characterize and track minute amounts of analytes\citep{WKa,LKq,QinL,LalS} (including single molecules \citep{SMNie,KKneipp,LRu}) for biology and medicine \citep{CiallaMayD}.

Most SERS measurements have been successfully interpreted within classical or semi-classical theories \citep{MMoskovits,ECLRu}, but recent experiments using well-controlled metallic nanostructures and precise positioning of molecules \citep{WZhu,ALombardi-0,RZhang} might allow to reach conditions where the quantum nature of the molecular vibration-plasmon interaction becomes relevant. In the last few years, a cavity quantum electrodynamics (QED) description of SERS has been developed \citep{RoelliP,MKSchmidt,MKSchmidt-1,MKDezfouli-1} by using second-quantization to model both photonic and vibrational excitations. This description is formally analogue to the one typically used in cavity optomechanics \citep{MAspelmeyer}, but with orders-of-magnitude larger losses and coupling strength.  This approach is able to  predict not only the population of the molecular vibrations,	 the Stokes and anti-Stokes SERS signal in standard situations, but also many other intriguing effects, such as Raman-induced plasmon resonance shifts, higher-order Stokes scattering, complex Raman photon correlations, heat-transfer between molecules and a strongly non-linear scaling of the Stokes and anti-Stokes signal with laser intensity that can even lead to a divergent Raman scattering (known as parametric instability in cavity optomechanics)\citep{RoelliP,MKSchmidt,FBenz,MKDezfouli-1,SMAshrafi}. 

While previous works focused mostly on single molecules, we provide here a thorough study of SERS when many molecules are present. Qualitativley different behaviors arise when the SERS is studied by the optomechanical description and by the standard classical treatment. The latter typically assumes that the molecules can be considered as independent, i.e. without interaction among them, so that the signal from $N$ identical molecules simply corresponds to $N$ times the signal from a single molecule. On the other hand, the optomechanical description suggests that the molecules can interact with each other via their coupling to the plasmonic structure, leading to collective effects under adequate conditions. For example, it has been pointed out theoretically \citep{RoelliP} that the presence of many molecules can facilitate reaching the parametric instability at lower laser intensity. The collective response has also been invoked  in the design of a photon up-conversion device based on SERS \citep{PRolli-1} and to explain a recent experiment\citep{ALombardi} that reveals a non-linear dependence of the Stokes SERS signal on the pulsed laser intensity. In other related contexts, collective interactions have been studied in Raman experiments with exquisitely controlled atoms at ultra-low temperatures and are now applied routinely to study a variety of interesting phenomena, such as superradiant Raman lasing \citep{GVrijsen}, spin-squeezing \citep{ASSorensen} and quantum phase transitions \citep{JKlinder}. In these systems the Raman signal can scale quadratically with the number of atoms \citep{JGBohnet}. 

\begin{figure*}[!ht]
\begin{centering}
\includegraphics[scale=0.65]{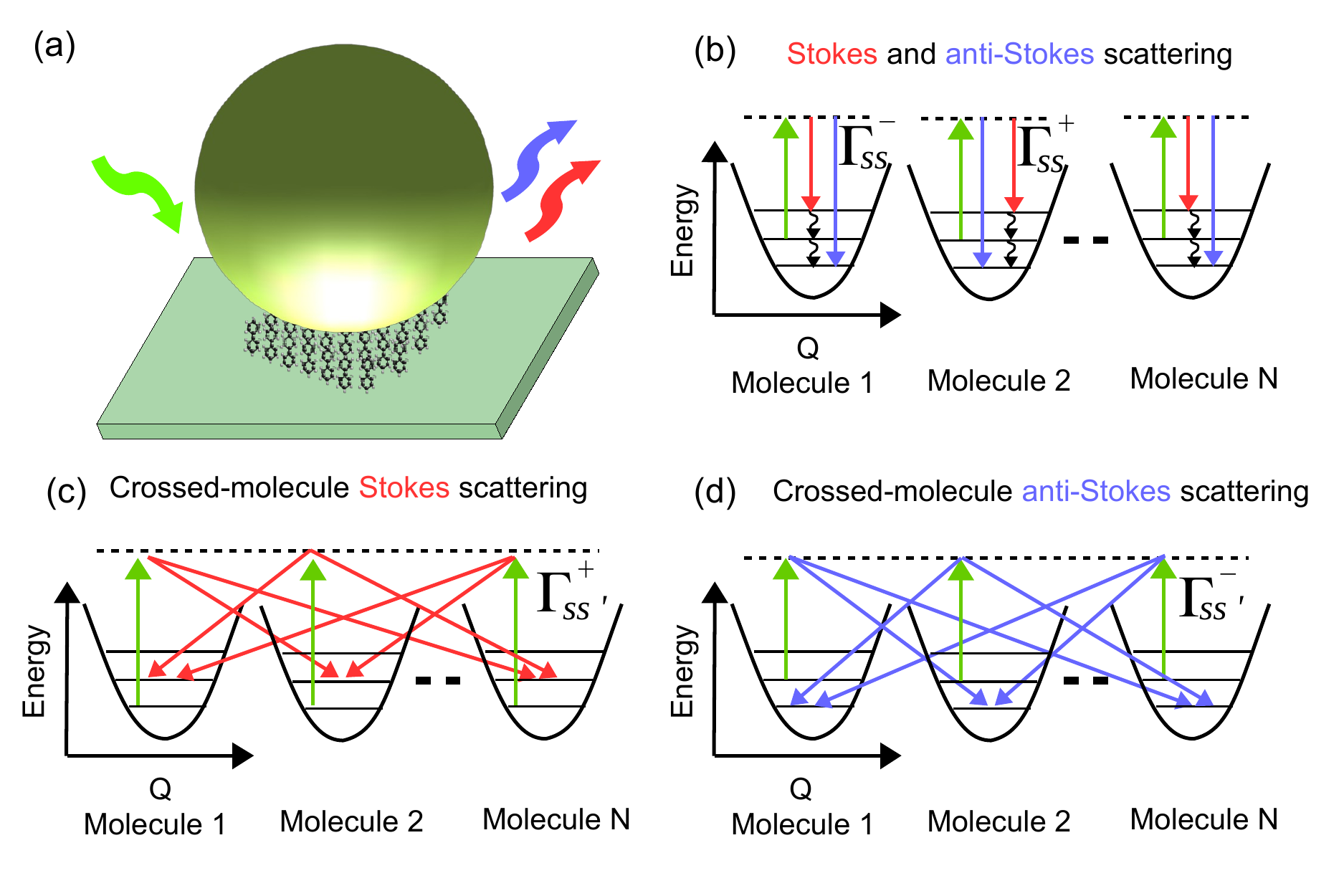}
\par\end{centering}
\caption{\label{fig:structures-processes} (a) Schematics of the SERS system. An ensemble of molecules are located in a gap between a gold nanostructure and a gold substrate. The molecules interact with the plasmonic system, which is excited by an external laser (green arrow), and the emission rate of Stokes (red arrow) and anti-Stokes (blue arrow) Raman photons is enhanced by the plasmonic response of the metallic nanostructure. (b) Diagram of the Raman process when the (plasmon-enhanced) interactions between different molecules are assumed to be absent. The vibrational mode of each molecule is modeled as a harmonic oscillator with equally-spaced energy levels in a parabolic potential energy surface (with respect to the normal mode coordinate $Q$). The plasmon-enhanced Stokes (vertical red lines) and anti-Stokes (vertical blue lines) scattering occur via virtual levels (horizontal dashed lines), excited by the plasmon-enhanced local field (green vertical lines). Phonon decay of individual molecules are also included  (vertical wavy arrows). (c) and (d) represent Raman processes associated with the collective effects in the (c) Stokes and (d) anti-Stokes scattering from several molecules, which occur due to the molecule-molecule correlations established by the plasmon-mediated interaction between different molecules. The parameters $\Gamma_{ss}^{\pm}$ and $\Gamma_{ss'}^{\pm}$ ($s=s'$ or $s\neq s'$) characterizing the processes in (b,c,d) are described in the text.}
\end{figure*}

In short, the optomechanical description suggests that novel collective effects can emerge in experiments, but most SERS measurements are regularly interpreted without considering these effects.  Motivated by this appealing opportunity, in this paper, we study under which conditions the collective effects can emerge in realistic SERS experiments. With this objective in mind, we extend the  molecular optomechanical description of non-resonant Raman to the case of many molecules (see sketch in Figure \ref{fig:structures-processes}a), which naturally incorporates the quantum correlations between different molecules that are the origin of these collective effects. We first focus on a simple system that consists of identical molecules and derive analytic expressions to identify two kinds of collective effects: (i) a quadratic increase of the Stokes and, more significantly, the anti-Stokes signal with an increasing number of molecules $N$ and (ii) a decrease of the laser power required to observe the parametric instability or a saturation of the vibrational population for an increasing $N$. The latter is connected with the cooling of mechanical oscillations that is often observed in other optomechanical systems \cite{MAspelmeyer}. In addition, we find that these collective effects are robust to the homogeneous broadening of molecules caused by, for example,  loss-induced dephasing, and also to the inhomogeneous broadening due to slight variations in the vibrational frequency of different molecules.  

\section{System and Model}

We study the Raman scattering from an arbitrary number of molecules that interact with a plasmonic nanostructure, as sketched in Figure \ref{fig:structures-processes}a. We consider biphenyl-4-thiol (BPT) molecules as canonical molecular species coupled to an optimized plasmonic system, such as a metallic nano-particle on a mirror configuration \citep{ALombardi-0} or a metallic STM tip over a metallic substrate \citep{RZhang}. Our model assumes that the molecules are sufficiently far apart so that they interact with each other only via their coupling with the plasmonic excitation of the nanostructure, and thus the model is more suitable for systems where the molecules are not closely packed. 

We consider the vibrational mode of the BPT molecule with energy $\hbar\omega_{s}=196.5$ meV (frequency $1580$ cm$^{-1}$) due to its strong Raman activity\citep{ALombardi} $R^2_{s} = 10^5 \epsilon_{0}\mathring{\mathrm{A}}^{4}/\mathrm{amu}$. Here, $\epsilon_0$ and $\mathrm{amu}$ are the vacuum permittivity and the atomic mass unit, respectively. The large value of $R^2_{s}$ is due to not only the intrinsic properties of the molecule but also to its chemical interaction with the metallic surfaces (chemical Raman enhancement). For simplicity, we neglect any possible infrared activity of the molecular vibrations, so that different molecules couple only with each other via Raman processes. The label $s$ distinguishes between molecules and this 
is useful for molecules with different vibrational frequencies as considered later on. We consider non-resonant SERS and thus do not include the electronic excited states of the molecule explicitly. In addition, we assume that the potential energy surface (of the electronic ground state) depends quadratically on the normal mode coordinates and thus the vibrations can be modeled as harmonic oscillators via the Hamiltonian $H_{vib}=\sum_{s}\hbar\omega_{s}b_{s}^{\dagger}b_{s}$, where  $b_{s}^{\dagger},b_{s}$ are the bosonic creation and annihilation operator of the vibrational excitation, respectively \citep{EricLR,MKSchmidt} and $\hbar$ is Planck's reduced constant. The incoherent coupling of the molecular vibrations with the environment results in a (small) phonon decay rate $\hbar\gamma_{s}=0.07$ meV \citep{FBenz,ALombardi} (except when otherwise stated), a thermal phonon population $n_s^{th} = [e^{\hbar\omega_{s}/k_{B}T}-1]^{-1}\approx 10^{-3} $ at temperature $T= 290 K$ ($k_{B}$ is the Boltzmann constant), and a vibrational pure-dephasing rate (homogeneous broadening) $\chi_s$  \citep{YZhao}. We set $\chi_s$ initially to zero and analyze its influence on the system later on. The incoherent processes are included in our description via Lindblad terms (see below). 

We assume that the metallic nanostructure is surrounded by vacuum and that its plasmonic response is dominated by a single Lorentzian-like cavity mode \citep{EstebanR,MKDezfouli}, characterized by an energy $\hbar\omega_{c}=1.722$ eV (wavelength $720$ nm), a damping rate $\hbar\kappa=200$ meV and an effective mode volume $V_{eff}=327$ nm$^3$. This volume is significantly below the diffraction limit but is large enough to accommodate many molecules and is well within values achievable with plasmonic structures \citep{RChikkaraddy,UrbietaM}.  We model this cavity mode within the canonical quantization scheme \citep{MKSchmidt} as a harmonic oscillator characterized by the Hamiltonian $H_{cav}=\hbar\omega_{c}a^{\dagger}a$, where $a^{\dagger}$ and $a$ are bosonic creation and annihilation operator of the plasmonic excitation, respectively. This model can be extended in a straightforward manner to a system with an arbitrary plasmonic response \citep{MKDezfouli,FrankeS}. In addition, the plasmonic cavity is excited by a laser of angular frequency $\omega_{l}$ as described by the Hamiltonian $H_{las}=i\hbar\Omega\left(a^{\dagger}e^{-i\omega_{l}t}-ae^{i\omega_{l}t}\right)$ in the rotating wave approximation (RWA). $\Omega=\frac{\kappa}{2}\sqrt{\frac{\epsilon_{0}V_{eff}}{2\hbar\omega_{c}}} K \sqrt{\frac{2I_{las}}{\epsilon_0 c}}$ is the coupling strength \citep{MKSchmidt,EstebanR} with $K = 206$  the maximum enhancement of the electric field amplitude at resonance, $I_{las}$ the laser intensity and $c$  the speed of light in vacuum \citep{ALombardi}.  

The molecular vibrations interact with the plasmonic mode via the molecular optomechanical coupling\citep{MKSchmidt,MKSchmidt-1,RoelliP} $H_{int}=-a^{\dagger}a\sum_{s}\hbar g_{s}\left(b_{s}^{\dagger}+b_{s}\right)$. The optomechanical coupling strength $g_{s}=f_s\sqrt{\frac{\hbar}{8\omega_s}}\frac{R_{s}\omega_{c}}{\varepsilon_0 V_{eff}}$ depends on the properties of the molecular vibrations, such as the Raman amplitude $R_{s}$, and those of the plasmon, such as the effective mode volume $V_{eff}$ \citep{MKSchmidt,MKSchmidt-1,RoelliP}. The factor $f_{s}\leq 1$  accounts for the position and orientation of the molecule and is one in the optimal case. In our system, we obtain $\hbar g_{s}= 0.084$ meV (using $f_{s}=1$) as a representative value, which is much stronger than the values in standard cavity optomechanical systems but is still relatively moderate in the context of molecular optomechanics \citep{MKSchmidt-1}. 

We model the dynamics of this lossy system with the standard quantum master equation \citep{HPBreuer} for the reduced density operator $\rho$ with the full Hamiltonian $H=H_{vib}+H_{cav}+H_{las}+H_{int}$ describing the coherent dynamics,  and the Lindblad superoperators incorporating incoherent processes, such as plasmon damping, phonon decay, thermal pumping and dephasing of molecular vibrations. Further on, we can simplify the solution of the master equation dramatically by adiabatically eliminating the plasmonic degree of freedom after linearizing the Hamiltonian $H_{int}$. As a result, we obtain an effective master equation for the reduced density operator $\rho_{v}$ of the molecular vibrations, which describes the dynamics associated with the vibrational (incoherent) noise operator $\delta b_{s}=b_{s}-\beta_{s}$, where $\beta_{s}=\mathrm{tr}\left\{ b_{s}\rho\right\}$ is the coherent amplitude. From the effective master equation we obtain the equations for the incoherent phonon population $n_s \equiv \bigl\langle\delta b_{s}^{\dagger}\delta b_{s}\bigr\rangle=\mathrm{tr}\left\{\delta b_{s}^{ \dagger}\delta b_{s}\rho_v \right\}$ and the noise correlations $c_{ss'} \equiv \bigl\langle\delta b_{s}^{\dagger}\delta b_{s'}\bigr\rangle$ ($s\neq s'$) as well as for  the noise amplitudes $\bigl\langle\delta b_{s}\bigr\rangle$ (or $\bigl\langle\delta b_{s}^\dagger \bigr\rangle$). We show in  Section S5.1 of the Supporting Information that the incoherent phonon population dominates over the coherent value  $\mid\beta\mid^2$ except for extremely intense lasers. The Raman spectra can be obtained by 
applying the quantum regression theorem \citep{PMeystre}, with the use of the equations for $\bigl\langle\delta b_{s}\bigr\rangle$ and $\bigl\langle\delta b_{s}^\dagger \bigr\rangle$. Significantly, we find that the Stokes and anti-Stokes scattering are affected not only by the dynamics of individual molecules but also by the molecule-molecule correlations.  More details on the derivation, the exact equations and the involved approximations are provided in the Methods Section and in Section S1 and S2 of the Supporting Information.

In our model the plasmonic mode acts as a structured reservoir and affects the vibrational dynamics by introducing: i)  a shift of the vibrational frequencies, in a similar way as the Lamb shift \citep{MOScully,ZYao}; ii)  plasmon-mediated coherent coupling between each pair  of molecules; iii) incoherent pumping of each vibration at rate $\Gamma_{ss}^{+}$; iv) incoherent damping at rate $\Gamma_{ss}^{-}$ and v) plasmon-mediated incoherent coupling of the vibrations at rate $\Gamma_{ss'}^{+}$, $\Gamma_{ss'}^{-}$ ($s\neq s'$). The last four sets of parameters (corresponding to iii-v) are particularly important for the phenomena discussed in this work. Here, the superscript "$+$" and "$-$" indicate that the parameters are evaluated with the spectral density of the plasmon at the Stokes $\omega_{l}-\omega_{s}$ ($+$) and anti-Stokes  $\omega_{l}+\omega_{s} $ ($-$) frequencies. The exact expressions of these parameters are given in Section S1.2 of the Supporting Information, but we note that all of them are proportional to the laser intensity. As an example, for two molecules with same vibrational frequency $\omega_s = \omega_{s'}$ but different optomechanical coupling to the plasmonic cavity $g_{s} \neq g_{s'}$, the rates $\Gamma_{ss'}^{+}$ and $\Gamma_{ss'}^{-}$ follow expressions similar to those of single molecules \citep{MKSchmidt,MKSchmidt-1} as
\begin{eqnarray}
\Gamma_{ss'}^{\pm} \propto
\frac{I_{las}   g_{s}  g_{s'} }{\left(\omega'_{c}-\omega_l\right)^{2}+\left(\kappa/2\right)^{2}}
\frac{\kappa  }{\left(\omega'_{c}-\omega_l\pm\omega_s\right)^{2}+\left(\kappa/2\right)^{2}}, \label{eq:optomechanical_rates}
\end{eqnarray}
with $\omega'_{c}= \omega_c - 2 \sum_s {\rm Re} \beta_s $ the plasmonic cavity frequency accounting for the slight shift  $2 \sum_s {\rm Re} \beta_s $ induced by the vibrations (analogue to the Lamb shift). The advantage of this approach is that it results in a closed set of equations solvable for many molecules (see Section S2 in Supporting Information). 

To conclude, Figure \ref{fig:structures-processes}b,c,d sketch an intuitive picture of the parameters $\Gamma_{ss'}^{+}$,$\Gamma_{ss'}^{-}$. More precisely, $\Gamma_{ss}^{+}$ ($\Gamma_{ss}^{-}$) corresponds to the transition rates from lower (higher) to higher (lower) vibrational states of {\it individual molecules} that are already present in the absence of any collective effect, as represented by the red (blue) arrows in Figure \ref{fig:structures-processes}b and discussed in previous works on single-molecule optomechanical SERS \citep{MKSchmidt,MKSchmidt-1}. On the other hand, $\Gamma_{ss'}^{+}$ and $\Gamma_{ss'}^{-}$ (with $s \neq s'$) emerge from the full collective situation and describe interference effects due to the plasmon-mediated interaction between molecule $s$ and $s'$ and introduce additional paths to excite or de-excite vibrational states, as represented by the red arrows in Figure \ref{fig:structures-processes}c and the blue arrows in Figure \ref{fig:structures-processes}d, respectively.

\begin{figure*}[!ht]
\begin{centering}
\includegraphics[scale=0.85]{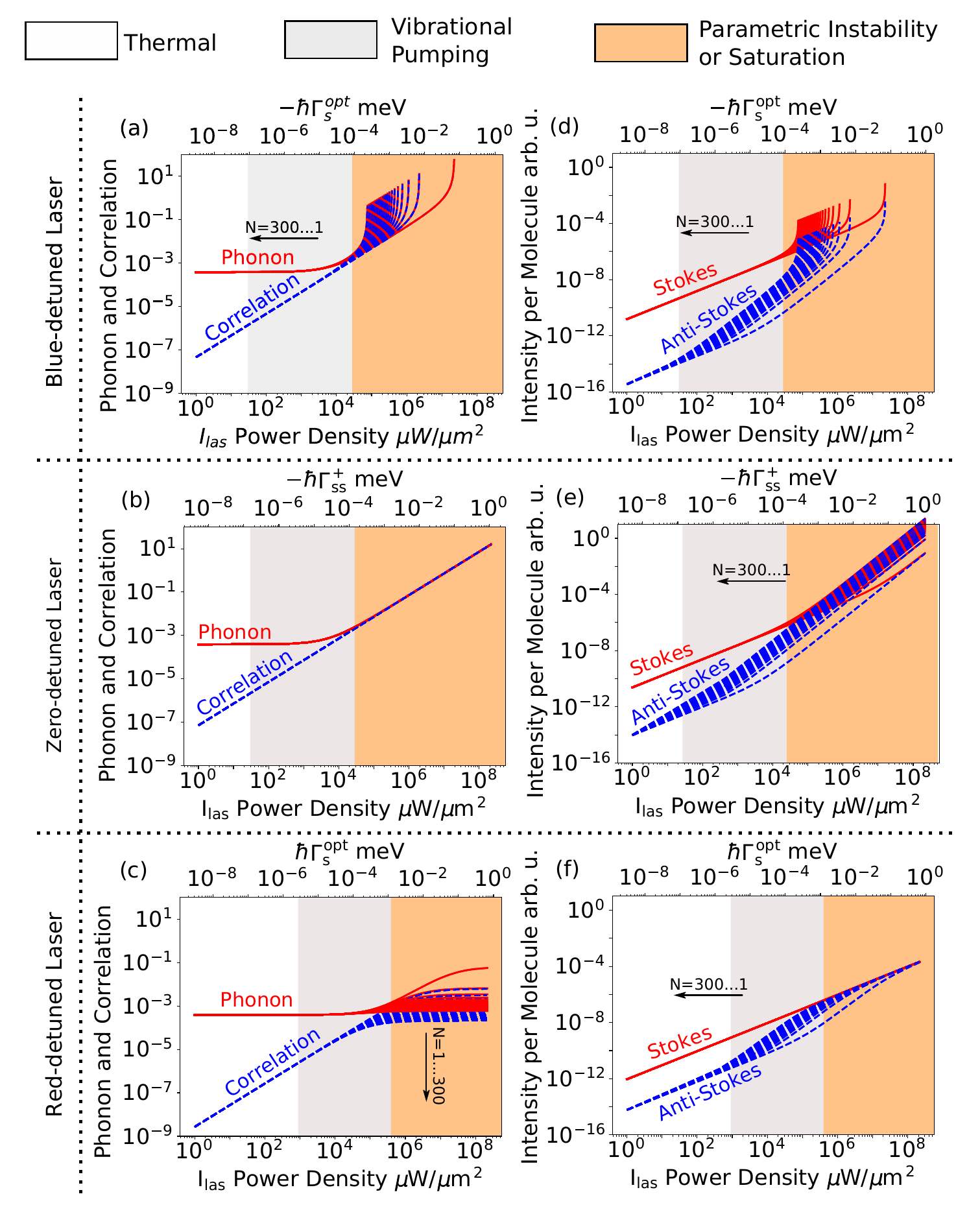}
\par\end{centering}
\caption{\label{fig:molecular-number-power} Influence of the number of identical molecules $N$ (increases as $1,10,20,...300$) on (a,b,c) the incoherent phonon population $n_s$ (red solid lines) and the noise correlation $c_{ss'}$ (blue dashed lines) and on (d,e,f) the frequency-integrated intensity of the Stokes (red solid lines) and anti-Stokes  (blue dashed lines) scattering \emph{per molecule} for increasing laser power density $I_{las}$. For reference we indicate in the upper axis the values of $\hbar\Gamma_{s}^{opt}$ or $\hbar\Gamma_{ss}^{+}$ for each $I_{las}$. Notice that the noise correlations are not defined for $N=1$. We consider systems illuminated with a laser (a,c) blue-, (b,d) zero-, and (e,f) red-detuned with respect to the shifted plasmon resonance [$\hbar \Delta \omega = \hbar (\omega_{l} - \omega_c')=$ $-236,0,236$ meV, respectively]. The white and gray shaded area indicate the thermal and vibrational pumping regime, respectively, and the orange shaded area signals the regime showing (a,d) parametric instability  or (c,f) saturation of the phonon population. We consider temperature $T=290 K$ and no homogeneous broadening $2\chi_{s}=0$. For other parameters see the text.}
\end{figure*}

\section{Collective Effects in Raman Scattering of Identical Molecules}

In this section, we focus on the simplest case of $N$ identical molecules and no homogeneous broadening, to identify under which conditions collective effects can emerge. Throughout the paper, the term {\it identical molecules} implies not only that the intrinsic properties of the molecular vibrations are the same, but also that they couple to the plasmon with same strength. To study this situation, we show in Figure \ref{fig:molecular-number-power} the incoherent phonon population and the noise correlation (a,b,c), and the (frequency-integrated) Stokes and anti-Stokes intensity (d,e,f) for different number of molecules $N$ ( $N=1,10,20,...300$) in the cavity, as a function of laser intensity $I_{las}$ (from $1$ $\mu W/\mu m^{2}$ to $10^{8}$ $\mu W/\mu m^{2}$, corresponding to $\hbar\Omega$ from $1.28$ meV to $12.8\times10^{3}$ meV). The Raman signal in (d,e,f) is normalized by $N$, i.e. the scattering per molecule, so that the collective effects are manifested by a change of this quantity with increasing $N$. We also consider different frequency detunings, $\Delta \omega = \omega_{l} - \omega_c'$, between the laser excitation and the plasmonic resonance to show that, consistent with the work on single molecules \citep{MKSchmidt}, different trends are observed when the strong laser illumination is blue-detuned ($\hbar \Delta \omega = {\rm 236 meV} >0$, Figure \ref{fig:molecular-number-power}a,d), zero-detuned ($\Delta \omega=0$, Figure \ref{fig:molecular-number-power}b,e) and red-detuned ($\hbar \Delta \omega = - 236 {\rm meV} <0$, Figure \ref{fig:molecular-number-power}c,f). To simplify the discussion, in all the calculations we fix the detuning  $\omega_{l} - \omega_c'$ with respect to the shifted plasmon resonance $\omega_c'$ by slightly shifting $\omega_l$ as the laser intensity is increased (see Section S5.5 in Supporting Information for results with $\omega_l$ fixed). 

To understand the results in Figure \ref{fig:molecular-number-power}, we derive analytical expressions by taking advantage of the permutation symmetry of identical molecules (see Section S3 in Supporting Information). We find that the intensity integrated over the Stokes and anti-Stokes lines can be expressed as
\begin{align}
 & I^{st}   \propto (\omega_l- \omega_s)^4 \Gamma_{ss}^{+} \left[N\left(1+n_s \right)+N\left(N-1\right)c_{ss'} \right],\label{eq:st-intensity}\\
 & I^{as} \propto (\omega_l + \omega_s)^4 \Gamma_{ss}^{-} \left[Nn_s +N\left(N-1\right)c_{ss'} \right],\label{eq:as-intensity}
\end{align}
where the factor $N$ and $N\left(N-1\right)$  originate from the sum over all identical molecules and all identical molecular pairs, respectively.  The latter leads to the emergence of  the collective effects when the noise correlations $c_{ss'}$ are sufficiently large. The factor $\omega^4$ originates from the frequency-dependence of dipolar emission and, for simplicity, is ignored in the following. 

Further, the noise correlation $c_{ss'}$ and the incoherent phonon population $n_s$ are given by
\begin{align}
c_{ss'}  & = \frac{\Gamma_{ss}^{+}-\Gamma_{s}^{opt}n_{s}^{th}}{\gamma_{s}+ N \Gamma_{s}^{opt}},\label{eq:correlation-analytical} \\
n_s & = n_{s}^{th}+\frac{\Gamma_{ss}^{+}-\Gamma_{s}^{opt}n_{s}^{th}}{\gamma_{s}+ N \Gamma_{s}^{opt}}=n_{s}^{th}+c_{ss'}
,\label{eq:phonon-number-analytical} 
\end{align}where eq \ref{eq:correlation-analytical} is defined for $N>1$  and we have defined the optomechanical damping rate $\Gamma_{s}^{opt}=\Gamma_{ss}^{-}-\Gamma_{ss}^{+}$ of single molecule  \citep{MKSchmidt,ALombardi,MAspelmeyer}  (with $\Gamma_{ss}^{\pm}\propto I_{las}$, see eq \ref{eq:optomechanical_rates}). The denominator, $\gamma_{s}+N\Gamma_{s}^{opt}$, in these expressions can be understood as a modification of the effective phonon decay rate due to the optomechanical damping rate.  We observe that the incoherent phonon population $n_s$ is equal to the noise correlation $c_{ss'}$ plus the thermal population $n_{s}^{th}$  and the noise correlation is built through the plasmon-mediated molecule-molecule interaction (notice $\Gamma_{ss'}^{\pm}=\Gamma_{ss}^{\pm}$ for identical molecules). In addition, we note that eqs \ref{eq:st-intensity}-\ref{eq:phonon-number-analytical} can also be derived within a collective oscillator model \cite{KipfT}, as detailed in Section S4 of the Supporting Information. 
 
Equations \ref{eq:optomechanical_rates}-\ref{eq:phonon-number-analytical} allow for understanding the collective effects revealed by Figure \ref{fig:molecular-number-power}. To this end, it is useful to distinguish three regimes as identified previously for single molecules \cite{MKSchmidt} (coded with different background colors in Figure \ref{fig:molecular-number-power}) for different laser intensity. 

\subsection{Weak and Moderate Laser Illumination: Thermal and Vibrational Pumping Regimes}
For weak and moderate laser intensity $I_{las}$, the vibrational damping and pumping rates given by eq \ref{eq:optomechanical_rates} for $s=s'$ (and thus the optomechanical damping rate) are small enough so that they do not affect the effective phonon decay rate for any $N$, i.e. $\gamma_{s}+\Gamma_{s}^{opt}N\approx \gamma_{s}$, and thus $c_{ss'}\approx \Gamma^+_{ss}/\gamma_s\propto I_{las}$ (for $n^{th}_s\ll 1$). As a consequence, the phonon population (eq \ref{eq:phonon-number-analytical}) adopts a very simple form, $n_s\approx n_s^{th}+\Gamma^+_{ss}/\gamma_s$, with a constant thermal population $n^{th}_{s}$ and a term $\Gamma^+_{ss}/\gamma_s\propto I_{las}$ proportional to the laser intensity that accounts for the creation of phonon by Stokes scattering, also known as vibrational pumping \citep{KKneipp-1,MaherR,ECLRu}. The noise correlations (eq \ref{eq:correlation-analytical}) follow the same linear dependence with $I_{las}$, but do not depend on the thermal population, i.e. $c_{ss'}\approx \Gamma^+_{ss}/\gamma_s\propto I_{las}$. 

We can now identify the first two regimes. When the laser intensity $ I_{las}$ is small enough, the thermal contribution dominates the incoherent phonon population, $n_s\approx n_s^{th}$, and we are thus in the so-called thermal regime (white-shaded area in Figure \ref{fig:molecular-number-power}). On the other hand, for moderate $I_{las}$  the incoherent phonon population is largely induced by the vibrational pumping rate $\Gamma^+_{ss}$ ($n_s\approx n_s^{th}+\Gamma^+_{ss}/\gamma_s$), and the system is in the vibrational pumping regime (grey shaded area in Figure \ref{fig:molecular-number-power}) \citep{MKSchmidt,ECLRu}. The noise correlations follow the same expression ($c_{ss'}\approx \Gamma^+_{ss}/\gamma_s\propto I_{las}$) for these weak and moderate laser intensities, but become significantly larger in the vibrational pumping regime. Notably, these expressions and the results in Figure \ref{fig:molecular-number-power}a-c demonstrate that neither the incoherent phonon number nor the noise correlation depends on the number of molecules,  and thus they do not manifest any collective effect neither in the thermal nor in the vibrational pumping regime. Furthermore, all the trends discussed here are independent of the laser detuning. 

We can now use the above analysis of the noise correlation and the incoherent phonon population to explain the evolution of the Raman signal in Figure \ref{fig:molecular-number-power}d,e,f for weak and moderate $I_{las}$. The number of molecules can affect the Raman signal per molecule due to its influence on the incoherent phonon population, or via the noise correlation that characterizes the molecule-molecule interaction (term scaling as $N^2$ in eqs \ref{eq:st-intensity} and \ref{eq:as-intensity}). Focusing first on the thermal regime, we found that both effects are negligible.  As a consequence, the integrated Stokes (red lines) and anti-Stokes (blue-lines) intensity normalized by $N$ in Figure \ref{fig:molecular-number-power} are independent of the number of molecules and they scale linearly with laser intensity, as observed directly from eqs \ref{eq:st-intensity},\ref{eq:as-intensity}, which become $I^{st}/N \propto \Gamma_{ss}^{+} \left(1+n_s \right)$ and $I^{as}/N  \propto \Gamma_{ss}^{-} n_s$ (notice $\Gamma^+_{ss}\propto I_{las}$ and $\Gamma^-_{ss}\propto I_{las}$). Thus, in the thermal regime the Raman scattering does not show the signature of collective effects.

On the other hand, in the vibrational pumping regime it is not possible to neglect the effect of the correlations $c_{ss'}$ on the Raman scattering or the linear dependence of the incoherent phonon population $n_s$ on the laser intensity $I_{las}$. As identified previously for single molecules \citep{MKSchmidt,FBenz}, the linear dependence of $n_s$ leads to a quadratic dependence of the anti-Stokes scattering on $I_{las}$. Further, eq \ref{eq:as-intensity} also indicates that as  $c_{ss'}$ becomes larger the integrated anti-Stokes intensity acquires a contribution that scales quadratically with the number of molecules $I^{as}\propto N^2$. The anti-Stokes intensity {\it per molecule} $I^{as}/N$ thus becomes dependent on the number of molecules for all laser detunings, as clearly revealed by the blue lines in Figure \ref{fig:molecular-number-power}d,e,f, which is the signature of the first collective effect. The $N^2$ scaling of the scattering corresponds to a superradiant behavior, similar to the superradiant Raman scattering of cold atoms \citep{JGBohnet,AndreevAV} or the single-photon superradiance from molecular or atomic electronic transitions \citep{NPVitaliy1,MOScully1}. This superradiant SERS can be understood as the result of the constructive interference of the anti-Stokes scattering from different molecules, which become in phase as a consequence of the increased noise correlations between the molecules in the vibrational pumping regime. Equivalently, we can attribute this effect to the $N^2$ paths of the anti-Stokes scattering shown in Figure \ref{fig:structures-processes}c,d that become relevant for sufficiently large noise correlations between different molecules.
  
It is also worthwhile to note that the term proportional to $N^2$ in eq \ref{eq:as-intensity} also scales with $I^2_{las}$ (for $c_{ss'}\propto I_{las}$). Thus, the quadratic dependence of the anti-Stokes signal with the laser intensity $I_{las}$ becomes easier to observe in Figure \ref{fig:molecular-number-power}d,e,f as $N$ is increased. Last, the Stokes intensity also acquires a contribution scaling as $N^2 I^2_{las}$ (eq \ref{eq:st-intensity}). The absolute strength of this superradiant scattering is similar to the one found for the anti-Stokes signal. However, this quadratic term adds to the linear contribution that dominates the scattering in the thermal regime (the terms proportional to $N$ in eqs \ref{eq:st-intensity}-\ref{eq:as-intensity}), which is significantly larger for the Stokes than for the anti-Stokes signal (because of $n_{th}\ll1$). Thus, it is harder to appreciate this quadratic contribution in the Stokes signal in the figure.  

We can quantify the different behavior of the Stokes and anti-Stokes signal more rigorously by estimating quantitatively the laser intensity above which the collective (or superradiant) $N^2$ scaling becomes relevant. The terms scaling with $N^2 I^2_{las}$ in eqs \ref{eq:st-intensity} and \ref{eq:as-intensity} become relevant when the conditions $N\Gamma_{ss}^{+}/\gamma_{s} = 1$ and  $N\Gamma_{ss}^{+}/\gamma_{s} = n_s^{th}$ are fulfilled for the Stokes and anti-Stokes intensity, respectively.  In the derivation of these conditions, we have used the simplified expressions $c_{ss'}\approx \Gamma_{ss}^{+}/\gamma_{s}$ and $n_s\approx n_s^{th} + \Gamma_{ss}^{+}/\gamma_{s}$. Because of $n_s^{th}\approx 10^{-3}$ the estimated threshold of the laser intensity is about three orders of magnitude smaller for the anti-Stokes scattering than for the Stokes scattering, as easily observed in Figure \ref{fig:molecular-number-power}b,e. In our system, the quadratic scaling of the anti-Stokes signal appears at $I_{las}=2 \times 10^4$ $\mu W /\mu m^2$ for a single molecule but could appear at only $60$ $\mu W /\mu m^2$ for $300$ molecules. The latter intensity is achievable with both CW\citep{FBenz} and pulsed laser\citep{ALombardi} and can be further reduced by working at low temperature (by reducing $n_s^{th}$).    

\subsection{Strong Laser Illumination}

In the following, we analyze the regime of strong laser illumination $I_{las}$ (orange-shaded area in Figure \ref{fig:molecular-number-power}). In contrast to previous regimes,  there are qualitative differences between the results obtained when the strong laser illumination is (a,d) blue-detuned, (b,e) zero-detuned and (c,f) red-detuned with respect to the plasmonic resonance. As discussed for single molecules \citep{MKSchmidt-1}, the key to understand these differences is that the optomechanical damping rate $\Gamma_{s}^{opt}$ becomes comparable to the intrinsic phonon decay $\gamma_{s}$, so that depending on the sign of  $\Gamma_s^{opt}$ the effective phonon decay $\gamma_{s}+\Gamma_{s}^{opt}N$ (i.e. the denominator in eqs \ref{eq:correlation-analytical},\ref{eq:phonon-number-analytical}) becomes larger or smaller than $\gamma_{s}$.  

\subsubsection{Blue-detuned Laser Illumination: Parametric Instability}

For blue-detuned illumination ($\hbar \omega_l= \hbar \omega_c'+236$ meV), the pumping rate $\Gamma_{ss}^{+}$ is larger than the damping rate $\Gamma_{ss}^{-}$, leading to a negative value of the optomechanical damping rate $\Gamma_{s}^{opt}<0$ (see eq \ref{eq:optomechanical_rates} and  Section S1.3 in the Supporting Information for the dependence of $\Gamma_{s}^{opt}$ on laser frequency). As a result, the effective phonon decay rate $\gamma_s+N\Gamma_{s}^{opt}<\gamma_s$ reduces with increasing laser intensity $I_{las}$ (notice $\Gamma_{s}^{opt}\propto I_{las}$) and this leads to larger incoherent phonon population and noise correlation (see eqs \ref{eq:correlation-analytical},\ref{eq:phonon-number-analytical}). For sufficiently large $I_{las}$, the (negative) optomechanical damping rate becomes comparable to $\gamma_s$ and the effective phonon decay rate approaches zero (i.e the denominator in eq \ref{eq:correlation-analytical},\ref{eq:phonon-number-analytical} becomes vanishingly small). In this case, the incoherent phonon population $n_s$ and the noise correlation $c_{ss'}$ become strongly non-linear with $I_{las}$ and finally diverge, as shown by the blue and red lines in Figure \ref{fig:molecular-number-power}a, respectively.  This divergence is known as parametric instability in cavity optomechanics\citep{MAspelmeyer},  and it is also seen in the Raman scattering\citep{MKSchmidt,RoelliP} (blue and red lines in Figure \ref{fig:molecular-number-power}d) because the Raman depends on $n_s,c_{ss'}$ (eqs \ref{eq:st-intensity} and \ref{eq:as-intensity}). In addition, we show in Section S5.2 of the Supporting Information that, in this regime, the Raman lines become also narrower and shifted. 

We can define the laser threshold intensity $I_{thr}$ to achieve the parametric instability as the value for which the Raman scattering diverges ($\Gamma_s^{opt}=-\gamma_s /N$). Taking into account that $\Gamma_s^{opt}=\Gamma_{ss}^{-}-\Gamma_{ss}^{+}\propto I_{thr}$ (eq \ref {eq:optomechanical_rates}), we obtain immediately that $I_{thr}$ is reduced as the number of molecules increases, i.e. the second collective effect, which is clearly shown in Figure \ref{fig:molecular-number-power}a,d. This collective effect can be understood as the consequence of coupling the plasmonic mode with the collective bright mode of the molecules, with a coupling strength that scales\citep{RoelliP} as $\sqrt{N}g_s$ (Section S4 in the Supporting Information). $I_{thr}$ is about $5 \times 10^7 \mu W/\mu m^2$ for a single molecule, but reduces to $1.5\times 10^5 $ $\mu W/\mu m^2$ for $300$ molecules. We note that such large intensities are difficult to reach in practise and can lead to effects not included here  (for example, it may even destroy the molecular sample \citep{ALombardi}).  Furthermore, for illumination larger than about $1.8 \times 10^6$ $ \mu W/\mu m^2$ the coupling strength with the driving laser, $\Omega$, becomes comparable to the plasmon frequency $\omega_c$ and the validity of the RWA approximation used in our model is compromised. 

\subsubsection{Zero-detuned Laser Illumination: Superradiant Stokes Scattering}

In Figure \ref{fig:molecular-number-power}b,e, where the laser is resonant with the plasmonic mode ($\omega_l=\omega'_c$), the vibrational damping and pumping  rate are equal, i.e.  $\Gamma_{ss}^{-} = \Gamma_{ss}^{+}$, leading to a vanishing optomechanical damping rate $\Gamma_{s}^{opt}=0$ (eq \ref{eq:optomechanical_rates}). The outcome of this situation is that the response maintains the trends in the vibrational pumping regime (where $\Gamma_{s}^{opt}$ is negligible because of the small laser intensity $I_{las}$): the noise correlations and the incoherent phonon populations exhibit identical linear scaling with the laser intensity  $c_{ss'}\approx n_s \approx \Gamma_{ss}^{+}/\gamma_{s}\propto I_{las}$  (Figure \ref{fig:molecular-number-power}b and eqs \ref{eq:phonon-number-analytical},\ref{eq:correlation-analytical}), and the integrated Stokes and anti-Stokes signal increase quadratically with both the laser intensity and the number of molecules (Figure \ref{fig:molecular-number-power}e and eqs \ref{eq:st-intensity},\ref{eq:as-intensity}). It is indeed in the situation with zero-detuned laser illumination  where the superradiant quadratic scaling of the Stokes scattering is easier to appreciate. We discuss in Section S5.4 of the Supporting Information how extra features appear for very large laser intensities if the laser frequency is detuned to the original plasmonic cavity frequency $\omega_c$ instead of the shifted one $\omega'_c$. 

\subsubsection{Red-detuned Laser Illumination: Phonon Saturation}

If we illuminate the system with a red-detuned laser  ($\hbar\omega_l =\hbar \omega_c'-236$ meV), the vibrational damping rate is larger than the pumping rate  $\Gamma_{ss}^{-}>\Gamma_{ss}^{+}$. Thus, the optomechanical damping rate is thus positive $\Gamma_{s}^{opt}>0$, and  the effective phonon decay rate becomes larger $\gamma_s+N\Gamma_{s}^{opt}>\gamma_s$. For sufficiently strong illumination, the larger loss compensates the linear increase of the vibrational pumping rate with increasing laser intensity. As a result, the incoherent phonon population and noise correlation saturate towards $n_s^{th}+\Gamma_{ss}^{st}/(N \Gamma_{s}^{opt})$ and $\Gamma_{ss}^{st}/(N \Gamma_{s}^{opt})$, respectively (Figure \ref{fig:molecular-number-power}c), which can be achieved by considering the limit of large laser intensity in eqs \ref{eq:correlation-analytical},\ref{eq:phonon-number-analytical} (with $\Gamma_{ss}^{st}\propto I_{las}$,$\Gamma_{s}^{opt}\propto I_{las}$ and $n_s^{th}\ll 1$). Because of the saturation, the Stokes and anti-Stokes signal become again linearly dependent on the laser intensity (Figure \ref{fig:molecular-number-power}f and eqs \ref{eq:st-intensity},\ref{eq:as-intensity}). 

In a similar manner as for the parametric instability, the saturation  becomes significant for $ N\Gamma_{s}^{opt} \approx \gamma_s$, so that a larger number of molecules allow for reaching this effect for weaker (but still very strong) laser intensity. This effect is again due to the coupling with the collective bright mode of the molecules (Section S4 in the Supporting Information). Furthermore, the expressions derived above indicate also that larger $N$ leads to a $\approx 1/N$ decrease of the saturated value of the noise correlation, as shown by Figure \ref{fig:molecular-number-power}c. The incoherent phonon population remains nonetheless larger than $n_s^{th}$. 

The dependence of the noise correlation and the phonon population on $N$ is a signature of collective effects. However, we see that the integrated Stokes intensity per molecule does not depend on $N$ for any laser intensity and  the anti-Stokes signal per molecule becomes independent of $N$ for very strong illumination (Figure \ref{fig:molecular-number-power}f). These behaviors occur because the superradiant contribution to the SERS signal that scales as $N^2$ is compensated by the $ 1/N$ decrease of the incoherent phonon and the noise correlation, so that the signal becomes proportional to the number of molecules (i.e. constant after normalization by $N$). In fact, the presence of collective effects for strong $I_{las}$ and red-detuned illumination may be more easily demonstrated by studying the change of the Raman lines, which would become broader and shifted as the laser becomes more intense (see Section S5.2 in the Supporting Information). 

Last, we note that in typical cavity-optomechanical systems, characterized by low mechanical frequencies and thus large thermal population $n_s^{th}$, a positive value of $\Gamma_{s}^{opt}$ is often exploited to reduce the phonon population below the thermal value, i.e. to cool the sample\citep{MAspelmeyer}. In contrast, we have shown (Figure \ref{fig:molecular-number-power}f)  that in our system, which exhibits a much larger vibrational frequency, the incoherent phonon population remains always larger than $n_s^{th}$. This difference occurs because the phonon decay rate of the thermally activated molecules equals $\Gamma_{s}^{opt}n_{s}^{th}$ (corresponding to the negative term in the numerator of eqs \ref{eq:correlation-analytical},\ref{eq:phonon-number-analytical}), which scales with the thermal phonon population. For a large $n_{s}^{th}$, as typical in cavity-optomechanics, this decay rate will dominate over the incoherent pumping rate $\Gamma_{ss}^{+}$ and thus the cooling can occur. In contrast, in our system  $\Gamma_{ss}^{+}$ remains the larger of the two contributions and thus the system is rather heated, i.e.  $n_s>n^{th}_s$. Thus,  when the laser is red-detuned with respect to the plasmon, we refer to the regime of large intensities as the saturation regime, instead of the cooling regime as often referred in cavity optomechanics.

\begin{figure}
\begin{centering}
\includegraphics[scale=0.5]{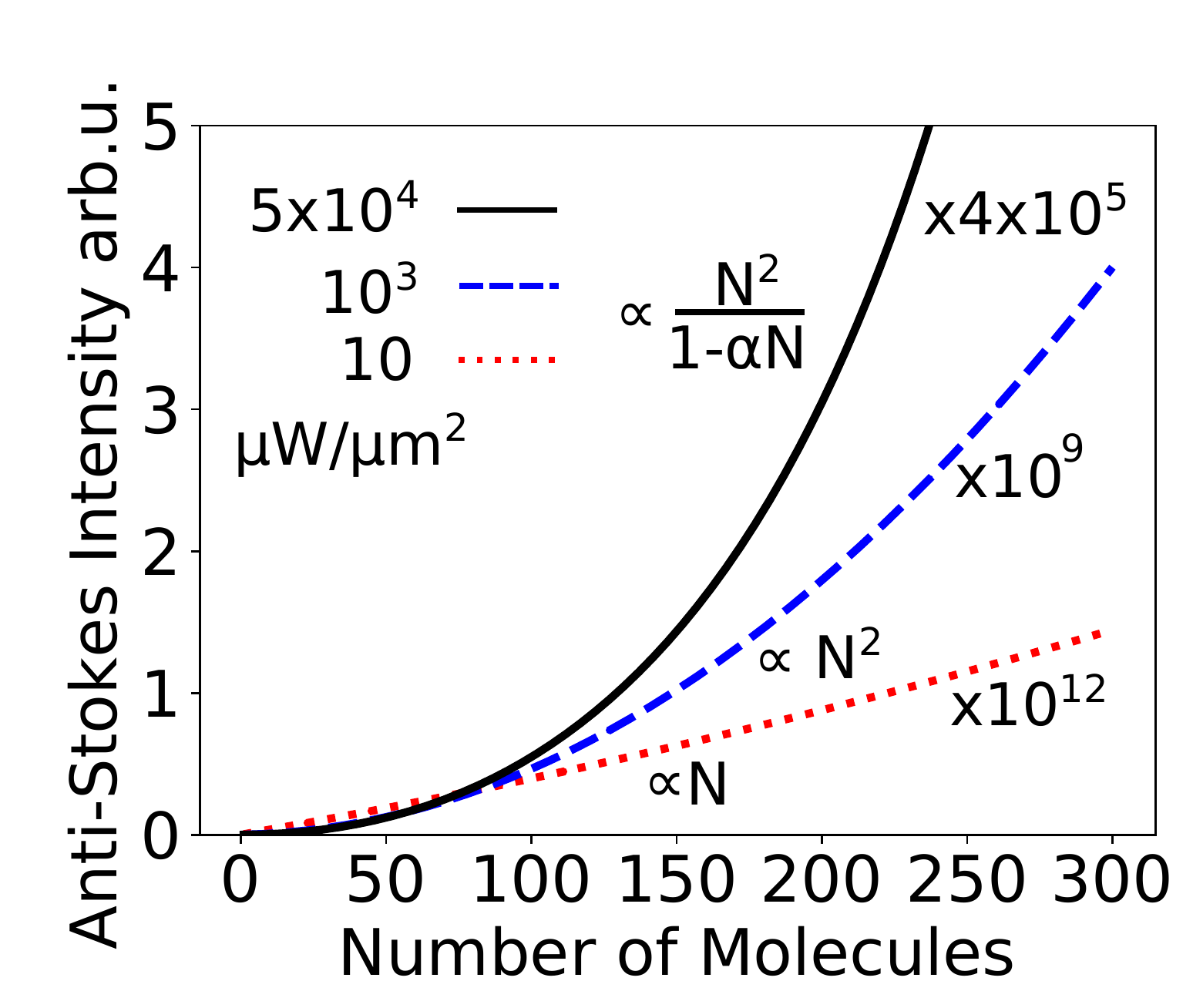}
\par\end{centering}
\caption{\label{fig:molecular-number} Integrated anti-Stokes intensity as a function of the number of molecules $N$ (from $1$ to $300$) for molecules in a plasmonic cavity (as in Figure \ref{fig:molecular-number-power}) illuminated by a blue-detuned laser [$\hbar (\omega_{l} - \omega_c') = 236$ meV] of the intensity (red dotted line) $I_{las}= 10$ $\mu$W/$\mu$m$^2$, (blue dashed line) $I_{las}= 10^3$ $\mu$W/$\mu$m$^2$ and (black solid line) $I_{las}= 5\times10^4$ $\mu$W/$\mu$m$^2$. The signal is scaled as indicated in the figure and we assume no homogeneous broadening $2\chi_{s}=0$, and temperature $T=290$ K.}
\end{figure}

\subsection{Collective Effects Landscape}

We summarize the collective effects in Figure \ref{fig:molecular-number}, where the integrated anti-Stokes signal is shown as a function of the number of molecules $N$ for blue-detuned laser illumination and different laser intensities $I_{las}$. The results for the  Stokes signal under a blue-detuned laser illumination, and for the anti-Stokes signal under red- and zero-detuned illumination are shown in Section S5.4 of the Supporting Information. Here, we plot the total signal from all the molecules and do not normalize them by $N$. For small $I_{las}$ (the thermal regime, red dotted line) the total signal scales linearly with $N$, as it should occur for independent molecules, which indicates the absence of collective effects. For intermediate $I_{las}$ (the vibrational pumping regime, blue dashed line), we find the first collective effect, namely the quadratic scaling of the anti-stokes SERS signal with $N$ that we have explained as a superradiant phenomenon. Last, for the strongest laser intensity  $I_{las}$ (the parametric instability regime, solid black line), the signal increases faster than $N^2$, which is a manifestation of the second collective effect, namely the influence of $N$ on the effective phonon decay rate and thus on the threshold laser intensity to achieve the parametric instability.  This can be understood as the consequence of as the result of coupling with the bright collective mode with a coupling strength $\sqrt{N}g_s$ or the multiple paths of the Stokes scattering in Figure \ref{fig:structures-processes}c.  More precisely, in this case,  the signal scales as  $N^2/(1-\alpha N)$, with $\alpha = |\Gamma_s^{opt}|/\gamma_s$ a constant proportional to $I_{las}$. Thus, for fixed $I_{las}$, an increasing number of molecules brings the laser illumination closer to the condition $\alpha N=1$ to achieve the parametric instability. In addition, in Section 5.3 of the Supporting Information we examine how the collective effects are affected by the Raman activity of the molecule.

\begin{figure*}
\begin{centering}
\includegraphics[scale=0.40]{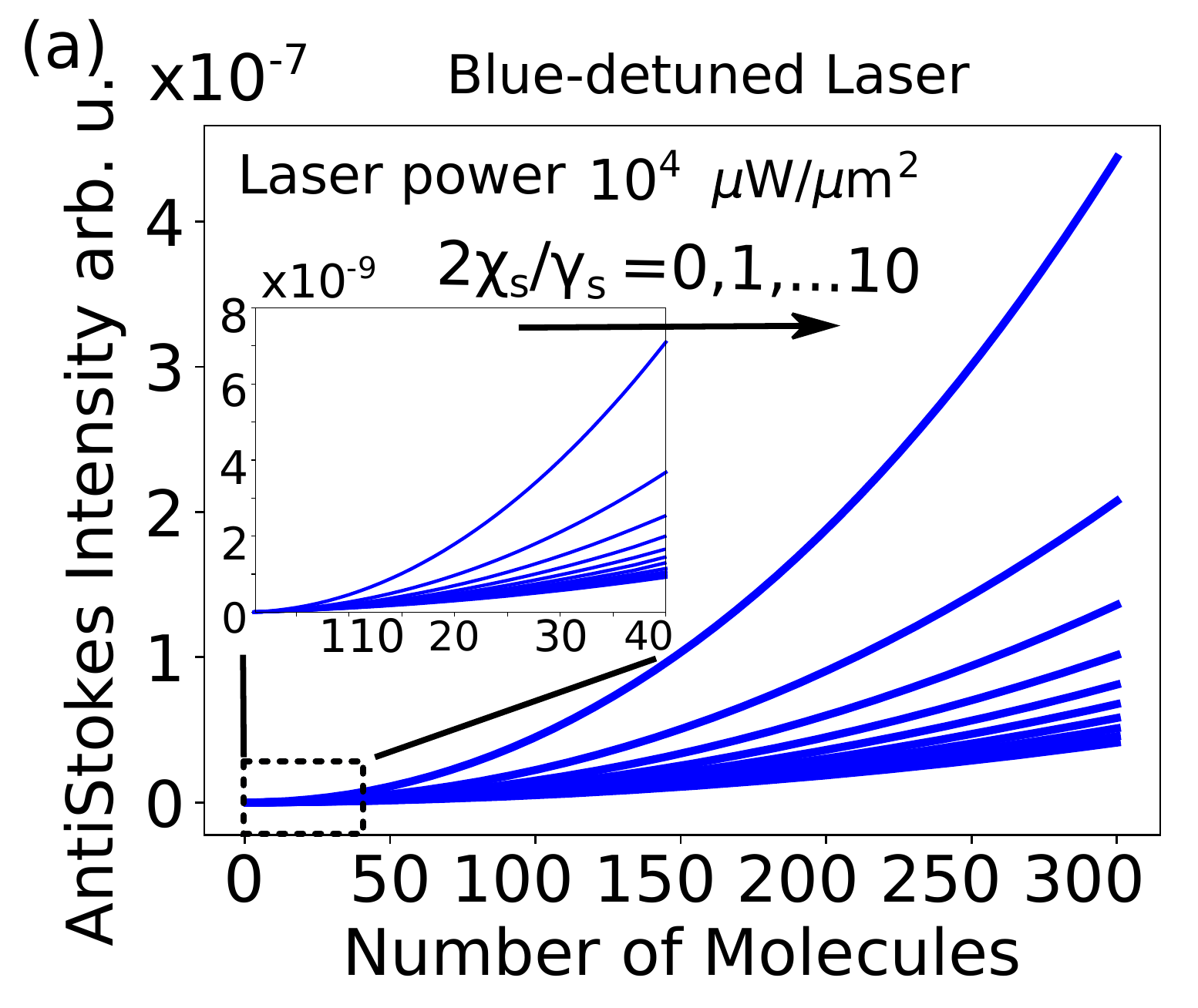}
\includegraphics[scale=0.40]{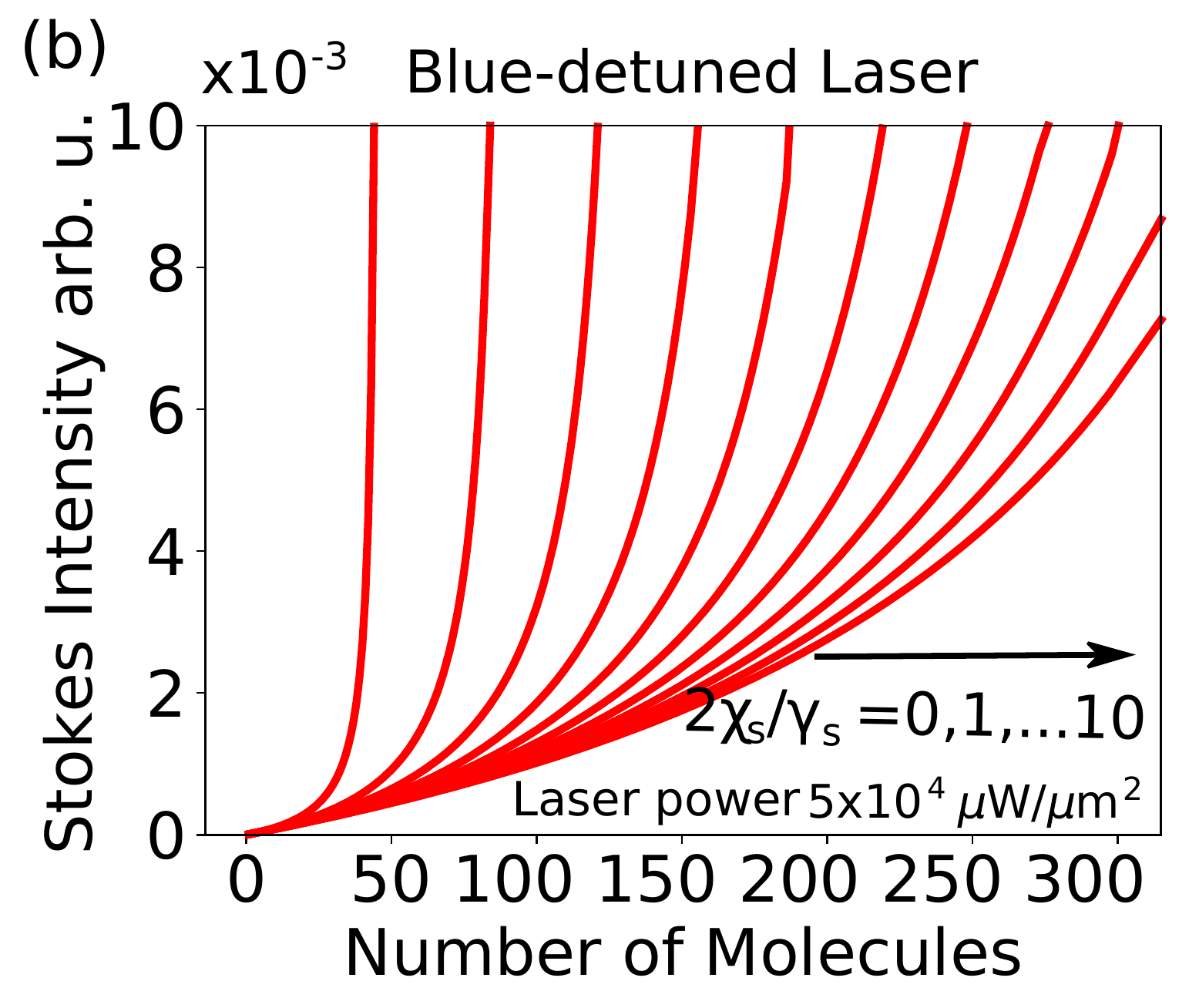}
\par\end{centering}
\centering{}\caption{\label{fig:homogeneous-dephasing} Influence of homogeneous broadening  $2\chi_{s}$ (from $\gamma_{s}$ to $10\gamma_{s}$) on the integrated (a) anti-Stokes and (b) Stokes intensity  from $N=1$ to $300$  molecules. The system is illuminated by a blue-detuned laser [$\hbar(\omega_{l} - \omega_c') = 236$ meV] of intensity (a) $I_{las}=10^{4}\mu W/\mu m^{2}$ and (b) $I_{las}=5\times10^{4}\mu W/\mu m^{2}$  intensity. In all panels,we assume $\hbar \gamma_{s}=0.07$ meV and temperature $T=290$ K.  The inset in (a) is a zoom-in to the region of small $N$.}
\end{figure*}
 
\section{Contributions to the Raman Linewidth}

We have so far focused on a simple system where the molecules are identical and the only loss mechanism experienced by them is the phonon decay. In real experiments, however, the situation can be more complex. For example, the molecules can show small variations of vibrational frequencies (inhomogeneous broadening) and the width of the Raman lines can be affected not only by the phonon decay but also by other phenomena, such as spectral wandering and collision-induced pure dephasing, (which leads to homogeneous broadening \citep{YZhao}).  To our knowledge, it is still not well understood to what extent the homogeneous and inhomogeneous broadening influence the vibrational dynamics. However, it has been shown that they can affect strongly the collective response of atomic ensembles \citep{AndreevAV}. Thus, it is important to examine their impact on the collective effects of SERS.

\subsection{Influence of Homogeneous Broadening \label{sec:homdephasing}}

We model the homogeneous broadening by a Lindblad term in the master equation with a dephasing rate $\chi_{s}$ (see Methods Section). Considering again identical molecules and exploiting the permutation symmetry, we obtain 
\begin{equation}
c_{ss'} =\frac{\gamma_{s}}{\gamma_{s}+2\chi_{s}}\frac{\Gamma_{ss}^{+}-\Gamma_{s}^{opt}n_{s}^{th}}{\gamma_{s}+\Gamma_{s}^{opt}\left(2\chi_{s}+N\gamma_{s}\right)/\left(\gamma_{s}+2\chi_{s}\right)},\label{eq:correlation-total_2}
\end{equation}
\begin{equation}
n_s =n_{s}^{th}+\frac{\Gamma_{ss}^{+}-\Gamma_{s}^{opt}n_{s}^{th}}{\gamma_{s}+\Gamma_{s}^{opt}\left(2\chi_{s}+N\gamma_{s}\right)/\left(\gamma_{s}+2\chi_{s}\right)},\label{eq:phonon-number-total_2}
\end{equation}
for the noise correlation and the incoherent phonon population, respectively. Comparing these equations with eqs \ref{eq:correlation-analytical} and \ref{eq:phonon-number-analytical}, we observe a change in the denominator that can be understood as a reduction of the effective number of molecules contributing to the collective response from $N$ to $N_{eff} = (2\chi_s + N\gamma_s)/(\gamma_s + 2\chi_s)$. In addition, the noise correlation is also reduced by $\gamma_{s}/\left(\gamma_{s}+2\chi_{s}\right)$ with respect to the value for $2\chi_s = 0$. The integrated Stokes and anti-Stokes intensity can be computed with eqs \ref{eq:st-intensity} and \ref{eq:as-intensity}, which do not depend explicitly on $\chi_s$, so that they are affected by the pure dephasing only due to their dependence on the noise correlation and incoherent phonon population. The derivation of all the expressions can be found in Section S3 in the Supporting Information.

We illustrate next the effect of the homogeneous broadening $\chi_s$ on the collective effects of systems under blue-detuned laser illumination and with the phonon decay rate $\hbar \gamma_s=0.07$ meV. Figure \ref{fig:homogeneous-dephasing}a demonstrates that, for moderate 
 illumination $I_{las}=10^4$ $\mu W/\mu m^2$ with a blue-detuned laser (the vibrational pumping regime), the evolution of the integrated anti-Stokes signal is dominated by the superradiant contribution that scales quadratically with the number of  molecules ($\propto N^2$). This contribution becomes weaker for increasing $\chi_s$ but remains significant for all values considered, which  indicates that the superradiant anti-Stokes scattering is robust to the homogeneous broadening. We can quantify this statement by inserting eqs \ref{eq:correlation-total_2} and \ref{eq:phonon-number-total_2} into eq  \ref{eq:as-intensity} to obtain the term scaling with $N^2$ as approximately $N^{2}\Gamma_{ss}^{-}\Gamma_{ss}^{+}/\left(\gamma_{s}+2\chi_{s}\right)$. In addition, this expression also indicates a quadratic scaling with laser intensity because $\Gamma_{ss}^{-} \propto I_{las}$ and $\Gamma_{ss}^{+} \propto I_{las}$.

Figure \ref{fig:homogeneous-dephasing}b shows that the larger the homogeneous broadening $2\chi_s$ the more molecules are required to observe the divergent Stokes signal at  strong laser illumination (here $I_{las}=5\times 10^4$ $\mu W/\mu m^2$), i.e. the parametric instability. The increase of number of molecules, however, is moderate and progressive. More precisely, the number of molecules  required to reach the divergence is  $N\approx \gamma_s (\gamma_s + 2\chi_s)/(\gamma_s |\Gamma_s^{opt}|)$ (obtained by setting the denominator in eq \ref{eq:phonon-number-total_2} as zero and assuming large $N$).  In short, the collective effect is robust to the homogeneous broadening.

\begin{figure*}[!ht]
\begin{centering}
\includegraphics[scale=0.35]{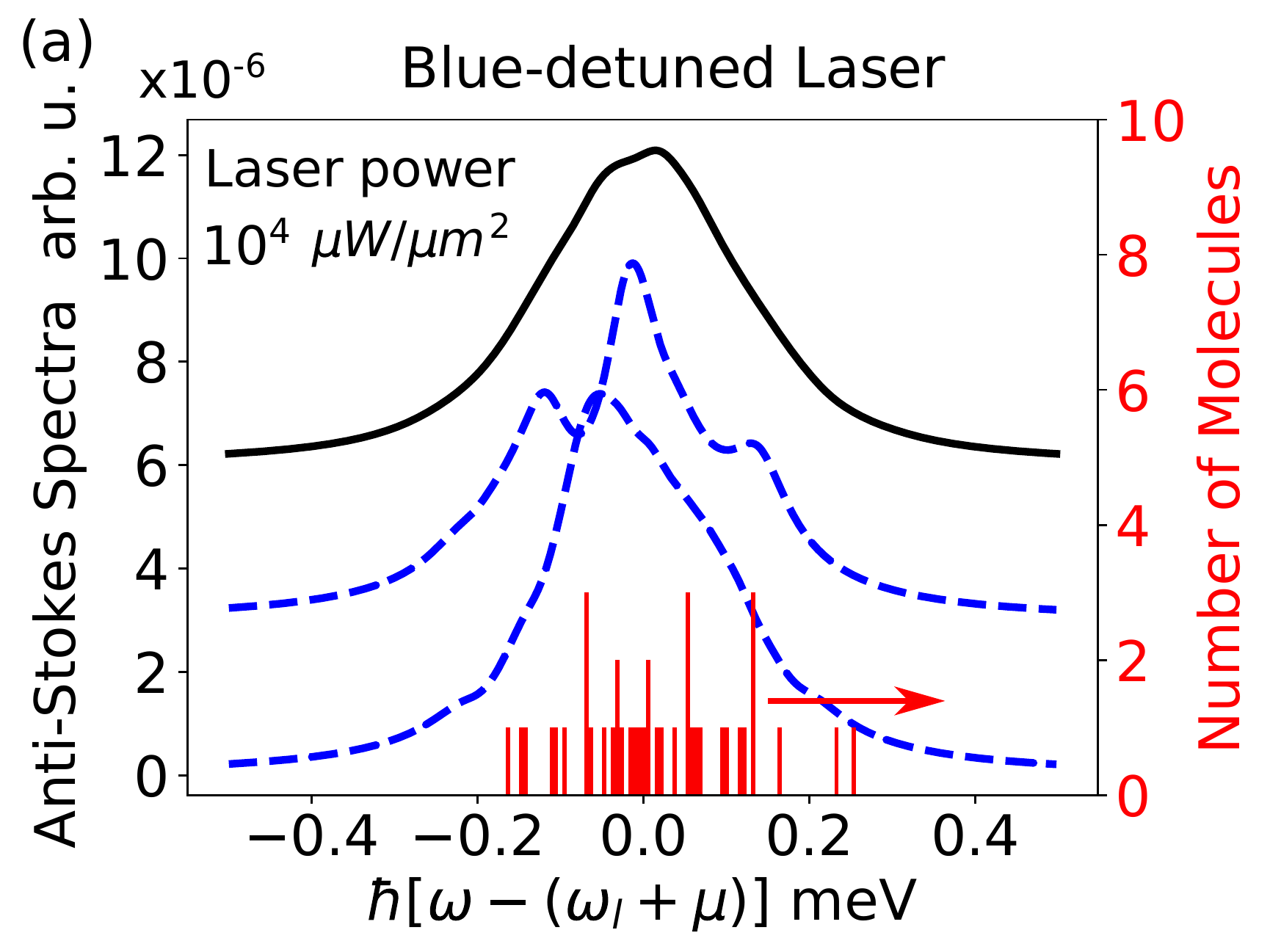} 
\includegraphics[scale=0.35]{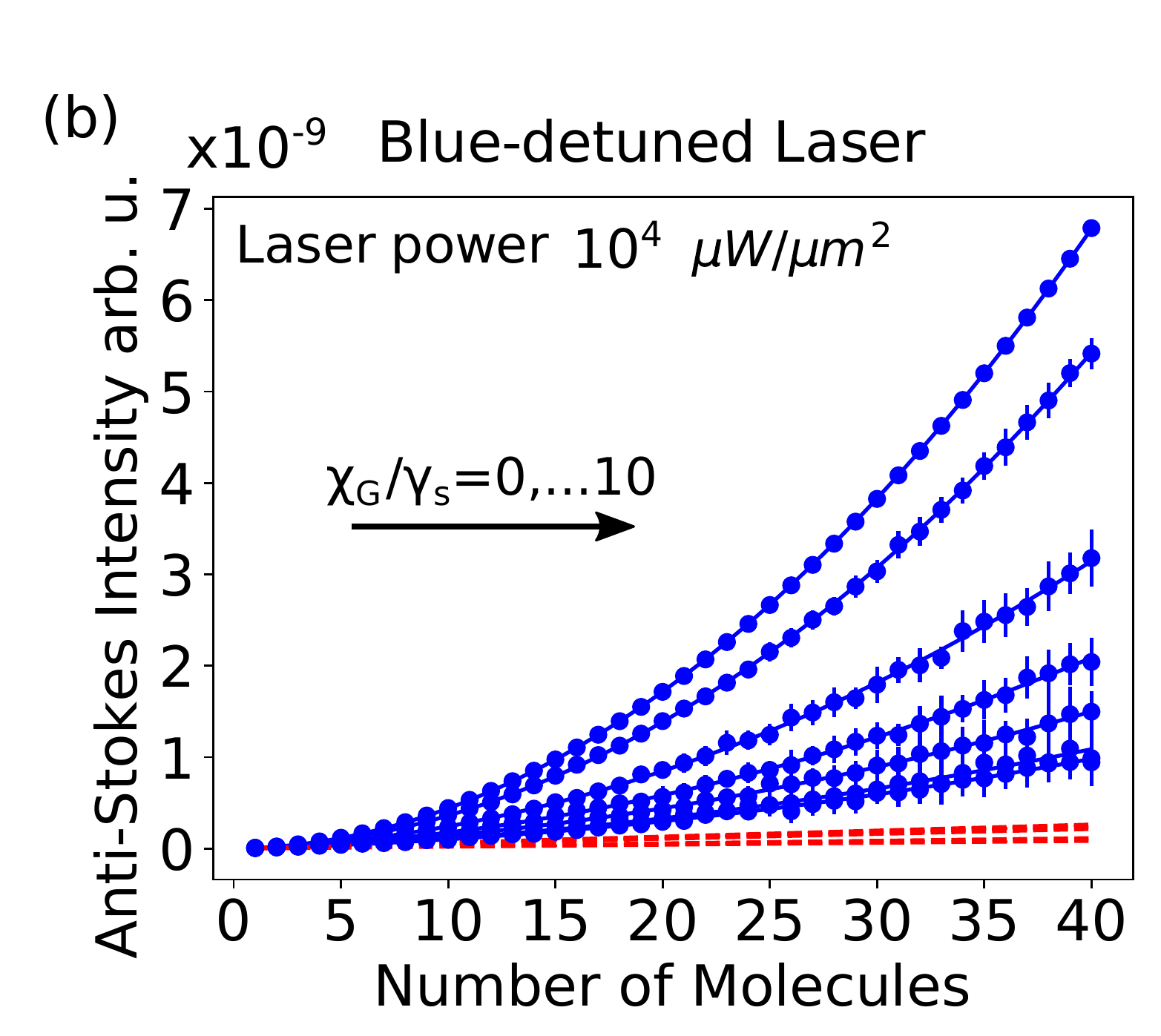}
\includegraphics[scale=0.35]{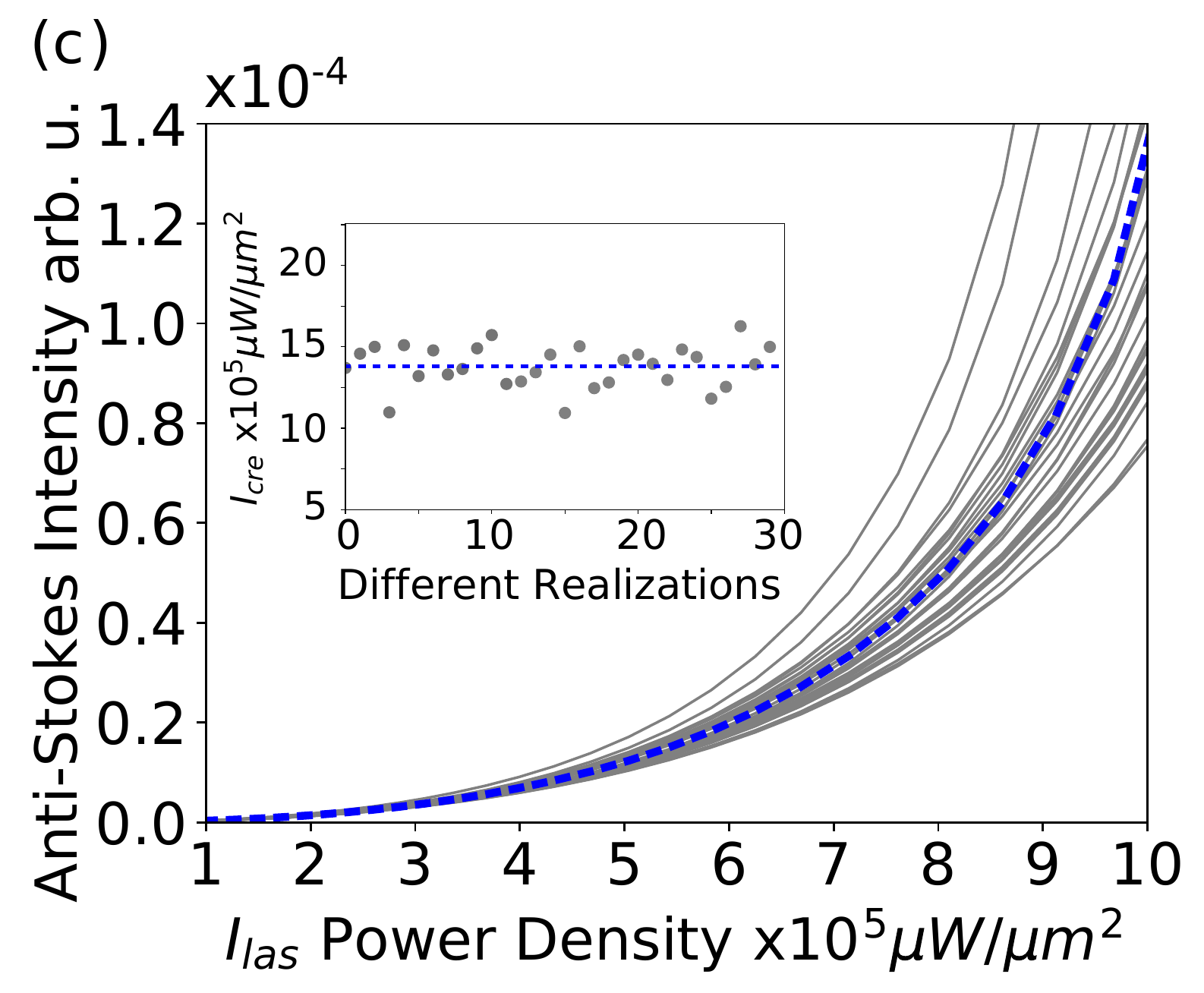}
\par\end{centering}
\centering{}\caption{\label{fig:inhomogeneous-dephasing} Influence of inhomogeneous broadening $\chi_{G}$ on the anti-Stokes spectra. (a) shows two examples of spectra (blue dashed curves, left axis) and the average over forty spectra (black solid curve, left axis) as well as one histogram of vibrational frequencies (red bars, right axis). The spectra are shifted vertically for visibility. (b) shows mean (blue dots) and 1-sigma standard deviation (blue bars) of the integrated anti-Stokes intensity versus number of molecules (from $N=1$ to $40$) for increasing $\chi_{G}$ (varying as $0,1,3,5,7,9,10 \gamma_{s}$). The statistics are obtained with thirty realizations. The average is fitted to $\alpha N + \beta N^2$ and the linear term $\alpha N$ is shown with red dashed lines. In (a) and (b) we consider moderate laser intensity $I_{las} = 10^{4}\mu W/\mu m^{2}$. (c) shows the integrated anti-Stokes intensity from $40$ molecules and $\hbar \chi_{G}=0.21$ meV as a function of $I_{las}$  (close to the parametric instability). The gray solid curves show thirty realizations and the blue dashed curve shows the average. The inset shows the laser threshold intensity $I_{thr}$ of the parametric instability. In all cases, we consider illumination by a  blue-detuned laser [$\hbar (\omega_l - \omega_c') = 236$ meV] and temperature $T=290$ K.}
\end{figure*}

\subsection{Influence of Inhomogeneous Broadening \label{sec:inhomdephasing}}

We consider next the inhomogeneous broadening due to slight variations of the vibrational frequencies in different molecules, which could be caused,  for example, by different Stark shifts induced by the local environment or by different chemical interaction with the metal atoms of the plasmonic system \citep{PGEtchegoin}. We model the inhomogeneous broadening with a Gaussian distribution $\left[\sigma\sqrt{2\pi}\right]^{-1}\exp\left\{ -\left(\omega_{s}-\mu\right)^{2}/\left(2\sigma^{2}\right)\right\}$ of the vibrational frequencies $\omega_s$, characterized by the mean $\hbar\mu=196.5$ $\mathrm{meV}$ and the standard deviation $\sigma$ (corresponding to a linewidth of the distribution $\chi_{G}=2\sqrt{2\mathrm{ln}2}\sigma$).  An example of the random frequency distribution is shown by the histogram in Figure \ref{fig:inhomogeneous-dephasing}a.  We compute the Raman spectra by solving numerically the equations given in Section S2 of the Supporting Information for systems with up to $40$ molecules. The blue dashed lines in Figure \ref{fig:inhomogeneous-dephasing}a show two examples of the anti-Stokes Raman spectra for $I_{las} = 10^{4}\mu W/\mu m^{2}$ and  $\chi_G=3\gamma_s$ ($\hbar \gamma_s = 0.07$ meV, spectra shifted for visibility). The average of such anti-Stokes spectra over thirty simulations is shown by the solid line and it shows a smooth single peak similar to those measured in typical experiments. For the parameters considered in Figure \ref{fig:inhomogeneous-dephasing}, the linewidth of the spectra is approximately $\chi_G$.

We show in Figure \ref{fig:inhomogeneous-dephasing}b the dependence of the integrated anti-Stokes spectra with the number of molecules $N$ for different values of  inhomogeneous broadening $\chi_{G}$ (relative to the phonon decay rate $\gamma_s=0.07$ meV). We show the mean and standard deviation of thirty realizations for laser illumination $I_{las}=10^4$ $\mu W/\mu m^2$ (in vibrational pumping regime). The standard deviation is relatively small and thus for given $\chi_{G}$ the results depend only weakly on the exact random distribution of the vibrational frequencies. As for the homogeneous broadening, increasing $\chi_{G}$ reduces the mean intensity but does not affect the quadratic scaling of the signal. To be more precise, we fit the mean intensity to $\alpha N + \beta N^2$ (with $\alpha,\beta$ as fitting parameters) and find that the linear contribution $\alpha N$ (red dashed lines) is negligible. The decrease of the quadratic contribution  with increasing $\chi_{G}$ is moderate and it becomes about three times smaller when the ratio $\chi_{G}/\gamma_s$ increases from zero (i.e. identical vibrational frequencies) to three. The latter $\chi_{G}/\gamma_s$ ratio is close to the value reported experimentally in ref \citep{PGEtchegoin}. We thus conclude that the superradiant $N^2$ scaling can survive in the presence of significant inhomogeneous broadening.

Last, Figure \ref{fig:inhomogeneous-dephasing}c shows the integrated anti-Stokes intensity for $N=40$ molecules, the inhomogeneous broadening $\chi_{G}/\gamma_s=3$ and increasing laser intensity $I_{las}$. The strongest intensities considered are close to the value leading to the divergent SERS signal (the parametric instability). The gray lines show thirty realizations and the solid blue line their average. The standard deviation of the results becomes larger as $I_{las}$ increases, but the qualitative behavior remains the same for all realizations. To characterize the variation quantitatively, we fit the results with the expression $I_{las}^2/(1- I_{las}/I_{thr})$, with $I_{thr}$ the threshold intensity at which the signal diverges. We plot the resulting $I_{thr}$ in the inset of Figure \ref{fig:inhomogeneous-dephasing}c. We obtain an average threshold of $1.38\times 10^6$ $ \mu W/\mu m^2$ and the threshold for different realizations differ from this average by a maximum of $\pm 21$ percent. In conclusion, we have seen that the strength of the collective effects is reduced by increasing inhomogeneous broadening, but that the change is gradual and moderate.

\begin{figure*}[!ht]
\begin{centering}
   \includegraphics[scale=0.40]{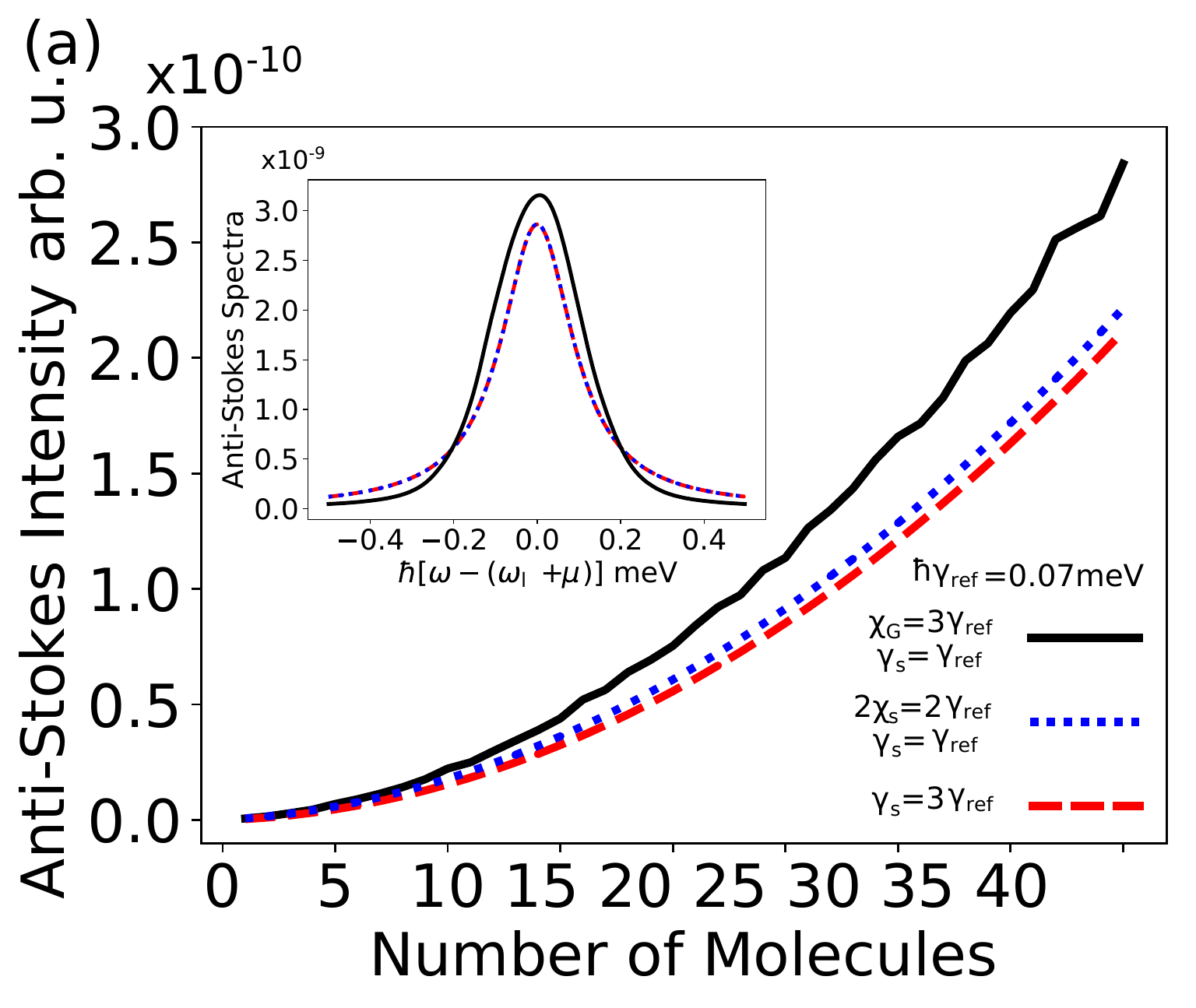} \includegraphics[scale=0.40]{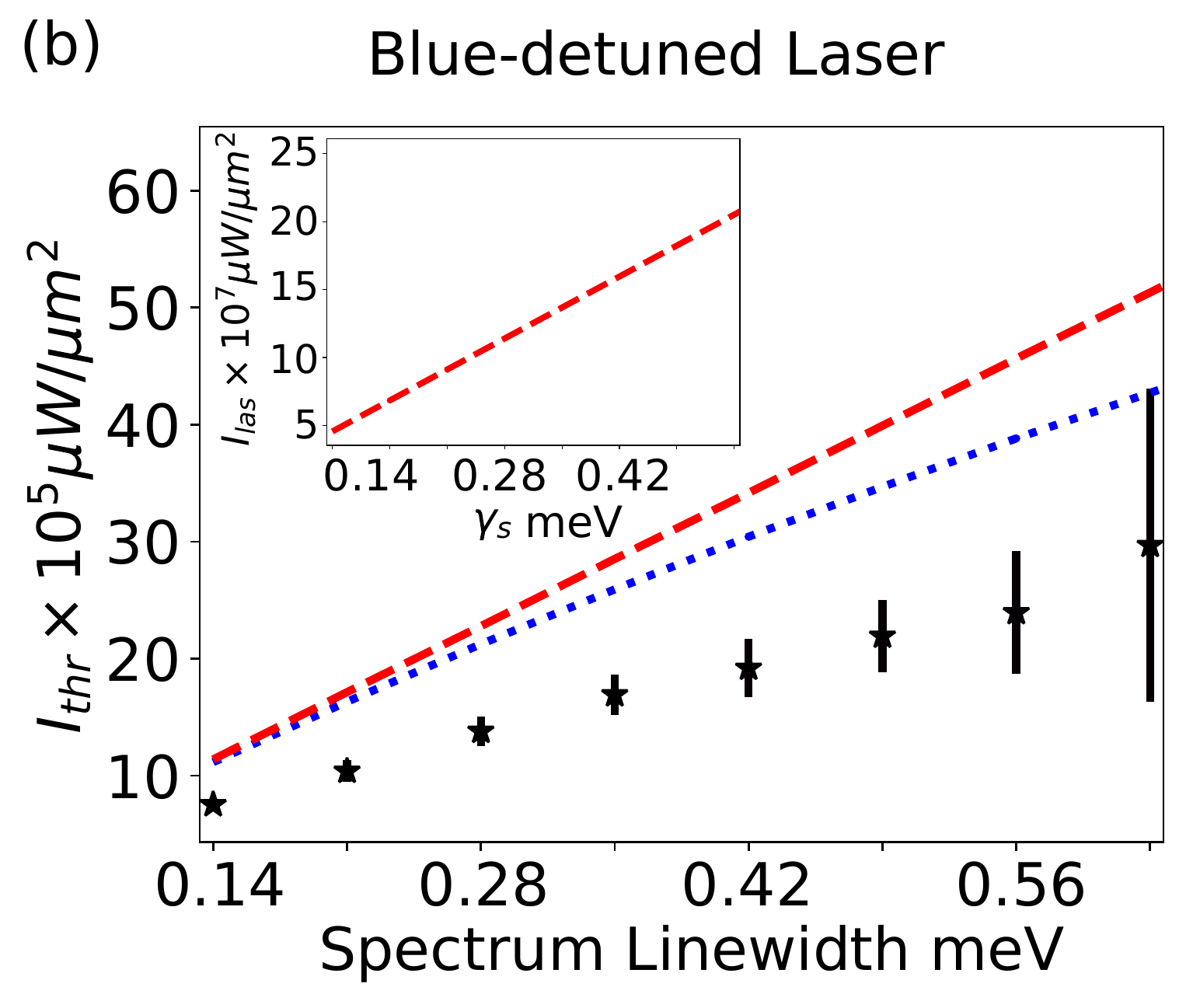}
\par\end{centering}
\caption{\label{fig:comparison-models} Comparison of the different origins of the Raman linewidth. 
(a) shows the integrated anti-Stokes intensity for increasing number of molecules, illuminated by a blue-detuned laser of intensity $I_{las}=10^4$ $\mu W/\mu m^2$ and for only phonon decay $\hbar \gamma_s = 0.21$ meV  (red dashed curves), homogeneous broadening  $\hbar 2\chi_s = 0.14 $ meV and decay rate $\hbar \gamma_s = 0.07$ meV  (blue dotted lines), and inhomogeneous broadening $\hbar \chi_G = 0.21$ meV and decay rate  $\gamma_s = 0.07$ meV (black solid line). The inset shows the corresponding anti-Stokes spectra for the three cases under weak laser illumination $I_{las}=10 \mu W/\mu m^2$. (b) Laser intensity threshold  $I_{thr}$ to achieve the parametric instability is plotted against the linewidth of the Raman lines $\gamma_{T}$ that would be obtained at low laser intensity. The linewidth for the three situations is estimated as $\gamma_{T}=\gamma_s$  (red dashed curves), $\gamma_{T}=2\chi_s + \gamma_{s}$   (blue dotted lines)  and $\gamma_{T}=\chi_G$  (black stars for the average and vertical bars for the 1-sigma standard deviation). In the latter two cases, $\gamma_{s}$ is fixed as $0.07$ meV and we modify $2\chi_s$ or $\chi_G$. The inset shows $I_{thr}$ for increasing $\gamma_{s}$ in the case of a single molecule. In all panels, we assume $\hbar (\omega_l - \omega_c') = 236$ meV and temperature $T=290$ K. The results plotted for $\chi_{G}>0$ are computed from thirty different realizations. }
\end{figure*}

\subsection{Equivalence of Homogeneous and Inhomogenous Contributions to the Collective SERS Signal}

In the previous sections we have considered three mechanisms (phonon decay, homogeneous broadening and inhomogeneous broadening) contributing to the width of the Raman lines. It is however not clear if collective effects depend on which of these mechanisms is present in a given experiment, or whether it is only the value of the linewidth (at low laser intensity $I_{las}$) that is important. We investigate this question  with a system illuminated by a blue-detuned laser [$\hbar (\omega_l - \omega_c') = 236$ meV]. As in the previous subsection, when the inhomogeneous broadening is present we average thirty realizations of molecules with slightly different (random) vibrational frequencies. 

Figure \ref{fig:comparison-models}a compares the integrated anti-Stokes intensity in the vibrational pumping regime ($I_{las}=10^4$ $\mu W/\mu m^2$) as a function of the number of molecules for  a situation where the Raman linewidth is only due to (i) the phonon decay rate $\hbar\gamma_{s}=0.21$ meV  (red dashed lines), and two other situations where the Raman linewidth is determined by (ii) a weaker decay rate $\hbar\gamma_{s}=0.07$ meV and a homogeneous broadening $2\chi_{s}=0.14$ meV  (blue dotted line) or (iii) an inhomogeneous broadening $\chi_{G}=0.21$ meV (solid black lines). These values are chosen because they lead to similar anti-Stokes spectra in the thermal regime, as demonstrated in the inset for $I_{las}=10$ $\mu W/\mu m^2$.  Notice that the spectra have Lorentzian shape in the first two cases but a Gaussian-shape in the last one. We observe that the three cases result in almost the same $N^2$ superradiant scaling. There is some difference in the results for systems with more molecules but this difference remains moderate.  

Further, we show in Figure \ref{fig:comparison-models}b the  laser threshold intensity $I_{thr}$ to achieve the parametric instability as a function of the total level of losses for the three situations under consideration. We quantify the losses by the approximate linewidth $\gamma_{T}$ that would be obtained for low laser intensity. For the three different situations under study, $\gamma_{T}$ corresponds to (i)  $\gamma_{T}=\gamma_{s}$ with $2\chi_{s}=\chi_{G}=0$  (red dashed lines) , (ii) $\gamma_{T}=2\chi_{s}+\gamma_{s}$ with $\chi_{G}=0$ (blue dotted line)   or (iii) $\gamma_{T}=\chi_{G}$ with $2\chi_{s}=0$  (black stars and black error bars). In the last situation, $\gamma_{T}=\chi_{G}$ is expected for sufficiently large $\chi_{G} \gtrsim \gamma_s$. For the first scenario, we vary the phonon decay rate $\gamma_s$, while for the other two we fix $\hbar  \gamma_s=0.07$ meV and vary either $2\chi_{s}$ or $\chi_{G}$. We obtain $I_{thr}$ from the theoretical expressions when the inhomogeneous broadening is absent, and otherwise from fitting the calculated results, as discussed in the previous sections.

The situation including inhomogeneous broadening ($\chi_G>0$) results in the smallest average $I_{thr}$ (black stars) for all $\gamma_{T}$, although the variation from realization to realization increases as the linewidth becomes larger (black error bars). For the first situation with only phonon decay $\gamma_s$ (dashed red line),  $I_{thr}$ scales linearly with the linewidth  $\gamma_{T}$ and is about two times larger than the situation with inhomogeneous broadening for the same $\gamma_{T}$. Last, the situation including homogeneous broadening ($2\chi_s>0$) leads to intermediate values of $I_{thr}$ for the largest $\gamma_{T}$ considered (dotted blue line), while for $\gamma_{T} \leqslant 0.21$ meV $I_{thr}$ becomes similar to the results with just the phonon decay. For reference, the inset of Figure \ref{fig:comparison-models}b gives the laser threshold for one single molecule and no broadening, which is is about $40$ times larger than those for $40$ molecules. 

We have thus shown that the mechanisms behind the width of the Raman lines can result in some differences in the SERS signal, but these differences are generally small or moderate. It thus seems possible  to predict the general impact of collective effects in an experiment even if the exact mechanism inducing the width of the Raman lines is not known.

\section{Summary and Discussion}

In summary, we have developed a model based on molecular optomechanics to describe surface-enhanced Raman scattering (SERS) from many molecules near a metallic nanostructure. The resulting equations can be solved analytically for identical molecules or numerically for more general systems.  

Our model indicates that the collective effects in SERS are mediated by the quantum correlation between molecules and it reveals the conditions under which the collective response could emerge in experiments. More precisely, we focus on two types of collective effects by analyzing the evolution of Raman scattering with increasing number of molecules $N$.

The first collective effect is a $1/N$ dependence of the threshold laser intensity required to observe: (i) the divergence and narrowing of the SERS lines (the parametric instability) for a laser blue-detuned with respect to the plasmonic resonance, and (ii) the saturation of the phonon population and the broadening of the SERS lines under a  red-detuned laser. The observation of these phenomena requires very intense illumination (likely a pulsed laser) even for optimized conditions (many molecules, vibrational modes with large Raman activity and low phonon decay rate).  The required intensity is so large that other mechanisms might affect the response of the system, such as the burning of the molecules and the presence of vibrational anharmonicities. Thus, the experimental demonstration of these phenomena would likely require very carefully designed systems. In addition, for such strong illumination, a more rigorous  treatment of the laser-plasmon coupling beyond the rotating wave-approximation may introduce some corrections to our results. 

As the second collective effect, we show that the SERS signal increases quadratically with the number of molecules for red-, blue- or zero-detuned laser illumination, i.e. we establish on a firm theoretical basis the effect of superradiant Raman scattering. The laser intensity to observe this effect in the anti-Stokes scattering at room temperature is about three orders of magnitude smaller that the intensity to observe the parametric instability, and phonon-population saturation or to observe superradiance in the Stokes scattering.  Further, the intensity required to observe the superradiant anti-Stokes scattering can be further reduced by working at low temperature. Thus, this collective effect seems particularly attractive for experimental demonstration with continuous or pulsed lasers.

To better understand the main features of collective effects in SERS, we have focused on a situation where molecules that support a single vibration interact with each other via their coupling to a single plasmonic mode, ignoring direct inter-molecular interaction \citep{SMAshrafi}. Our results are thus better suited, for example, for well-separated molecules and laser illumination with sufficiently low frequency so that the multiplicity of high-order electromagnetic modes (or pseudomodes \citep{ADelga}) do not contribute significantly. However, this model can be actually extended to describe more general situations that might involve direct inter-molecular interactions, multiple Raman-active vibrations, multiple plasmonic modes or infrared active vibrations.  Further, we initially considered a relatively simple situation of identical molecules with no decay channel beyond standard phonon decay, but we also demonstrated that the collective response survives in more complex scenarios. Specifically, we verify that the collective phenomena are affected only moderately by the presence of homogeneous and inhomogeneous broadening of the molecular vibrations, so that the effects reported here seem robust.

In conclusion, our results establish a general theoretical framework to study collective effects in SERS, and suggest that novel collective phenomena can be accessible to experiments under realistic laser illumination.

\section{Methods \label{sec:methods}}

We apply open quantum system theory \citep{HPBreuer} to describe SERS, including all relevant incoherent processes. In this description, the dynamics are governed by the quantum master equation for the density operator $\rho$:
\begin{align}
& \partial_{t}\rho =-\frac{i}{\hbar}\left[H,\rho\right]+\mathcal{D}\left[\rho\right] \label{eq:master-equation}
\end{align} 
with the full Hamiltonian $H=H_{vib}+H_{cav}+H_{las}+H_{int}$ and the following Lindblad terms
\begin{align}
& \mathcal{D}\left[\rho\right]=\left(\kappa/2\right)\mathcal{D}\left[a\right]\rho + \sum_{s}\chi_{s}\mathcal{D}\left[ b_{s}^{\dagger} b_{s}\right]\rho \nonumber \\
&  +\sum_{s}\left(\gamma_{s}/2\right)\left\{ \left(n_{s}^{th}+1\right)\mathcal{D}\left[b_{s}\right]\rho+n_{s}^{th}\mathcal{D}\left[b_{s}^{\dagger}\right]\rho\right\},
\end{align}
where we introduce the superoperator (for any operator $o$) $\mathcal{D}\left[o\right]\rho=2o\rho o^{\dagger}-\left(o^{\dagger}o\rho+\rho o^{\dagger}o\right)$ and the thermal phonon population  $n_{s}^{th}= [e^{\hbar\omega_{s}/k_{B}T}-1]^{-1}$ at temperature $T$ ($k_{B}$ is the Boltzmann constant). The first Lindblad term describes the damping of the plasmonic mode at rate $\kappa$, the second the homogeneous broadening due to the pure dephasing rate $\chi_s$, and the last two the phonon decay at rate $\gamma_s$ and the thermal pumping  of the molecular vibrations, respectively. 

To solve the master equation (eq \ref{eq:master-equation}), we first go to a frame rotating with the laser frequency, then linearize the optomechanical interaction $H_{int}$ and finally eliminate the plasmonic degree of freedom. In the end, we obtain the following effective master equation for the reduced density operator $\rho_{v}$ of the vibrational noise operator $\delta b_{s}=b_{s}-\beta_{s}$  (with coherent amplitudes $\beta_{s}=\mathrm{tr}\left\{ b_{s}\rho\right\}$):
\begin{equation}
\partial_{t}\rho_v =-\frac{i}{\hbar}\left[H_v,\rho_v\right]+\mathcal{D}_v\left[\rho_v\right]  \label{eq:effmeq}
\end{equation} 
with Hamiltonian 
\begin{equation}
H_{v}=\sum_{s}\hbar\omega_{s}\delta b_{s}^{\dagger}\delta b_{s}-\sum_{s,s'}(\hbar/2)\left(\Omega_{ss'}^{+}+\Omega_{s's}^{-}\right)\delta b_{s}^{\dagger}\delta b_{s'}
\end{equation}
and Lindblad terms
\begin{align}
 & \mathcal{D}_{v}\left[\rho_{v}\right]= \sum_{s} \chi_{s}\mathcal{D}\left[\delta b_{s}^{\dagger}\delta b_{s}\right]\rho_{v} \nonumber \\
 & +\sum_{s}\left(\gamma_{s}/2\right)\left\{ \left(n_{s}^{th}+1\right)\mathcal{D}\left[\delta b_{s}\right]\rho_{v}+n_{s}^{th}\mathcal{D}\left[\delta b_{s}^{\dagger}\right]\rho_{v}\right\} \nonumber \\
 & + (1/2)\sum_{s,s'}\left\{ \Gamma_{ss'}^{+}\mathcal{D}\left[\delta b_{s}^{\dagger},\delta b_{s'}\right]\rho_{v}+\Gamma_{ss'}^{-}\mathcal{D}\left[\delta b_{s},\delta b_{s'}^{\dagger}\right]\rho_{v}\right\}
\end{align}
for the superoperator $\mathcal{D}\left[o,p\right]\rho=2o\rho p - \left(po\rho +\rho po\right)$ (for any pair of operators $o,p$). The coherent amplitude of the vibration $\beta_{s}=|\alpha|^2/[\omega_s - i(\gamma_s/2+\chi_s)]$ can be computed from the coherent amplitude of the plasmon $\alpha = \Omega/[i(\omega'_c-\omega_l) + \kappa/2]$. The value of $\mid \beta\mid^2$ is shown in Section S5.1 in Supporting Information, and it is always  much smaller than the incoherent phonon population $n_s \equiv \bigl\langle\delta b_{s}^{\dagger}\delta b_{s}\bigr\rangle$ except for extremely strong laser intensity.  

The derivation of these expressions and the values of the different parameters are given in  Section S1 of the Supporting Information. Briefly, the parameters  $\Omega_{ss'}^{+},\Omega_{ss'}^{-}$ can be obtained from the real part of the spectral density $S_{ss'}(\omega)$ at the Stokes  $\omega=\omega_l - \omega_s$ and anti-Stokes lines $\omega=\omega_l + \omega_s$, respectively, and describe the plasmon-induced frequency shift ($s=s'$) and  the plasmon-mediated coherent coupling ($s \neq s'$). Similarly, the parameters $\Gamma_{ss'}^{+},\Gamma_{ss'}^{-}$ can be calculated from the imaginary  part of $S_{ss'}(\omega)$ and describe the plasmon-induced pumping $\Gamma_{ss}^{+}$ and damping $\Gamma_{ss}^{-}$ and the plasmon-mediated dissipative coupling ($\Gamma_{ss'}^{+}$ and $\Gamma_{ss'}^{-}$ with $s \neq s'$).  $S_{ss'}(\omega)$ depends on the optomechanical couplings $g_s g_{s'}$ of two distant  molecules, the frequency detuning between the laser and the plasmonic cavity mode, the plasmonic losses and the laser intensity. In Section S1.3 in Supporting Information, we show the dependence of $\Omega_{ss}^{\pm}$ and $\Gamma_{ss}^{\pm}$ on laser frequency, which is key to understand the collective effects under intense laser illumination.

From eq \ref{eq:effmeq} we can derive the equations $\partial_t \left<o\right>={\rm tr}\{o\partial_t\rho_v\}=\frac{i}{\hbar}\left<\left[H_v,o]\right> + {\rm tr}\left\{o\mathcal{D}_v\left[\rho_v \right]\right\}\right\}$ for the expectation values of the different operators ($\left<o\right> = {\rm tr}\{o\rho_v\}$, with $tr$ the trace). In particular, we derive a close set of equations for the  incoherent phonon number $\left<\delta b^{\dagger}_s \delta b_s \right> = n_s $ and the noise correlations $\left<\delta b^{\dagger}_s \delta b_{s'}\right> = c_{ss'}$ ($s\neq s'$). These equations can be solved for the system with a significant number of molecules. 

Last, we obtain the Stokes and anti-Stokes SERS signal from the correlations of the noise dynamics according to \citep{MKDezfouli} $S^{st,as}\left(\omega\right)\propto \omega^4 \sum_{ss'}\Gamma_{ss'}^{+} \mathrm{Re} S_{ss'}^{st,as} \left( \omega -\omega_l \right)$ with $ S_{ss'}^{st} \left(\omega \right)= \int_{0}^{\infty}d\tau e^{-i \omega\tau}\left\langle \delta b_{s}\left(\tau\right)\delta b_{s'}^{\dagger}\left(0\right)\right\rangle $ and $S^{as}_{ss'}\left(\omega\right) = \int_{0}^{\infty}d\tau e^{-i\omega\tau}\left\langle \delta b^{\dagger}_{s}\left(\tau\right)\delta b_{s'}\left(0\right)\right\rangle $, respectively. According to the quantum regression theorem \citep{PMeystre},  the two-time correlations  $\left\langle \delta b_{s}\left(\tau\right)\delta b_{s'}^{\dagger}\left(0\right)\right\rangle $
and $\left\langle \delta b_{s}^{\dagger}\left(\tau\right)\delta b_{s'}\left(0\right)\right\rangle $
follow the same equations
as $\left\langle \delta b_{s}\right\rangle $ and $\left\langle \delta b_{s}^{\dagger}\right\rangle $, but
with initial conditions $\left\langle \delta b_{s}\delta b_{s'}^{\dagger}\right\rangle_{ste} $
and $\left\langle \delta b_{s}^{\dagger}\delta b_{s'} \right\rangle_{ste}$. Here, the label "$ste$" refers to the steady-state and the equations
for $\left\langle \delta b_{s}\right\rangle $ and $\left\langle \delta b_{s}^{\dagger}\right\rangle$ are obtained from $\partial_t \left<o\right>={\rm tr}\{o\partial_t \rho_v\}$.

\vspace{1cm}
{\bf Supporting Information}

Supporting Information includes: derivation of effective master equation, equations for incoherent phonon population and noise correlation, expression for SERS spectrum, analytic expressions for systems with identical molecules, collective oscillator model and supplemental results (coherent phonon population, SERS line shift, line narrowing and broadening, laser threshold for molecules with different Raman activity, collective effects landscape under blue-, zero- and red-detuned laser illumination, and influence of phonon-induced plasmon shift).  

\vspace{1cm}
{\bf Acknowledgement}

We would like to thank Mikolaj K. Schmidt and Jeremy J. Baumberg for  fruitful discussions.  We acknowledge the project FIS2016-80174-P from the Spanish Ministry of Science, Innovation and Universities, the project PI2017-30 of the Department of Education of the Basque Government, the project H2020-FET Open ``THOR'' Nr. 829067 from the European Commission, and grant IT1164-19 for consolidated groups of the Basque University, through the Department of Universities of the Basque Government, and the NSFC-DPG joint project Nr. 21961132023.

\newpage
\clearpage
\onecolumngrid
\renewcommand{\theequation}{S\arabic{equation}}
\renewcommand{\thesection}{S\arabic{section}}
\renewcommand{\thefigure}{S\arabic{figure}}
\renewcommand*{\citenumfont}[1]{S#1}
\renewcommand*{\bibnumfmt}[1]{[S#1]}

\renewcommand{\thepage}{S\arabic{page}}  
\setcounter{page}{1}
\setcounter{equation}{0}
\setcounter{figure}{0}
\setcounter{section}{0}

\begin{center}
 {\bf {\Large  Supporting information to:

 Optomechanical Collective Effects in Surface-Enhanced Raman Scattering from Many Molecules}}
\end{center}
\section{Effective quantum master equation for molecular vibrations  \label{sec:derivation-qme}}

\subsection{System and model}
In the main text, we have outlined the procedure to obtain the effective master equation for the molecular vibrations. In this section of the Supporting Information, we describe the derivation, the approximations involved and the final equations in more detail.

For easier reference, we first reintroduce the Hamiltonians involved. The total Hamiltonian is  $H=H_{vib}+H_{cav}+H_{int}+H_{las}$. Here, $H_{vib}=\sum_{s}\hbar\omega_{s}b_{s}^{\dagger}b_{s}$
describes molecular vibrations with angular frequencies $\omega_{s}$ and bosonic creation $b_{s}^{\dagger}$ and annihilation operators $b_{s}$.
$H_{cav}=\hbar\omega_{c}a^{\dagger}a$ describes the single plasmonic cavity
mode with a frequency $\omega_{c}$, a bosonic creation $a^{\dagger}$ and annihilation
operator $a$, and $H_{int}=-a^{\dagger}a\sum_{s}\hbar g_{s}\left(b_{s}^{\dagger}+b_{s}\right)$
corresponds to the optomechanical interaction with strength $g_{s}$. Last, $H_{las}=i\hbar\Omega\left(a^{\dagger}e^{-i\omega_{l}t}-ae^{i\omega_{l}t}\right)$ describes the plasmon excitation by a laser of angular frequency $\omega_{l}$ within the rotating wave approximation. 

The strength of the optomechanical interaction $g_{s}=f_s R_{s}Q_{s}^{0}\omega_{c}/\left(2\epsilon_{0}V_{eff}\right)$ is determined by  the amplitude of Raman tensor $R_{s}$, the zero-point
amplitude $Q_{s}^{0}=\sqrt{\hbar/2\omega_{s}}$ of the molecular vibrational modes, and $\epsilon_{0}$ the vacuum permittivity. The effective mode volume is given by $V_{eff} = W/(\epsilon_0 |{\bf E}_m|^2)$, where $W$ is the total electromagnetic field energy and $|{\bf E}_m|$ is the maximum of the local electric field at resonance.  The factor $f_{s} $ accounts for the position and orientation of the molecules, with $f_{s}=1$ for a molecule at the position of maximum local field  and with an optimal orientation, and $f_{s}<1$ otherwise \citep{SI-MKSchmidt-1}. The coefficient $\Omega$ describes the efficiency of the plasmon  excitation by the incoming laser and follows $\Omega=\frac{\kappa}{2}\sqrt{\frac{ \epsilon_{0}V_{eff}}{2\hbar\omega_{c}}} K \left| E_{0}\right|$ with $\kappa$ the plasmon damping rate, $K=|{\bf E}_m|/|E_0|$ the local-field enhancement factor  and $|E_0|$ the laser amplitude. In all the paper, we assume that the molecules are placed in vacuum, and do not consider explicitly the (relatively small) correction on the local fields due to the off-resonant molecule polarizability (i.e. the optical contrast between the molecules and the surrounding vacuum). The derivation of these expressions and a longer discussion can be found in Ref \citenum{SI-MKSchmidt,SI-MKSchmidt-1}. 
We describe the system dynamics with the following quantum master equation 
\begin{align}
\frac{\partial}{\partial t}\rho & =-\frac{i}{\hbar}\left[\tilde H,\rho\right]+\frac{1}{2}\kappa\mathcal{D}\left[a_l\right]\rho+\sum_{s}\chi_{s}\mathcal{D}\left[b_{s}^{\dagger}b_{s}\right]\rho\nonumber \\
 & +\frac{1}{2}\sum_{s}\gamma_{s}\left\{ \left(n_{s}^{th}+1\right)\mathcal{D}\left[b_{s}\right]\rho+n_{s}^{th}\mathcal{D}\left[b_{s}^{\dagger}\right]\rho\right\} .\label{eq:master-equation-single-mode}
\end{align}
To concentrate on the slowly varying dynamics, we work in a frame that rotates with the laser frequency $\omega_{l}$ so that $\tilde{H}=e^{iH_{0}t}\left(H-H_{0}\right)e^{-iH_{0}t}$ with $H_{0}=\hbar\omega_{l}a^{\dagger}a$. In this case, we have $\tilde{H}=\tilde{H}_{vib}+\tilde{H}_{cav}+\tilde{H}_{int}+\tilde{H}_{las}$ with $\tilde{H}_{vib}=H_{vib}$, $\tilde{H}_{cav}=\hbar\left(\omega_{c}-\omega_{l}\right)a_{l}^{\dagger}a_{l}$, $\tilde{H}_{int}=-\hbar a_{l}^{\dagger}a_{l}\sum_{s}g_{s}\left(b_{s}^{\dagger} + b_{s}\right)$, $\tilde{H}_{las}=i\hbar\Omega\left(a_{l}^{\dagger}-a_{l}\right)$, and the slowly varying operators $a_{l}^{\dagger}=a^{\dagger}e^{-i\omega_{l}t}$, $a_{l}=a e^{i\omega_{l}t}$. The Lindblad terms $\mathcal{D}[o]\rho=2o\rho o^\dagger-o^\dagger o\rho-\rho o^\dagger o $ account for possible dissipative processes associated with an operator $o$. In our system, we include the plasmon damping with rate $\kappa$, and the dephasing, decay and thermal pumping of the vibrational modes with rate  $\chi_{s}$, $\gamma_s$ (and thermal phonon population $n_{s}^{th}$), respectively. 

\subsection{Effective master equation \label{sec:effectivemeq}}
To proceed, we separate the coherent amplitudes $\alpha=\left\langle a_{l}\right\rangle =\mathrm{tr}\left\{ a_{l}\rho\right\}$ and $\beta_s = \left\langle b_{s}\right\rangle =\mathrm{tr}\left\{ b_{s}\rho\right\} $  from  the noise operators $\delta a_{l}$ and $\delta b_{s}$ according to $a_{l}= \alpha  +\delta a_{l}$
and $b_{s}= \beta_s  +\delta b_{s}$ (${\rm tr}$ indicates the trace).  The equations for the coherent amplitudes can be obtained from the equation $\partial_t \left\langle o\right\rangle = \partial_t \mathrm{tr}\left\{ o\rho\right\} =\mathrm{tr}\left\{ o \partial_t \rho \right\}$ with $o=a_l,b_s$ and eq \ref{eq:master-equation-single-mode}:
\begin{align}
\frac{\partial}{\partial t} \alpha  & \approx-\left[i\left(\omega'_{c}-\omega_{l}\right)+\kappa/2 \right]
\alpha +\Omega,\label{eq:dynamic_a} \\
\frac{\partial}{\partial t}\beta_s  & \approx-\left(i\omega_{s}+\gamma_{s}/2 + \chi_s\right)\beta_s +i g_s \left|\alpha\right|^{2},\label{eq:dynamic_b}
\end{align}
where we have introduced $\omega'_{c}=\omega_{c}-\sum_{s}g_{s}2\mathrm{Re}\beta_{s}$
and ignored the contributions $\left\langle \delta a_{l}\delta b_{s}\right\rangle $
and $\left\langle \delta a_{l}\delta b_{s}^{\dagger}\right\rangle $. The above equations indicate that the coherent dynamics are not affected by the noise dynamics. The steady-state solution is simply  \citep{SI-MKSchmidt,SI-MKSchmidt-1}  $\alpha =\Omega/\left[i\left(\omega'_{c}-\omega_{l}\right)+\kappa/2\right]$ and $\beta_{s} =\left|\alpha \right|^{2}/\left[\omega_{s}-i(\gamma_{s}/2 + \chi_s)\right]$.

Applying $a_{l}= \alpha +\delta a_{l}$
and $b_{s}=\beta_s +\delta b_{s}$ directly to the Hamiltonian $\tilde{H}$ in the rotating framework and dropping again the negligible terms $\delta a_{l}\delta b_{s}$,$\delta a_{l}\delta b_{s}^{\dagger}$, we can approximate the Hamiltonian as the linearized Hamiltonian $\tilde{H'}=\tilde{H'}_{vib}+\tilde{H}_{c}^{'}+\tilde{H}'_{int}$ with $\tilde{H'}_{vib}=\hbar\sum_{s}\omega_{s}\delta b_{s}^{\dagger}\delta b_{s}$, $\tilde{H}_{cav}^{'}\approx\hbar\left(\omega'_{c}-\omega_{l}\right)\delta a_{l}^{\dagger}\delta a_{l}$ and 
\begin{equation}
\tilde{H}'_{int}=-\left(\alpha_{l}^{*}\delta a_{l}+\alpha_{l}\delta a_{l}^{\dagger}\right)\sum_{s}\hbar g_{s}\left(\delta b_{s}^{\dagger}+\delta b_{s}\right).\label{eq:linearzed-Hamiltonian}
\end{equation}
After this approximation the Lindblads remain identical as in eq \ref{eq:master-equation-single-mode} (but applied to the noise operators). The Hamiltonian $\tilde{H}_{las}$ describing the plasmon excitation by laser does not appear explicitly, and the effect of the laser is included by the values of $\alpha$ and $\beta_s$.

Once this linearized Hamiltonian has been obtained,  we can then treat the plasmon as a reservoir that acts a source of incoherent pumping and losses and eliminate it from the master equation and finally obtain the effective master equation for the reduced density operator $\tilde \rho_v={\rm tr_R} \{ \tilde \rho\}$ of the vibrational noise operators (${\rm tr}_{R}$ indicates the trace over the plasmonic reservoir). The derivation is similar to that found in the formulation of open-quantum systems \citep{SI-HPBreuer,SI-TNeuman1}. We sketch the main steps in the following. 

We take $\tilde{H'}_{vib}$ as the system Hamiltonian, $\tilde{H}'_{cav}$ as the reservoir Hamiltonian and $\tilde{H}'_{int}$ as the system-reservoir interaction Hamiltonian. To work in the interaction picture, we apply the transformation $\tilde{H}'_{int}=e^{i\tilde{H}_{0}t}\tilde{H}'_{int}e^{-i\tilde{H}_{0}t}$ with $\tilde{H}_{0}=\tilde{H'}_{vib}+\tilde{H}'_{cav}$ and get the new interaction Hamiltonian 
\begin{align}
 & \tilde{H}'_{int}\left(t\right)=-\alpha_{l}^{*}\delta a_{l}e^{-i(\omega'_{c}-\omega_{l})t}\sum_{s}\hbar g_{s} \left(\delta b_{s}e^{-i\omega_{s}t}+\delta b_{s}^{\dagger}e^{i\omega_{s}t}\right)+\mathrm{h.c.} \label{eq:interaction}
\end{align}
We then treat this interaction as a perturbation in second order and apply the Born-Markov approximation to obtain  equation of motion for the reduced density operator $\tilde{\rho}_v$ of the system (molecular vibrations) \citep{SI-HPBreuer} 
\begin{equation}
\frac{\partial}{\partial t}\tilde{\rho}_v=-\frac{1}{\hbar^{2}}\int_{0}^{\infty}d\tau\mathrm{tr}_{R}\left\{ \left[\tilde{H}'_{int}\left(t\right),\left[\tilde{H}'_{int}\left(t-\tau\right),\tilde{\rho}_v\tilde{\rho}_{R}\right]\right]\right\}. \label{eq:standard-master-equation}
\end{equation}
On the right side of the above equation, we have decomposed the total density operator $\tilde{\rho}$ as the product $\tilde{\rho}_v\tilde{\rho}_{R}$ of that of the molecular vibrations $\tilde{\rho}_v$ and of the plasmon $\tilde{\rho}_R$. 

We can now insert eq \ref{eq:interaction} into eq \ref{eq:standard-master-equation} and evaluate the emerging terms. Since the plasmonic noise operators $\delta a^\dagger_l, \delta a_l$ describe the noise dynamics after removing the coherent plasmonic excitation $\alpha$,  they follow similar dynamics as  a harmonic oscillator of same loss $\kappa$ on the ground state. Thus, we can then treat the plasmon as a reservoir by assuming  $\mathrm{tr}_R \left\{\delta a_{l}^{\dagger}\delta a_{l} \tilde{\rho}_R \right\} =0$ and $\mathrm{tr}_R \left\{ \delta a_{l}\delta a_{l}^{\dagger} \tilde{\rho}_R \right\} =1$ (satisfied by the harmonic oscillator). Since this reservoir decays exponentially at the rate of the plasmonic loss $\kappa$, we introduce the damping term $e^{-(\kappa/2)t}$ for  the noise operators $a^\dagger_l,\delta a_l  \propto  e^{-(\kappa/2)t}$ in eq \ref{eq:interaction}. 

Inserting the resulting $\tilde{H}_{int}(t)$ into  eq \ref{eq:standard-master-equation}, we encounter the integrals
$\int_{0}^{\infty}d\tau e^{-i\left(\omega'_{c}-\omega_{l}\pm\omega_{s}\right)\tau - (\kappa/2)\tau }$ and 
$\int_{0}^{\infty}d\tau e^{i\left(\omega'_{c}-\omega_{l}\pm\omega_{s}\right)\tau - (\kappa/2)\tau}$. We can solve these integrals analytically to obtain  $\left\{ i\left[\omega'_{c}-\left(\omega_{l}\mp\omega_{s}\right)\right]+\kappa/2\right\} ^{-1}$
and $\left\{ -i\left[\omega'_{c}-\left(\omega_{l}\mp\omega_{s}\right)\right]+\kappa/2\right\} ^{-1}$,
respectively. Finally, we arrive at the effective  master equation
\begin{equation}
\frac{\partial \rho_{v}}{\partial t}=-\frac{i}{\hbar}\left[H_{v},\rho_{v}\right]+\mathcal{D}_{v}\left[\rho_{v}\right]\label{eq:effective-master-equation}
\end{equation}
 for the reduced density operator $\rho_{v}$ of the molecular vibrations with Hamiltonian 
\begin{equation}
H_{v}=\sum_{s}\hbar\omega_{s}\delta b_{s}^{\dagger}\delta b_{s}-\hbar\sum_{s,s'}\frac{1}{2}\left(\Omega_{ss'}^{+}+\Omega_{s's}^{-}\right)\delta b_{s}^{\dagger}\delta b_{s'}
\end{equation}
and dissipation 
\begin{align}
 & \mathcal{D}_{v}\left[\rho_{v}\right]=\sum_{s}\chi_{s}\mathcal{D}\left[\delta b_{s}^{\dagger}\delta b_{s}\right]\rho_{v}  +\sum_{s}\frac{\gamma_{s}}{2}\left\{ \left(n_{s}^{th}+1\right)\mathcal{D}\left[\delta b_{s}\right]\rho_{v}+n_{s}^{th}\mathcal{D}\left[\delta b_{s}^{\dagger}\right]\rho_{v}\right\} \nonumber \\
 & +\frac{1}{2}\sum_{s,s'}\left\{ \Gamma_{ss'}^{+}\mathcal{D}\left[\delta b_{s}^{\dagger},\delta b_{s'}\right]\rho_{v}+\Gamma_{ss'}^{-}\mathcal{D}\left[\delta b_{s},\delta b_{s'}^{\dagger}\right]\rho_{v}\right\} .
\end{align}
Here, we have introduced  the superoperator $\mathcal{D}\left[o,p\right]\rho=2o\rho p - \left(po\rho +\rho po\right)$ (for any pair of operator $o,p$). The plasmon affects the vibrational dynamics through the parameters 
\begin{align}
\Omega_{ss'}^{\pm} & =-i\left[S_{ss'}^{+}\left(\omega_{l}\mp\omega_{s'}\right)-S_{ss'}^{-}\left(\omega_{l}\mp\omega_{s}\right)\right], \label{eq:omest}  \\
\Gamma_{ss'}^{\pm} & =S_{ss'}^{+}\left(\omega_{l}\mp\omega_{s'}\right)+S_{ss'}^{-}\left(\omega_{l}\mp\omega_{s}\right), \label{eq:gammast}  
\end{align}
which are determined by the spectral density
\begin{equation}
S_{ss'}^{\pm}\left(\omega\right)=\left|\alpha_{l}\right|^{2}g_{s'}g_{s}\frac{\kappa/2\pm i (\omega'_{c}-\omega )}{\left(\omega'_{c}-\omega\right)^{2}+\left(\kappa/2\right)^{2}}.\label{eq:apectraldensitypair}
\end{equation}
The parameters $\Omega_{ss'}^{+},\Omega_{ss'}^{-}$ describe the plasmon-induced frequency shift ($s=s'$) and  the plasmon-mediated coherent coupling ($s \neq s'$) while the parameters $\Gamma_{ss'}^{+},\Gamma_{ss'}^{-}$ describe the plasmon-induced pumping $\Gamma_{ss}^{+}$ and damping $\Gamma_{ss}^{-}$ and the plasmon-mediated incoherent coupling $\Gamma_{ss'}^{+}$ and $\Gamma_{ss'}^{-}$ ($s \neq s'$). Notice that the spectral density defined by eq \ref{eq:apectraldensitypair} can  depend  on the optomechanical couplings $g_s g_{s'}$ of two distant molecules. 

Using the expressions of $\alpha_{l},g_{s},\Omega$, we can also write explicitly the dependence of eq \ref{eq:apectraldensitypair} on the amplitude of Raman tensor and the effective mode volume of the plasmon as
\begin{align}
 & S_{ss'}^{\pm}\left(\omega\right)=\frac{1}{\hbar}I_{las}f_{s}f_{s'}\frac{K^{2}}{c\epsilon_{0}^{2}}\frac{R_{s}Q_s^0 R_{s'} Q_{s'}^0 \omega_c}{4  V_{eff}}\frac{\left(\kappa/2\right)^{2}}{\left(\omega_{c}'-\omega_{l}\right)^{2}+\left(\kappa/2\right)^{2}} \left[\frac{ \kappa/2 }{\left(\omega_{c}'-\omega\right)^{2}+\left(\kappa/2\right)^{2}}\pm\frac{i \left(\omega_{c}'-\omega\right)}{\left(\omega_{c}'-\omega\right)^{2}+\left(\kappa/2\right)^{2}}\right],\label{eq:spectral-density-single-mode}
\end{align}
where we have introduced  the laser power density $I_{las}=\frac{1}{2}\epsilon_{0}c\left|E_{0}\right|^{2}$.

\subsection{Dependence of plasmon-induced parameters on laser frequency \label{sec:parameters}}

\begin{figure}
\begin{centering}
\includegraphics[scale=0.4]{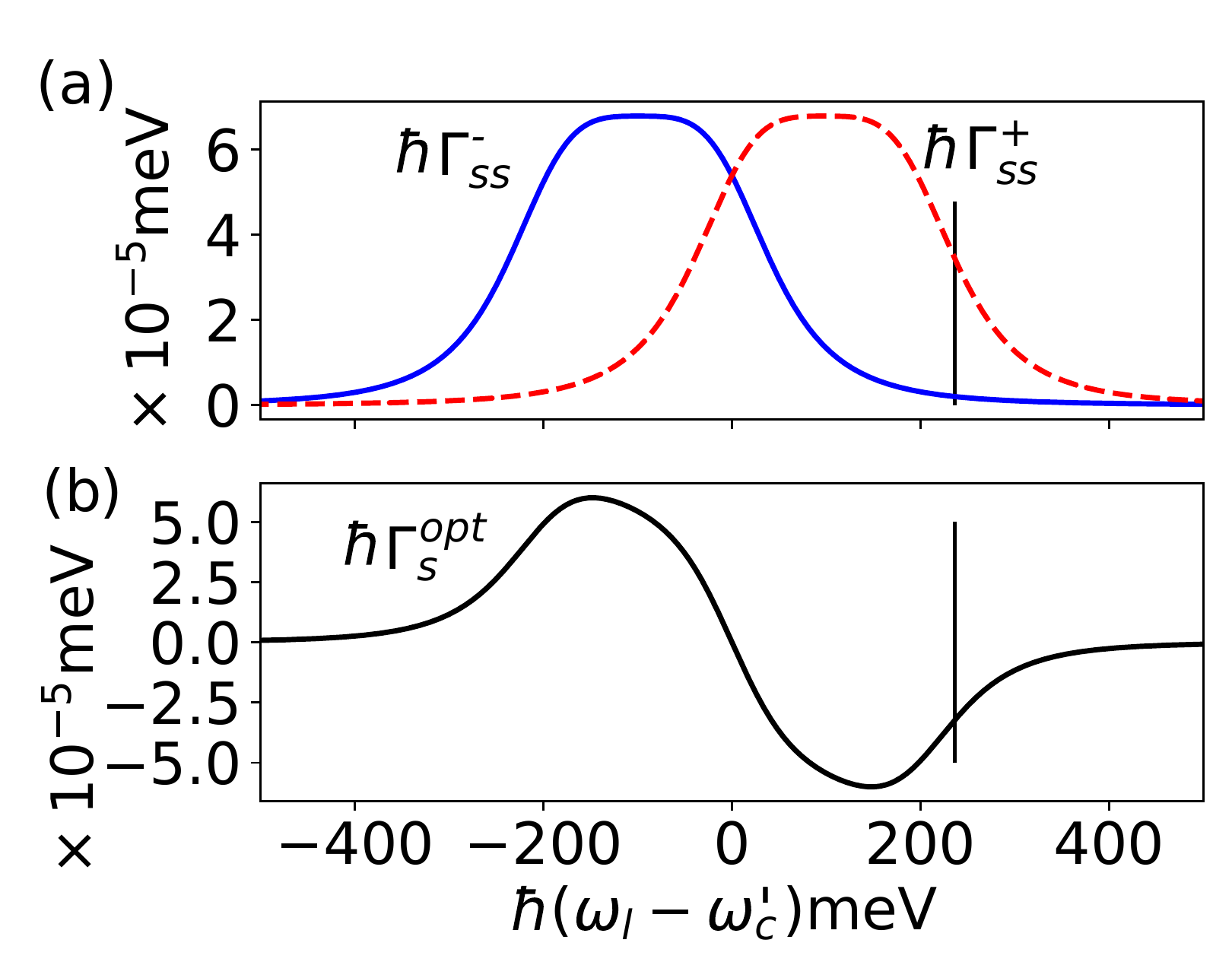}\includegraphics[scale=0.4]{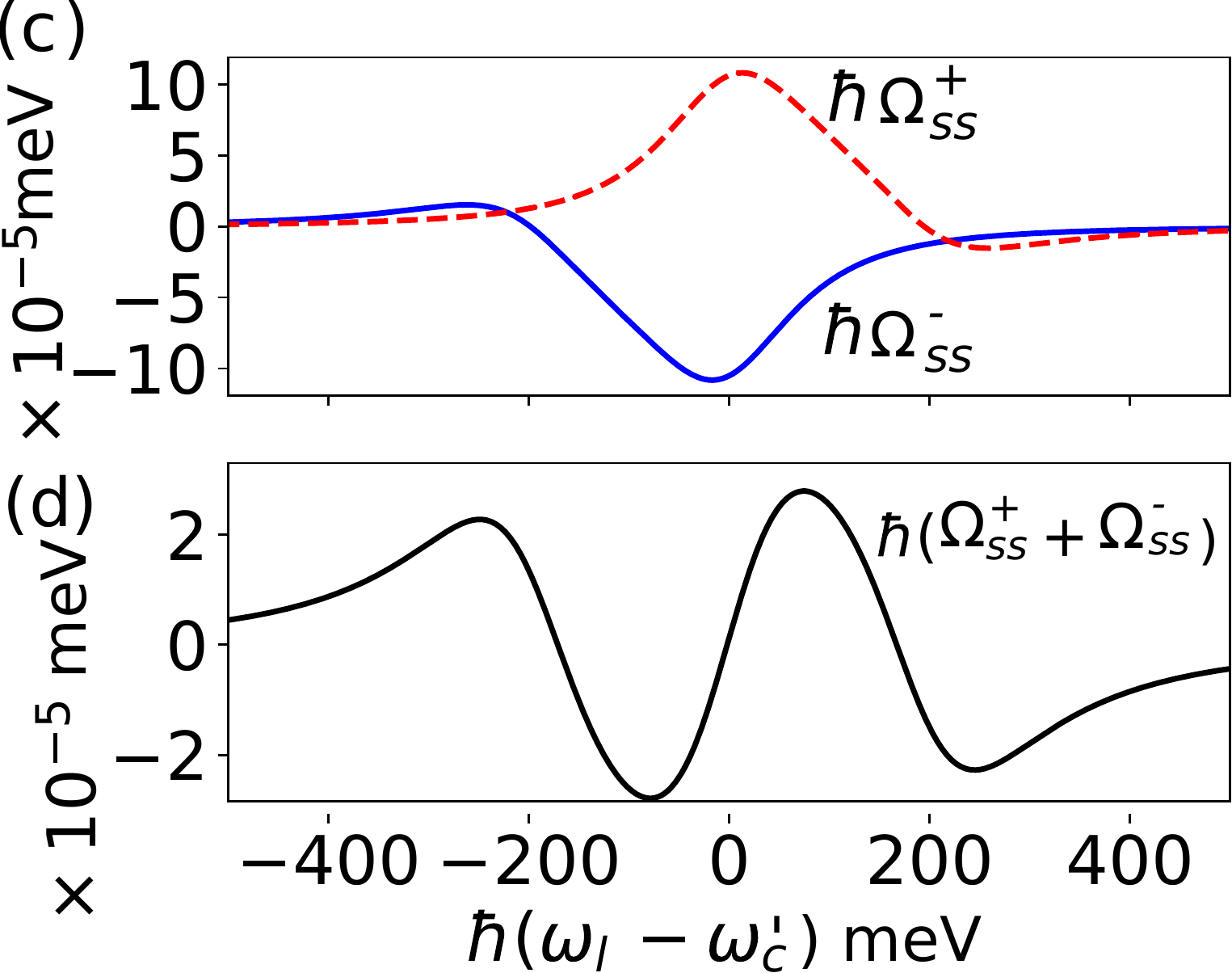}
\par\end{centering}
\caption{\label{fig:parameters} Laser-frequency 
dependence of the different terms describing the effect of the plasmon on the molecular vibrational dynamics. Spectral evolution of (a) the damping $\Gamma_{ss}^{-}$ and pumping $\Gamma_{ss}^{+}$ (blue solid and red dashed curve) and of (b)  the optomechanical damping $\Gamma_{s}^{opt} =\Gamma_{ss}^{-}-\Gamma_{ss}^{+}$. Spectral evolution of (c) the (twice) vibrational shifts $\Omega_{ss}^{-}$ , $\Omega_{ss}^{+}$ (blue solid and red dashed curve) and of (d) the total shift $\Omega_{ss}^{+}+\Omega_{ss}^{-}$.  In all figures, the laser power density is $I_{las}=10^{5}$ $\mu W/\mu m^{2}$ and the vertical lines indicate for reference the detuning of the laser used in the experiment in Ref. \citenum{SI-ALombardi}.  Other parameters are same as used in the main text.}
\end{figure}

We study next the evolution of the parameters $\Omega_{ss'}^{+},\Omega_{ss'}^{-},\Gamma_{ss'}^{+},\Gamma_{ss'}^{-}$ with laser frequency $\omega_l$, which describe the effect of the plasmon on the molecular vibrations (as discussed in the previous section). We  restrict our study to the parameters related to single molecules, i.e. $s=s'$, but the parameters related to two molecules, i.e. $s\neq s'$, behave similarly as far as the two molecules have similar vibrational frequencies and optomechanical coupling strengths, i.e. $\omega_s \approx \omega_{s'}, g_s \approx g_{s'}$. In general, all the parameters are linearly proportional to the laser intensity $I_{las}$ and here we consider $I_{las}=10^{5}$ $\mu W/\mu m^{2}$. 

Figure \ref{fig:parameters}a  shows that $\Gamma_{ss}^{+}$ and $\Gamma_{ss}^{-}$
have a maximum around $\omega_{l}=\omega_{c}'+\omega_{s}$ and $\omega_{l}=\omega_{c}'-\omega_{s}$,
respectively. These maxima present a $\approx 100$ meV-wide flat region because of the enhancement at the excitation and emission frequencies, which is resulted from two overlapping Lorentzian functions centered  at $\omega_c'$ and $\omega_c' + \omega_s$ for $\Gamma_{ss}^{+}$ or $\omega_c'$ and $\omega_c' - \omega_s$ for $\Gamma_{ss}^{-}$ 
(see eqs. \ref{eq:gammast}, \ref{eq:apectraldensitypair}). 
The resulting optomechanical damping rate  $\Gamma_{s}^{opt}=\Gamma_{ss}^{-}-\Gamma_{ss}^{+}$ is negative for blue-detuned laser illumination $\omega_{l}>\omega_{c}'$, zero for zero-detuned laser illumination $\omega_{l}=\omega_{c}'$  and positive for red-detuned laser illumination $\omega_{l}<\omega_{c}'$ (see Figure \ref{fig:parameters}b), which determines the response of the system  for strong laser illumination as discussed in the main text.

Similarly,  Figure \ref{fig:parameters}c shows that $\Omega_{ss}^{+}$ ($\Omega_{ss}^{-}$) approaches a positive maximum (negative minimum) around $\omega_{l}=\omega_{c}'$ and it becomes positive (negative) for $\omega_{l}<\omega_{c}'-\omega_{s}$ ($\omega_{l}>\omega_{c}'+\omega_{s}$). As a result, the total $\Omega_{ss}^{+}+\Omega_{ss}^{-}$ (twice the vibrational frequency shift) can be either positive [in the range $(-\infty,-160)$ meV and $(0,160)$ meV] or negative [in the range $(-160,0)$ meV and $(160,\infty)$ meV], see Figure \ref{fig:parameters}d. The maximum shift for the considered laser intensity $I_{las}=10^{5}$ $\mu W/\mu m^{2}$ reaches a value of  $\hbar\left|\Omega_{ss}^{+}+\Omega_{ss}^{-}\right|\approx 2.0\times10^{-5}$ meV. 

\section{Incoherent phonon population, noise correlation and SERS spectrum \label{sec:equation-observables}}

In this section, we present the effective equations for the physical observables describing the dynamics of the molecular vibrations and the SERS signal. The equation for the expectation value  $\left\langle o\right\rangle = {\rm tr } \left\{ o \rho_v \right\}$  of any operator $o$ can be obtained by applying $\partial_t \left\langle o\right\rangle =\mathrm{tr}\left\{ o \partial_t \rho_v \right\}$ (with $\partial_t \rho_v$ given by eq \ref{eq:effective-master-equation}) and using the cyclic property of the trace when necessary, e.g. ${\rm tr } \left\{ o \delta b_s \rho_v \delta b_{s'}^\dagger \right\} = {\rm tr } \left\{ \delta b_{s'}^\dagger  o \delta b_s \rho_v \right\} = \left \langle b_{s'}^\dagger  o b_s \right \rangle$.  Using $o=\delta b_{s}^{\dagger}\delta b_{s'}$ we obtain the following closed set of equations for the noise correlation $\left\langle \delta b_{s}^{\dagger}\delta b_{s'}\right\rangle $ ($s\neq s'$) and  the incoherent phonon population $\left\langle \delta b_{s}^{\dagger}\delta b_{s}\right\rangle $: 
\begin{align}
 & \frac{\partial}{\partial t}\left\langle \delta b_{s}^{\dagger} \delta  b_{s'}\right\rangle =-\kappa_{ss'}\left\langle \delta  b_{s}^{\dagger} \delta  b_{s'}\right\rangle +i\sum_{s''}\left\langle \delta  b_{s}^{\dagger} \delta  b_{s''}\right\rangle v_{s''s'}^{\left(1\right)}-i\sum_{s''}v_{ss''}^{\left(2\right)}\left\langle \delta  b_{s''}^{\dagger} \delta  b_{s'}\right\rangle  +\eta_{ss'} ,\label{eq:second-order-correlation}
\end{align}
where we have introduced the abbreviations $\kappa_{ss'}=i\left(\tilde{\omega}_{s'}-\tilde{\omega}_{s}^{*}\right)-\delta_{ss'}\left(\chi_{s}+\chi_{s'}\right)$, $\eta_{ss'}=\Gamma_{s's}^{+}+\delta_{ss'}n_{s}^{th}\gamma_{s}$, $v_{ss'}^{\left(1\right)}=\frac{1}{2}\left(\Omega_{s's}^{+}+\Omega_{ss'}^{-}\right)-i\frac{1}{2}\left(\Gamma_{s's}^{+}-\Gamma_{ss'}^{-}\right)$, $v_{ss'}^{\left(2\right)}=\frac{1}{2}\left(\Omega_{s's}^{+}+\Omega_{ss'}^{-}\right)+i\frac{1}{2}\left(\Gamma_{s's}^{+}-\Gamma_{ss'}^{-}\right)$ and $\tilde{\omega}_{s}=\omega_{s}-i\left(\frac{1}{2}\gamma_{s}+\chi_{s}\right)$. In the main text, for simplicity, we have used the symbol $n_s,c_{ss'}$ to represent $\left\langle \delta b_{s}^{\dagger}\delta b_{s}\right\rangle$  and $\left\langle \delta b_{s}^{\dagger}\delta b_{s'}\right\rangle $ ($s\neq s'$), respectively.  In the following, however, we keep the notation $\left\langle \delta b_{s}^{\dagger}\delta b_{s}\right\rangle$, $\left\langle \delta b_{s}^{\dagger}\delta b_{s'}\right\rangle $ in order to present the formulas in a compact way.  We can write the set of equations in eq \ref{eq:second-order-correlation} in a matrix form $\partial x/\partial t=-\left[\Gamma-i\left(V^{\left(1\right)}-V^{\left(2\right)}\right)\right]x+\lambda$, where the vectors $x,\lambda$ and the matrices $\Gamma,V^{\left(1\right)},V^{\left(2\right)}$
are defined with the elements $x_{\alpha}=\left\langle \delta b_{s}^{\dagger}\delta b_{s'}\right\rangle $,
$\lambda_{\alpha}=\eta_{ss'}$, $\Gamma_{\alpha\beta}=\delta_{\alpha\beta}\kappa_{ss'}$,
$V_{\alpha\beta}^{\left(1\right)}=\delta_{s's'''}v_{ss''}^{\left(a\right)}$,
$V_{\alpha\beta}^{\left(2\right)}=\delta_{s,s''}v_{s'''s'}^{\left(2\right)}$. The subindexes order the elements and follow $\alpha=sN+s'$, $\beta=s'' N+s'''$ with $N$ the total number of molecules. We then obtain the steady-state values of the incoherent phonon population  and the noise correlation  by simply calculating $x=\left[\Gamma-i\left(V^{\left(1\right)}-V^{\left(2\right)}\right)\right]^{-1}\lambda$.

On the other hand, setting $o=\delta b_{s}$ and $\delta b_{s}^{\dagger}$ we obtain another close set of equations for the noise amplitudes $\left\langle \delta b_{s}\right\rangle$:
\begin{align}
 & \frac{\partial}{\partial t}\left\langle \delta  b_{s}\right\rangle =-i\tilde{\omega}_{s}\left\langle  \delta  b_{s}\right\rangle +i\sum_{s'}v_{ss'}^{\left(1\right)}\left\langle \delta  b_{s'}\right\rangle ,\label{eq:bs}
 \end{align}
and the conjugate equations
 \begin{align}
 & \frac{\partial}{\partial t}\left\langle \delta  b_{s}^{\dagger}\right\rangle =i\tilde{\omega}_{s}^{*}\left\langle \delta  b_{s}^{\dagger}\right\rangle -i\sum_{s'}v_{ss'}^{\left(2\right)}\left\langle \delta  b_{s'}^{\dagger}\right\rangle .\label{eq:bs-dagger}
\end{align}
These equations allow us to obtain the emitted spectra, as we show below. The spectrum detected in the far field can be computed  \citep{SI-MKSchmidt,SI-MKSchmidt-1} as $S(\omega) \propto \omega^4 \int_{-\infty}^{\infty} d\tau e^{-i\omega\tau} \left\langle  a^\dagger \left(\tau\right) a\left(0\right)\right\rangle $ (with $\omega^4$ accounting for the frequency-dependence of dipolar emission). Using $a^\dagger=a_l^\dagger e^{i\omega_l t}$ and $a_l^\dagger = \alpha^* + \delta a_l^\dagger$ (and their conjugates) and focusing on the incoherent part of the spectrum (responsible for the Raman scattering), we obtain $S(\omega) \propto \omega^4 \int_{-\infty}^{\infty} d\tau e^{-i(\omega-\omega_l)\tau} \left\langle  \delta a^\dagger \left(\tau\right) \delta a\left(0\right)\right\rangle $. Further, using the relations $\delta a^\dagger_l   \propto \alpha \sum_s g_s (\delta b_s + \delta b_s^\dagger)$ (that can be derived from the 
master equation \ref{eq:master-equation-single-mode} with the linearized Hamiltonian given by eq \ref{eq:linearzed-Hamiltonian}) and focusing on the slowly varying terms (describing the dominant low-order Raman scattering), we finally obtain the Stokes and anti-Stokes spectrum as 
\begin{align}
S^{st}\left(\omega\right) & \propto \omega^4 \sum_{ss'}\Gamma_{ss'}^{+}\mathrm{Re}S_{ss'}^{st}\left(\omega-\omega_{l}\right), \label{eq:St-b} \\
S^{as}\left(\omega\right) & \propto \omega^4 \sum_{ss'}\Gamma_{ss'}^{-}\mathrm{Re}S_{ss'}^{as}\left(\omega-\omega_{l}\right)  \label{eq:As-b} 
\end{align}
with 
\begin{align}
S_{ss'}^{st}\left(\omega\right) & =\int_{-\infty}^{\infty}d\tau e^{-i\omega\tau}\theta\left(\tau\right)\left\langle \delta b_{s}\left(\tau\right)\delta b_{s'}^{\dagger}\left(0\right)\right\rangle, \\ 
S_{ss'}^{as}\left(\omega\right) & =\int_{-\infty}^{\infty}d\tau e^{-i\omega\tau}\theta\left(\tau\right)\left\langle \delta b_{s}^{\dagger}\left(\tau\right)\delta b_{s'}\left(0\right)\right\rangle .
\end{align}
Here, the step-function $\theta\left(\tau\right)$ accounts for the causality in the evolution of the two-time correlation functions, e.g. $\left\langle \delta b_{s}\left(\tau\right)\delta b_{s'}^{\dagger}\left(0\right)\right\rangle$, and the argument $\tau$ refers to the time difference relative to the
steady-state value (labeled by the argument $0$). 
 To evaluate the correlations we apply the quantum regression theorem \citep{SI-PMeystre}, which states that $\left\langle \delta b_{s}\left(\tau\right)\delta b_{s'}^{\dagger}\left(0\right)\right\rangle $
and $\left\langle \delta b_{s}^{\dagger}\left(\tau\right)\delta b_{s'}\left(0\right)\right\rangle $
 follow the same equations \ref{eq:bs}, \ref{eq:bs-dagger} as $\left\langle \delta b_{s}\right\rangle $ and $\left\langle \delta b_{s}^{\dagger}\right\rangle $, but  
with initial conditions $ \left\langle \delta b_{s}\left(0\right)\delta b_{s'}^{\dagger}\left(0\right)\right\rangle=\left\langle \delta b_{s}\delta b_{s'}^{\dagger}\right\rangle_{ste} $
and $ \left\langle \delta b_{s}^{\dagger}\left(0\right)\delta b_{s'}\left(0\right)\right\rangle=\left\langle \delta b_{s}^{\dagger} \delta b_{s'} \right\rangle_{ste}$. Here, "ste" stands for steady-state. Using these equations and defining $S_{ss'}^{st}\left(\tau\right)=\theta\left(\tau\right)\left\langle \delta b_{s}\left(\tau\right)\delta b_{s'}^{\dagger}\left(0\right)\right\rangle$, $S_{ss'}^{as}\left(\tau\right)=\theta\left(\tau\right)\left\langle \delta b_{s}^{\dagger}\left(\tau\right)\delta b_{s'}\left(0\right)\right\rangle$, we can derive 
\begin{align}
 & \frac{\partial}{\partial\tau}S_{ss'}^{st}\left(\tau\right)  = \delta(\tau) \left\langle \delta b_{s}\delta b_{s'}^{\dagger}\right\rangle_{ste} -i\tilde{\omega}_{s} S_{ss'}^{st}\left(\tau\right) +i\sum_{s''}v_{ss''}^{\left(1\right)} S_{s''s'}^{st}\left(\tau\right) ,\label{eq:Gssst} \\
 & \frac{\partial}{\partial\tau}S_{ss'}^{as}\left(\tau\right) = \delta(\tau) \left\langle \delta b_{s}^{\dagger} \delta b_{s'} \right\rangle_{ste} + i\tilde{\omega}_{s}^{*}S_{ss'}^{as}\left(\tau\right) -i\sum_{s''}v_{ss''}^{\left(2\right)}S_{s''s'}^{as}\left(\tau\right) .\label{eq:Gssas}
\end{align}
Using  eqs.\ref{eq:Gssst}-\ref{eq:Gssas} and the inverse Fourier transform $S_{ss'}^{k=st,as}\left(\tau\right)=\left(1/2\pi\right)\int_{-\infty}^{\infty}d\omega e^{i\omega\tau}S_{ss'}^{k}\left(\omega\right)$, we obtain  
\begin{align}
\sum_{s''}\left(\delta_{ss''}i\left(\omega+\tilde{\omega}_{s}\right)-iv_{ss''}^{\left(1\right)}\right)S_{s''s'}^{st}\left(\omega\right) & =\left\langle \delta b_{s}\delta b_{s'}^{\dagger}\right\rangle _{ste},\label{eq:G1omega}\\
\sum_{s''}\left(\delta_{ss''}i\left(\omega-\tilde{\omega}_{s}^{*}\right)+iv_{ss''}^{\left(2\right)}\right)S_{s''s'}^{as}\left(\omega\right) & =\left\langle \delta b_{s}^{\dagger}\delta b_{s'}\right\rangle _{ste}.\label{eq:G2omega}
\end{align}
Here, we have used $\delta\left(\tau\right)=\left(1/2\pi\right)\int_{-\infty}^{\infty}d\omega e^{i\omega\tau}$. 
To solve eqs \ref{eq:G1omega} and \ref{eq:G2omega} efficiently,
we transform them into matrix form $M^{k=st,as}x^{k=st,as;s'}=\lambda^{k=st,as;s'}$ for all $s'$  (notice that $M^{k}$ is independent of $s'$)  with the vectors $x^{st;s'},x^{as;s'},\lambda^{st;s'},\lambda^{as;s'}$ and the matrices $M^{st},M^{as}$ defined by the elements $x_{s}^{st;s'}=S_{ss'}^{st}\left(\omega\right),x_{s}^{as;s'}=S_{ss'}^{as}\left(\omega\right),\lambda_{s}^{st;s'}=\left\langle \delta b_{s}\delta b_{s'}^{\dagger}\right\rangle_{ste},\lambda_{s}^{as;s'}=\left\langle \delta b_{s}^{\dagger}\delta b_{s'}\right\rangle_{ste}$,
$M_{ss''}^{st}=\delta_{ss''}i\left(\omega+\tilde{\omega}_{s}\right)-iv_{ss''}^{\left(1\right)}$
and $M_{ss''}^{as}=\delta_{ss''}i\left(\omega-\tilde{\omega}_{s}^{*}\right)+iv_{ss''}^{\left(2\right)}$. We then obtain the solutions  $x^{k;s'}=\left[M^{k}\right]^{-1}\lambda^{k;s'}$ by direct matrix inversion. 

\section{Systems with identical molecules \label{sec:formulas-identical}}
In this section, we consider systems with $N$ identical molecules, i.e. all molecules have same vibrational frequency, same losses and couple with the plasmon in the same manner.  Under these conditions, all the observables are invariant if we permute any two molecules. More precisely, the diagonal elements $n_s = \left\langle \delta b_{s}^{\dagger}\delta b_{s}\right\rangle $ are the same for any molecule ($s$) and the off-diagonal elements $c_{ss'} = \left\langle \delta b_{s}^{\dagger}\delta b_{s'}\right\rangle $
($s\neq s'$) are the same for any molecular pair $\left(s,s'\right)$.
As a result, we can simplify eq \ref{eq:second-order-correlation}
to a system of only two equations
\begin{align}
 & \frac{\partial}{\partial t}\left\langle \delta b_{s}^{\dagger}\delta b_{s}\right\rangle =-\kappa_{ss}\left\langle \delta b_{s}^{\dagger}\delta b_{s}\right\rangle +\eta_{ss}  +i\left(v_{ss}^{\left(1\right)}-v_{ss}^{\left(2\right)}\right)\left[\left\langle \delta b_{s}^{\dagger}\delta b_{s}\right\rangle +\left(N-1\right)\left\langle \delta b_{s}^{\dagger}\delta b_{s'}\right\rangle \right], \label{eq:bspbs} \\
 & \frac{\partial}{\partial t}\left\langle \delta b_{s}^{\dagger}\delta b_{s'}\right\rangle =-\kappa_{ss'}\left\langle \delta b_{s}^{\dagger}\delta b_{s'}\right\rangle +\eta_{ss'}  +i\left(v_{ss}^{\left(1\right)}-v_{ss}^{\left(2\right)}\right)\left[\left\langle \delta b_{s}^{\dagger}\delta b_{s}\right\rangle +\left(N-1\right)\left\langle \delta b_{s}^{\dagger}\delta b_{s'}\right\rangle \right].\label{eq:bspbsprim}
 \end{align}
In this section, $s'\neq s$ is implied always. To obtain these equations, we have taken into account that, for identical  molecules, $\Gamma^{+}_{ss'}=\Gamma^{+}_{ss}$, $\Gamma^{-}_{ss'}=\Gamma^{-}_{ss}$, $\Omega^{+}_{ss'}=\Omega^{+}_{ss}$ and $\Omega^{-}_{ss'}=\Omega^{-}_{ss}$, as can be seen from eqs \ref{eq:omest}-\ref{eq:gammast}.  Thus, we also have $v_{ss'}^{\left(1\right)}=v_{ss}^{\left(1\right)}= \frac{1}{2}\left(\Omega_{ss}^{+}+\Omega_{ss}^{-}\right)-i\frac{1}{2}\left(\Gamma_{ss}^{+}-\Gamma_{ss}^{-}\right)$,  and $v_{ss'}^{\left(2\right)}=v_{ss}^{\left(2\right)}=\frac{1}{2}\left(\Omega_{ss}^{+}+\Omega_{ss}^{-}\right)+i\frac{1}{2}\left(\Gamma_{ss}^{+}-\Gamma_{ss}^{-}\right)$.  For the other relevant parameters we obtain $\kappa_{ss}=\gamma_{s},\kappa_{ss'}=\gamma_{s}+2\chi_{s},\eta_{ss}=\Gamma_{ss}^{+}+n_{s}^{th}\gamma_{s},\eta_{ss'}=\Gamma_{ss}^{+}$ (the original definition of these parameters can be found just below eq \ref{eq:second-order-correlation}).

Eqs. \ref{eq:bspbs}-\ref{eq:bspbsprim} can be solved analytically in the steady-state, which lead to:
\begin{equation}
\left\langle \delta b_{s}^{\dagger}\delta b_{s}\right\rangle_{ste} =\frac{\eta_{ss}-i\left(v_{ss}^{\left(1\right)}-v_{ss}^{\left(2\right)}\right)\left(N-1\right)\left(\eta_{ss}-\eta_{ss'}\right)/\kappa_{ss'}}{\kappa_{ss}-i\left(v_{ss}^{\left(1\right)}-v_{ss}^{\left(2\right)}\right)\left(1+\left(N-1\right)\kappa_{ss}/\kappa_{ss'}\right)},
\end{equation}
\begin{equation}
\left\langle \delta b_{s}^{\dagger}\delta b_{s'}\right\rangle_{ste} =\frac{\kappa_{ss}}{\kappa_{ss'}}\left\langle \delta b_{s}^{\dagger}\delta b_{s}\right\rangle_{ste} -\frac{\eta_{ss}-\eta_{ss'}}{\kappa_{ss'}},\label{eq:b1j2j}
\end{equation}
or
\begin{equation}
\left\langle \delta b_{s}^{\dagger}\delta b_{s}\right\rangle_{ste} =n_{s}^{th}+\frac{\Gamma_{ss}^{+}-\left(\Gamma_{ss}^{-}-\Gamma_{ss}^{+}\right)n_{s}^{th}}{\gamma_{s}+\left(\Gamma_{ss}^{-}-\Gamma_{ss}^{+}\right)\left(2\chi_{s}+N\gamma_{s}\right)/\left(\gamma_{s}+2\chi_{s}\right)},\label{eq:phonon-number-total}
\end{equation}
\begin{equation}
\left\langle \delta b_{s}^{\dagger}\delta b_{s'}\right\rangle_{ste} =\frac{\gamma_{s}}{\gamma_{s}+2\chi_{s}}\left(\left\langle \delta b_{s}^{\dagger}\delta b_{s}\right\rangle_{ste} -n_{s}^{th}\right).\label{eq:correlation-total}
\end{equation}
These equations are the same as eqs 6 and 7 in the main text (where we use $n_s$ and $c_{ss'}$ to represent the incoherent phonon population and the noise correlation). Similarly,  in the absence of the dephasing rate $\chi_s =0$, they correspond to eqs 4  and 5 in the main text.
 
Next, we consider the SERS spectrum  from many identical molecules.  For identical molecules the functions $S_{ss}^{k=st,as}\left(\omega\right)$ are the same for any molecule ($s$) and $S_{ss'}^{k}\left(\omega\right)$
($s\neq s'$) are identical for any molecular pair $\left(s,s'\right)$.
As a result, we can compute the Stokes spectrum from 
\begin{equation}
 S^{st}\left(\omega\right)\propto \omega^4 \Gamma_{ss}^{+}\bigl[N\mathrm{Re}S_{ss}^{st}\left(\omega-\omega_{l}\right) +N\left(N-1\right)\mathrm{Re}S_{ss'}^{st}\left(\omega-\omega_{l}\right)\bigr],\label{eq:Stokes-signal}
\end{equation}
where we have used once more  $\Gamma_{ss'}^{+}=\Gamma_{ss}^{+}$.   Notice that $N\left(N-1\right)$ corresponds to the number of molecular pairs for the $N$ identical molecules. We then use eq \ref{eq:G1omega} to obtain the following equations for $S_{ss}^{st}\left(\omega\right)$ and $S_{ss'}^{st}\left(\omega\right)$:
\begin{align}
 & i\left(\omega+\tilde{\omega}_{s}\right)S_{ss}^{st}\left(\omega\right)-iv_{ss}^{\left(1\right)}\Bigl[S_{ss}^{st}\left(\omega\right) +\left(N-1\right)S_{ss'}^{st}\left(\omega\right)\Bigr]=\left\langle \delta b_{s}\delta b_{s}^{\dagger}\right\rangle_{ste}, \label{eq:Sss_system} \\ 
 & i\left(\omega+\tilde{\omega}_{s}\right)S_{ss'}^{st}\left(\omega\right)-iv_{ss}^{\left(1\right)}\Bigl[S_{ss}^{st}\left(\omega\right) +\left(N-1\right)S_{ss'}^{st}\left(\omega\right)\Bigr]=\left\langle \delta b_{s} \delta b_{s'}^{\dagger} \right\rangle_{ste}.\label{eq:Sss'_system}
\end{align}
Subtracting the two equations and using $\left \langle \delta b_s \delta b^\dagger_s \right \rangle=1+ \left \langle  \delta b_s^\dagger \delta b_s \right \rangle$ and $\left \langle \delta b_s \delta b^\dagger_{s'}\right \rangle=\left \langle \delta b_{s'}^\dagger \delta b_s \right \rangle$ ($s\neq s'$), we obtain 
\begin{align}
 S_{ss'}^{st}\left(\omega\right)=S_{ss}^{st}\left(\omega\right)+\frac{i}{\omega+\tilde{\omega}_{s}}  \left(1+\left\langle \delta b_{s}^{\dagger}\delta b_{s}\right\rangle_{ste} -\left\langle \delta b_{s}^{\dagger}\delta b_{s'}\right\rangle_{ste} \right).\label{eq:Gst-2-1}
\end{align}
Then, inserting this equation into eq \ref{eq:Sss_system} we obtain 
\begin{align}
 & S_{ss}^{st}\left(\omega\right)=\frac{1}{i\left(\omega+\tilde{\omega}_{s}-Nv_{ss}^{\left(1\right)}\right)} \left[1+\left\langle \delta b_{s}^{\dagger}\delta b_{s}\right\rangle_{ste} -\frac{\left(N-1\right)v_{ss}^{\left(1\right)}}{\omega+\tilde{\omega}_{s}}\left(1+\left\langle \delta b_{s}^{\dagger}\delta b_{s}\right\rangle_{ste} -\left\langle \delta b_{s}^{\dagger}\delta b_{s'}\right\rangle_{ste} \right) \right] .
\end{align}
Last, we just need to insert this result into eq \ref{eq:Stokes-signal} to get
\begin{equation}
 S^{st}\left(\omega\right)\propto \frac{\omega^4 \Gamma_{ss}^{+} \Gamma_{st}/2}{\left(\omega-\omega_{st}\right)^{2}+\left(\Gamma_{st}/2\right)^{2}} \left[N\left(1+\left\langle \delta b_{s}^{\dagger} \delta b_{s}\right\rangle_{ste} \right)+N\left(N-1\right)\left\langle \delta b_{s}^{\dagger} \delta b_{s'}\right\rangle_{ste} \right],\label{eq:Stokes-Spectrum}
\end{equation}
where we have introduced the frequency $\omega_{st}=\omega_{l}-\omega_{s}+N\left(\Omega_{ss}^{+}+\Omega_{ss}^{-}\right)/2$
and the linewidth $\Gamma_{st}=\gamma_{s}+2\chi_{s}+N\left(\Gamma_{ss}^{-}-\Gamma_{ss}^{+}\right)$. 

The anti-Stokes spectrum can be computed in the same way. We start from
\begin{equation}
 S^{as}\left(\omega\right)\propto \omega^4 \Gamma_{ss}^{-}\bigl[N\mathrm{Re}S_{ss}^{as}\left(\omega-\omega_{l}\right)  +N\left(N-1\right)\mathrm{Re}S_{ss'}^{as}\left(\omega-\omega_{l}\right)\bigr].\label{eq:antiStokes-signal}
\end{equation}
The equations for the functions $S_{ss}^{as}\left(\omega\right)$ and $S_{ss'}^{as}\left(\omega\right)$ are obtained from eq \ref{eq:G2omega} and have the form 
\begin{align}
 & i\left(\omega-\tilde{\omega}_{s}^{*}\right)S_{ss}^{as}\left(\omega\right)+iv_{ss}^{\left(2\right)}\Bigl[S_{ss}^{as}\left(\omega\right) 
  +\left(N-1\right)S_{ss'}^{as}\left(\omega\right)\Bigr]=\left\langle \delta b_{s}^{\dagger}\delta b_{s}\right\rangle_{ste},  \label{eq:Ass_system}  \\
 & i\left(\omega-\tilde{\omega}_{s}^{*}\right)S_{ss'}^{as}\left(\omega\right)+iv_{ss}^{\left(2\right)}\Bigl[S_{ss}^{as}\left(\omega\right) 
  +\left(N-1\right)S_{ss'}^{as}\left(\omega\right)\Bigr]=\left\langle \delta b_{s}^{\dagger}\delta b_{s'} \right\rangle_{ste} . \label{eq:Ass'_system}
\end{align}
Subtracting the two equations, we obtain 
 \begin{equation}
S_{ss'}^{as}\left(\omega\right)=S_{ss}^{as}\left(\omega\right)+\frac{i}{\omega-\tilde{\omega}_{s}^{*}}  \left(\left\langle \delta b_{s}^{\dagger}\delta b_{s}\right\rangle_{ste} - \left\langle \delta b_{s}^{\dagger}\delta b_{s'}\right\rangle_{ste} \right).\label{eq:Gas2-Gas1}
\end{equation}

Inserting this expression back to eq \ref{eq:Ass_system}, we obtain 
\begin{align}
 & S_{ss}^{as}\left(\omega\right)=\frac{1}{i\left(\omega-\tilde{\omega}_{s}^{*}+Nv_{ss}^{\left(2\right)}\right)}\Bigl[\left\langle \delta b_{s}^{\dagger}\delta b_{s}\right\rangle_{ste}  - \left(N-1\right)\frac{v_{ss}^{\left(2\right)}}{\omega-\tilde{\omega}_{s}^{*}}\left(\left\langle \delta b_{s}^{\dagger}\delta b_{s}\right\rangle_{ste} - \left\langle \delta b_{s}^{\dagger}\delta b_{s'}\right\rangle_{ste} \right)\Bigr].
\end{align}
Using eq \ref{eq:antiStokes-signal}, we get 
\begin{equation}
  S^{as}\left(\omega\right)\propto \frac{\omega^4 \Gamma_{ss}^{-} \Gamma_{as}/2}{\left(\omega-\omega_{as}\right)^{2}+\left(\Gamma_{as}/2\right)^{2}} \left[N\left\langle \delta b_{s}^{\dagger} \delta b_{s}\right\rangle_{ste} +N\left(N-1\right)\left\langle \delta b_{s}^{\dagger} \delta b_{s'}\right\rangle_{ste} \right].\label{eq:Anti-Stokes-Spectrum}
\end{equation}
with frequency $\omega_{as}=\omega_{l}+\omega_{s}-N\left(\Omega_{ss}^{+}+\Omega_{ss}^{-}\right)/2$ and linewidth $\Gamma_{as}=\Gamma_{st}$. 
Integrating eqs \ref{eq:Stokes-Spectrum} and \ref{eq:Anti-Stokes-Spectrum} with respect to the frequency $\omega$, we obtain eqs 2 and 3 in the main text. In the integration, we assume the $\omega^4$ prefactor to be constant, as the Raman lines are spectrally very narrow. As before, in the main text we use the notation $n_s$ and $c_{ss'}$ to represent the incoherent phonon population and the noise correlation. 

\section{Collective oscillator model \label{sec:collective-oscillator-model}}
 Ref.\citenum{SI-RoelliP}  proposed that the vibration of many molecules can form collective oscillator modes with a bright  mode that couples to the plasmonic mode with an enhanced strength $g_{bright}\propto \sqrt{N} g_s$ (for $N$ identical molecules).   Following this idea, we demonstrate in this section that the collective oscillator model (in the absence of vibrational dephasing $\chi_{s}=0$)  leads to the same results as what we obtained in the previous section. To this end, we first introduce the collective modes by the collective operators $\delta B_{\beta}=\sum_{s}c_{\beta s}\delta b_{s}$, with $\delta b_{s}$ the noise operators of the individual molecules. The coefficients $c_{\beta s}$ define $N$ orthonormal vectors, e.g. $\left(c_{\beta1},...,c_{\beta N}\right)$, and satisfy $\sum_{s}c_{\beta s}c_{\beta' s}^*=\delta_{\beta\beta'}$ \citep{SI-TNeuman,SI-RoelliP,SI-TKipf}. The inverse expression is $\delta b_{s}=\sum_{\beta}c_{\beta s}^*\delta B_{\beta}$.  We can then rewrite the linearized interaction Hamiltonian in eq \ref{eq:linearzed-Hamiltonian}
as $\tilde{H}'_{int}=-\hbar\left(\alpha^{*}\delta a+\alpha\delta a^{\dagger}\right)\sum_{\beta} G_{\beta}\left(\delta B_{\beta}^{\dagger}+\delta B_{\beta}\right)$
with coefficients $G_{\beta}=\sum_{s}c_{\beta s}^* g_{s}$. Taking into account that $g_{s}$ is real and choosing
$c_{1s}=\frac{1}{\sqrt{\sum_{s'}g_{s'}^{2}}}g_{s}$ for $\beta=1$, we obtain $G_{1}=\sqrt{\sum_{s}g_{s}^{2}}$ and $G_{\beta>1}=0$, so that we can rewrite the Hamiltonian as $\tilde{H}'_{int}=-\hbar\left(\alpha^{*}\delta a+\alpha\delta a^{\dagger}\right)G_{1}\left(\delta B_{1}^{\dagger}+\delta B_{1}\right)$.
This Hamiltonian shows that the plasmon couples only with the first collective  mode, which thus can be called  the bright mode \citep{SI-RoelliP}. The other collective operators correspond to the dark modes.  

We next apply the  transformation $\delta b_{s}=\sum_{\beta}c_{\beta s}^*\delta B_{\beta}$ to the vibrational Hamiltonian $\hbar\sum_{s}\omega_{s}\delta b_{s}^{\dagger}\delta b_{s}$
and the Lindblad terms $\sum_{s}\left(\gamma_{s}/2\right)\left\{ \left(n_{s}^{th}+1\right)\mathcal{D}\left[\delta b_{s}\right]\rho+n_{s}^{th}\mathcal{D}\left[\delta b_{s}^{\dagger}\right]\rho\right\} $ in eq \ref{eq:effective-master-equation} governing the effective vibrational dynamics. We obtain terms of the type
$\sum_{\beta,\beta'}\left(\sum_{s}c_{\beta s}c_{\beta' s}^* \omega_{s}\right)\delta B_{\beta}^{\dagger}\delta B_{\beta'}$ or $\sum_{\beta,\beta'}\left(\sum_{s}c_{\beta s}c_{\beta' s}^* \gamma_{s}n_{s}^{th}\right)\delta B_{\beta}\rho_{v}\delta B_{\beta'}^{\dagger}$, which can  become diagonal ($\omega_{s}\sum_{\beta}\delta B_{\beta}^{\dagger}\delta B_{\beta}$ and $\gamma_{s}n_{s}^{th}\sum_{\beta}\delta B_{\beta}^{\dagger}\rho_{v}\delta B_{\beta}$, respectively) due to the orthogonality of the different collective  modes  only for identical molecules ($\omega_s$ and $\gamma_s n_s^{th}$ are independent of the molecular index $s$).  In the rest of this section, $g_{s},\omega_{s},\gamma_{s},n_{s}^{th}$ denote respectively the  optomechanical coupling, vibrational frequency, phonon decay rate and phonon thermal population for all identical molecules. 
Since $g_s$ is identical for all the molecules, we have the coefficient $c_{1s}=1/\sqrt{N}$ and $\delta B_{1}=\sum_{s}c_{1 s}\delta b_{s} = \sqrt{N} \delta b_s $.

 Using the quantum noise approach to eliminate the plasmon in the same way as in Section \ref{sec:derivation-qme}, we arrive at the effective master equation in terms of the collective modes
\begin{align}
 & \frac{\partial}{\partial t}\rho_{v}=-i\sum_{\beta}\left[\omega_{s}- \delta_{\beta,1}\frac{1}{2}N \left(\Omega_{ss}^{+} +\Omega_{ss}^{-}\right)\right]\left[\delta B_{\beta}^{\dagger}\delta B_{\beta},\rho_{v}\right]\nonumber \\
 & +\frac{\gamma_{s}}{2}\sum_{\beta}\left\{ \left(n_{s}^{th}+1\right)\mathcal{D}\left[\delta B_{\beta}\right]\rho_{v}+n_{s}^{th}\mathcal{D}\left[\delta B_{\beta}^{\dagger}\right]\rho_{v}\right\} \nonumber \\
 & +\frac{1}{2}\left(N \Gamma^{-}_{ss} \mathcal{D}\left[\delta B_{1}\right]\rho_{v}+N \Gamma^{+}_{ss}\mathcal{D}\left[\delta B_{1}^{\dagger}\right]\rho_{v}\right), \label{eq:master-equation-collective}
\end{align}
where  the parameters $\Gamma_{ss}^{-},\Gamma_{ss}^{+},\Omega_{ss}^{-},\Omega_{ss}^{+}$ were already introduced in eqs \ref{eq:omest}-\ref{eq:gammast} in Section \ref{sec:derivation-qme}. Notice that the Lindblad terms in this master equation depend only on the operators of a single collective mode. In contrast, the superoperator $\mathcal{D}$ in eq \ref{eq:effective-master-equation} involves the operators of two different molecular vibrations.

From this master equation, we can derive the equations for the phonon population of the collective modes 
\begin{equation}
\frac{\partial}{\partial t}\left\langle \delta B_{\beta}^{\dagger}\delta B_{\beta}\right\rangle =\delta_{\beta,1} N \Gamma^{+}_{ss} +n_{s}^{th}\gamma_{s} -\left[\gamma_{s}+\delta_{\beta,1} N \left(\Gamma^{-}_{ss}-\Gamma^{+}_{ss}\right)\right]\left\langle \delta B_{\beta}^{\dagger}\delta B_{\beta}\right\rangle .\label{eq:B+B}
\end{equation}
with the steady-state solution 
\begin{equation}
\left\langle \delta B_1^{\dagger}\delta B_1\right\rangle _{ste}=n_{s}^{th}+ N \frac{\Gamma^{+}_{ss}-\left(\Gamma^{-}_{ss}-\Gamma^{+}_{ss} \right)n_{s}^{th}}{\gamma_{s}+ N \left(\Gamma^{-}_{ss}-\Gamma^{+}_{ss}\right)}\label{eq:expectation-collective}
\end{equation}
for the bright collective mode and $\left\langle \delta B_\beta^{\dagger}\delta B_\beta\right\rangle_{ste} =n_{s}^{th}$ for  the dark modes ($\beta \neq 1$). 

The population of the collective modes can be also transformed into the incoherent phonon populations of the individual molecules $\left\langle \delta b_{s}^{\dagger}\delta b_{s}\right\rangle $. 
Using the transformation $\delta b_{s}=\sum_{\beta}c_{\beta s}^* \delta B_{\beta}$, the orthonormality condition $\sum_{s}c_{\beta s}c_{\beta' s}^*=\delta_{\beta\beta'}$ and the equality $\left\langle \delta b_{s}^{\dagger}\delta b_{s}\right\rangle =\frac{1}{N}\sum_{s}\left\langle \delta b_{s}^{\dagger}\delta b_{s}\right\rangle $  (because  all molecules are identical), we obtain 
\begin{equation}
 \left\langle \delta b_{s}^{\dagger}\delta b_{s}\right\rangle_{ste} =\frac{1}{N}\sum_{\beta}\left\langle \delta B_{\beta}^{\dagger}\delta B_{\beta}\right\rangle_{ste}  =n_{s}^{th}+ \frac{\Gamma^{+}_{ss}-\left(\Gamma^{-}_{ss}-\Gamma^{+}_{ss}\right)
 n_{s}^{th}}{\gamma_{s}+N\left(\Gamma^{-}_{ss}-\Gamma^{+}_{ss}\right)}.\label{eq:phonon-number}
\end{equation}
Using the coefficient $c_{1s}=1/\sqrt{N}$ for identical molecules, we obtain the relation $\left\langle \delta B_{1}^{\dagger}\delta B_{1}\right\rangle =\frac{1}{N}\sum_{s',s}\left\langle \delta b_{s}^{\dagger}\delta b_{s'}\right\rangle$. Since  $\left\langle \delta b_{s}^{\dagger}\delta b_{s}\right\rangle$, $\left\langle \delta b_{s}^{\dagger}\delta b_{s'}\right\rangle$  ($s\neq s'$) are identical for all the molecules and all the molecular pairs, respectively, we can further write the relation as $\left\langle \delta B_{1}^{\dagger}\delta B_{1}\right\rangle =\left\langle \delta b_{s}^{\dagger}\delta b_{s}\right\rangle +\left(N-1\right)\left\langle \delta b_{s}^{\dagger}\delta b_{s'}\right\rangle $, which allows us to evaluate the noise correlation of any molecular pair
\begin{equation}
\left\langle \delta b_{s}^{\dagger}\delta b_{s'}\right\rangle_{ste} = 
\frac{\Gamma^{+}_{ss}-\left(\Gamma^{-}_{ss}-\Gamma^{+}_{ss}\right)
 n_{s}^{th}}{\gamma_{s}+N\left(\Gamma^{-}_{ss}-\Gamma^{+}_{ss}\right)} .\label{eq:correlation}
\end{equation}
Thus, we have reproduced eqs \ref{eq:phonon-number-total}
and \ref{eq:correlation-total} (for $\chi_{s}=0$). 

Since only the bright mode couples with the plasmon, the Stokes and anti-Stokes signal are determined by this mode. If all the molecules are identical, we have $\delta B_{1}=\sum_{s}c_{1 s}\delta b_{s} = \sqrt{N} \delta b_s $ (with $c_{1s}=1/\sqrt{N}$) for this mode and thus can compute the Stokes and anti-Stokes spectrum as given by eq \ref{eq:St-b} and \ref{eq:As-b} with  $S^{st}\left(\omega\right)\propto \omega^4 N \Gamma^{+}_{ss}\mathrm{Re}\int_{0}^{\infty}d\tau e^{-i\left(\omega-\omega_{l}\right)\tau}\left\langle \delta B_{1}\left(\tau\right)\delta B_{1}^{\dagger}\left(0\right)\right\rangle $ and $S^{as}\left(\omega\right)\propto \omega^4 N \Gamma^{-}_{ss} \mathrm{Re}\int_{0}^{\infty}d\tau e^{-i\left(\omega-\omega_{l}\right)\tau}\left\langle \delta B_{1}^{\dagger}\left(\tau\right)\delta B_{1}\left(0\right)\right\rangle $. Applying the quantum regression theorem \citep{SI-PMeystre}, the two-time correlations  $\left\langle \delta B_{1}\left(\tau\right)\delta B_{1}^{\dagger}\left(0\right)\right\rangle$, $\left\langle \delta B_{1}^{\dagger}\left(\tau\right)\delta B_{1}\left(0\right)\right\rangle $ satisfy the same equations as the amplitudes of the collective modes $\left\langle \delta B_{1}\right\rangle $,$\left\langle \delta B_{1}^{\dagger}\right\rangle $. The equations for the amplitudes can be derived from eq \ref{eq:master-equation-collective} and have the form
\begin{align}
 & \frac{\partial}{\partial t}\left\langle \delta B_1 \right\rangle =-i\Bigl[\omega_{s}-\frac{1}{2} N \left(\Omega^{+}_{ss}+\Omega^{-}_{ss}\right)  -i\frac{1}{2} \left(\gamma_{s}+N (\Gamma^{-}_{ss}-\Gamma^{+}_{ss})\right)\Bigr]\left\langle \delta B_1 \right\rangle, \\
 & \frac{\partial}{\partial t}\left\langle \delta B_1^{\dagger}\right\rangle = i\Bigl[\omega_{s}-\frac{1}{2} N \left(\Omega^{+}_{ss} + \Omega^{-}_{ss}\right) 
  +i\frac{1}{2}\left(\gamma_{s}+N(\Gamma^{-}_{ss}-\Gamma^{+}_{ss})\right)\Bigr]\left\langle \delta B_1^{\dagger}\right\rangle.
\end{align}
Thus, the equations for the correlations are 
\begin{align}
 & \frac{\partial}{\partial t}\left\langle \delta  B_{1}\left(\tau\right)\delta B_{1}^{\dagger}\left(0\right) \right\rangle =-i\Bigl[\omega_{s}-\frac{1}{2} N \left(\Omega^{+}_{ss}+\Omega^{-}_{ss}\right) 
 -i\frac{1}{2} \left(\gamma_{s}+N (\Gamma^{-}_{ss}-\Gamma^{+}_{ss})\right)\Bigr]\left\langle \delta  B_{1}\left(\tau\right)\delta B_{1}^{\dagger}\left(0\right) \right\rangle ,\\
 & \frac{\partial}{\partial t}\left\langle \delta B_{1}^{\dagger}\left(\tau\right)\delta B_{1}\left(0\right) \right\rangle = i\Bigl[\omega_{s}-\frac{1}{2} N \left(\Omega^{+}_{ss} + \Omega^{-}_{ss}\right) 
  +i\frac{1}{2}\left(\gamma_{s}+N(\Gamma^{-}_{ss}-\Gamma^{+}_{ss})\right)\Bigr]\left\langle \delta  B_{1}^{\dagger}\left(\tau\right)\delta B_{1}\left(0\right) \right\rangle.
\end{align}
 We can solve the above equations with the initial conditions $1+\left\langle \delta B_{1}^{\dagger}\delta B_{1}\right\rangle_{ste},\left\langle \delta B_{1}^{\dagger}\delta B_{1}\right\rangle_{ste}$ and obtain the following expressions for the Stokes and anti-Stokes spectrum 
\begin{equation}
S^{st}\left(\omega\right)\propto\frac{\omega^4 \Gamma^{+}_{ss} \left(\Gamma_{st}/2\right)}{\left(\omega-\omega_{st}\right)^{2}+\left(\Gamma_{st}/2\right)^{2}} N \left(1+\left\langle \delta B_{1}^{\dagger}\delta B_{1}\right\rangle_{ste} \right), \label{eq:stokes-spectrum-collective}
\end{equation}
\begin{equation}
S^{as}\left(\omega\right)\propto\frac{\omega^4 \Gamma^{-}_{ss} \left(\Gamma_{as}/2\right)}{\left(\omega-\omega_{as}\right)^{2}+\left(\Gamma_{as}/2\right)^{2}} N \left\langle \delta B_{1}^{\dagger}\delta B_{1}\right\rangle _{ste},\label{eq:anti-stokes-spectrum-collective}
\end{equation}
with $\omega_{st}=\omega_{l}-\omega_{s}+\frac{N}{2}\left(\Omega^{+}_{ss}+\Omega^{-}_{ss}\right)$
, $\omega_{as}=\omega_{l}+\omega_{s}-\frac{N}{2}\left(\Omega^{+}_{ss} +\Omega^{-}_{ss}\right)$, $\Gamma_{st}=\Gamma_{as}=\gamma_{s}+N(\Gamma^{-}_{ss}-\Gamma^{+}_{ss})$, and the steady-state phonon populations given by eq \ref{eq:expectation-collective}. These results are identical to eqs \ref{eq:Stokes-Spectrum} and \ref{eq:Anti-Stokes-Spectrum} in the Section \ref{sec:formulas-identical} for $\chi_s=0$.

Last, we consider briefly the influence of the vibrational dephasing on
the dynamics of the collective modes  (for identical molecules). To do so, we use the transformation
$\delta b_{s}=\sum_{\beta}c_{\beta s}^*\delta B_{\beta}$ to rewrite the Lindblad term
$\chi_{s}\sum_{s}\mathcal{D}\left[\delta b_{s}^{\dagger}\delta b_{s}\right]\rho_{v}$
for the vibrational dephasing. As a result, we obtain terms like $\sum_{\alpha',\alpha}\sum_{\beta',\beta}\left[\sum_{s}c_{\alpha's} c_{\alpha s}^*c_{\beta' s} c_{\beta s}^*\right]\delta B_{\alpha'}^{\dagger}\delta B_{\alpha}\rho\delta B_{\beta'}^{\dagger}\delta B_{\beta}$
and $\sum_{\alpha',\alpha}\sum_{\beta',\beta}\left[\sum_{s}c_{\alpha's} c_{\alpha s}^*c_{\beta' s} c_{\beta s}^*\right]\delta B_{\alpha'}^{\dagger}\delta B_{\alpha}\delta B_{\beta'}^{\dagger}\delta B_{\beta}\rho$. Since there are four coefficients in the sums in the brackets, we cannot use the orthogonal condition to eliminate the non-diagonal terms. Therefore, the vibrational dephasing couples the different collective modes and the equations become complicated to solve. Thus, we found it more convenient to obtain the general solution including dephasing by working in the base of individual molecules as discussed in Section \ref{sec:formulas-identical}, instead of using the collective base  as considered in this section.

\section{Supplementary numerical results\label{sec:SNumRes}}

\begin{figure}
\begin{centering}
\includegraphics[scale=0.32]{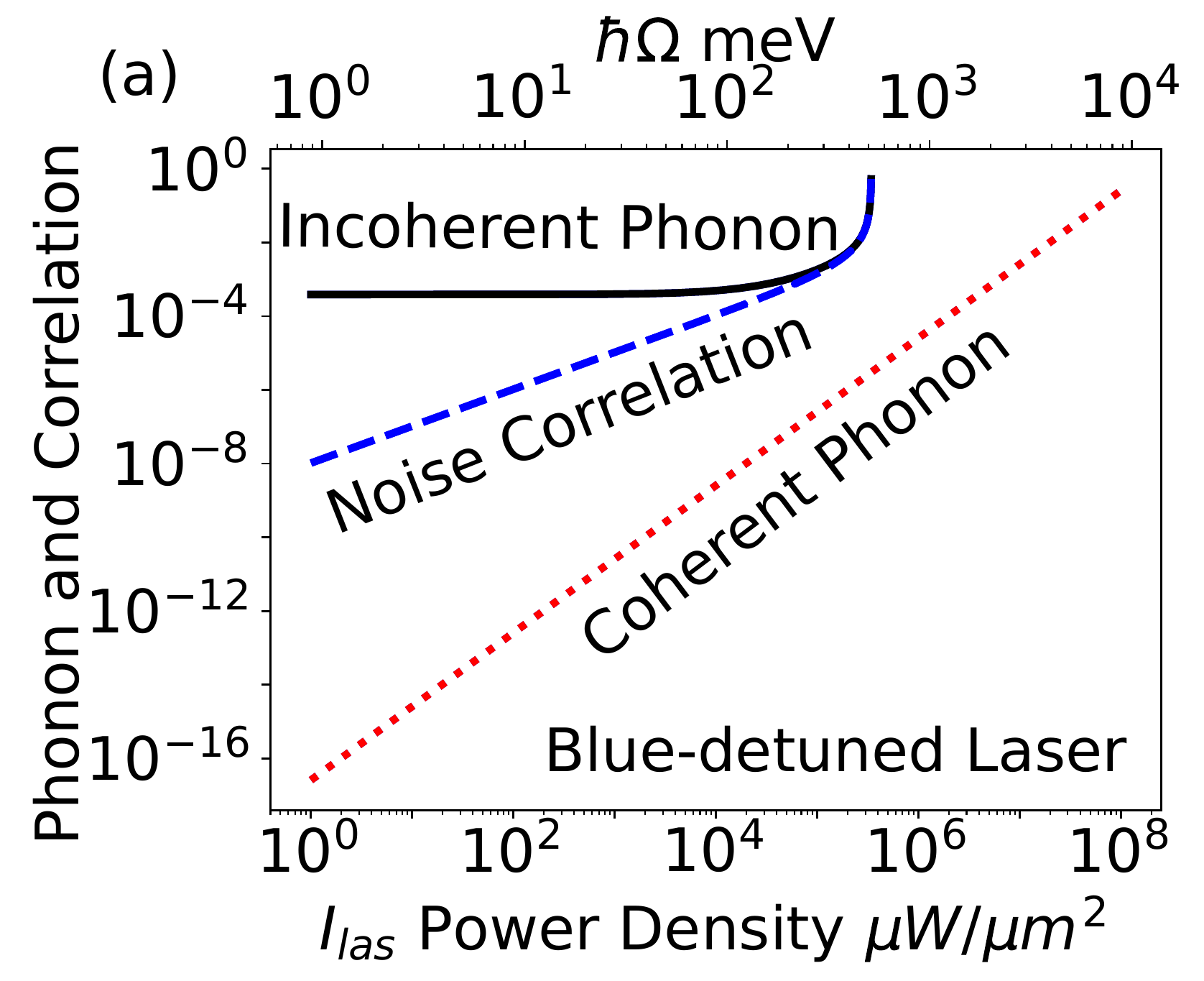}
\includegraphics[scale=0.32]{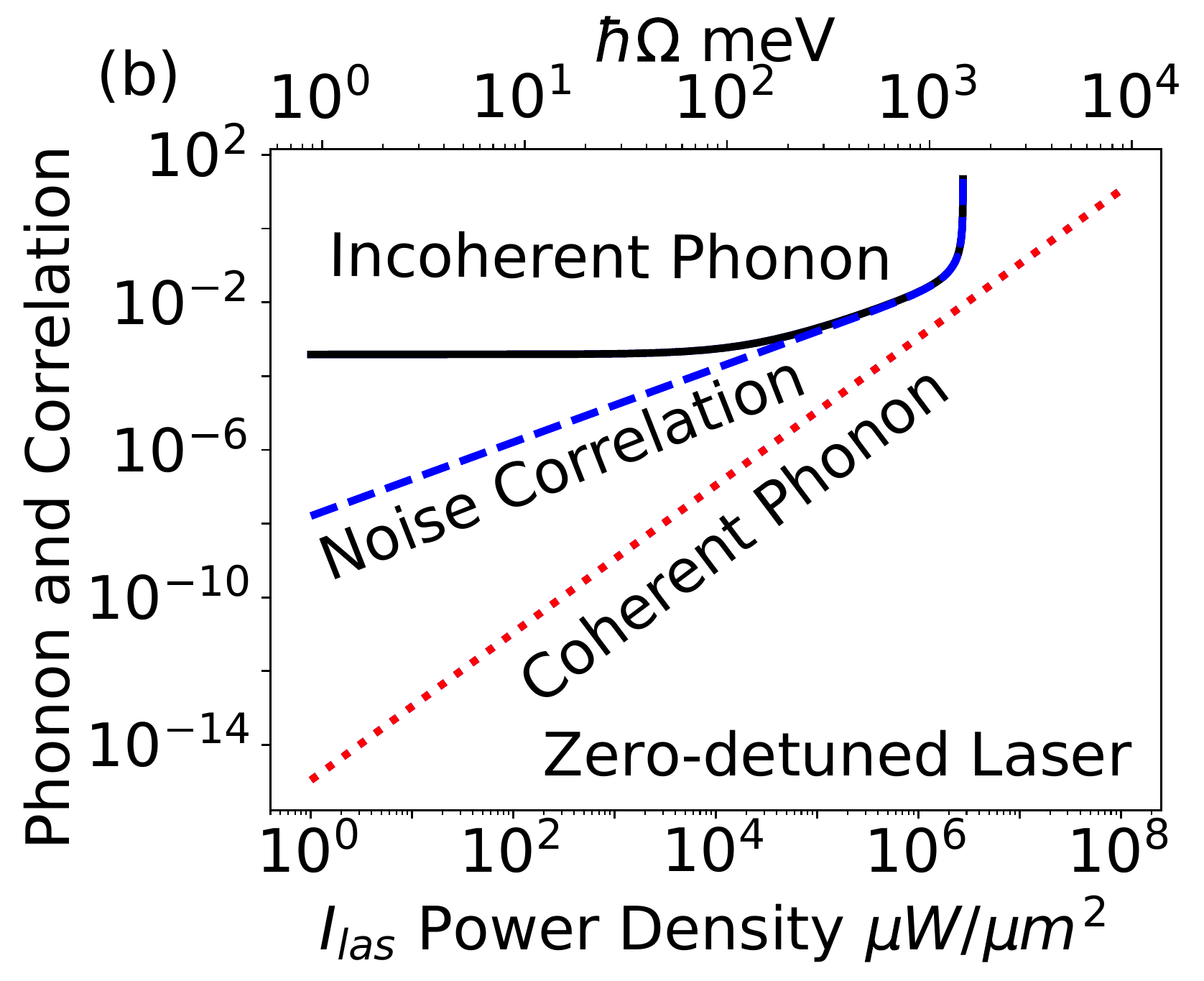}\includegraphics[scale=0.32]{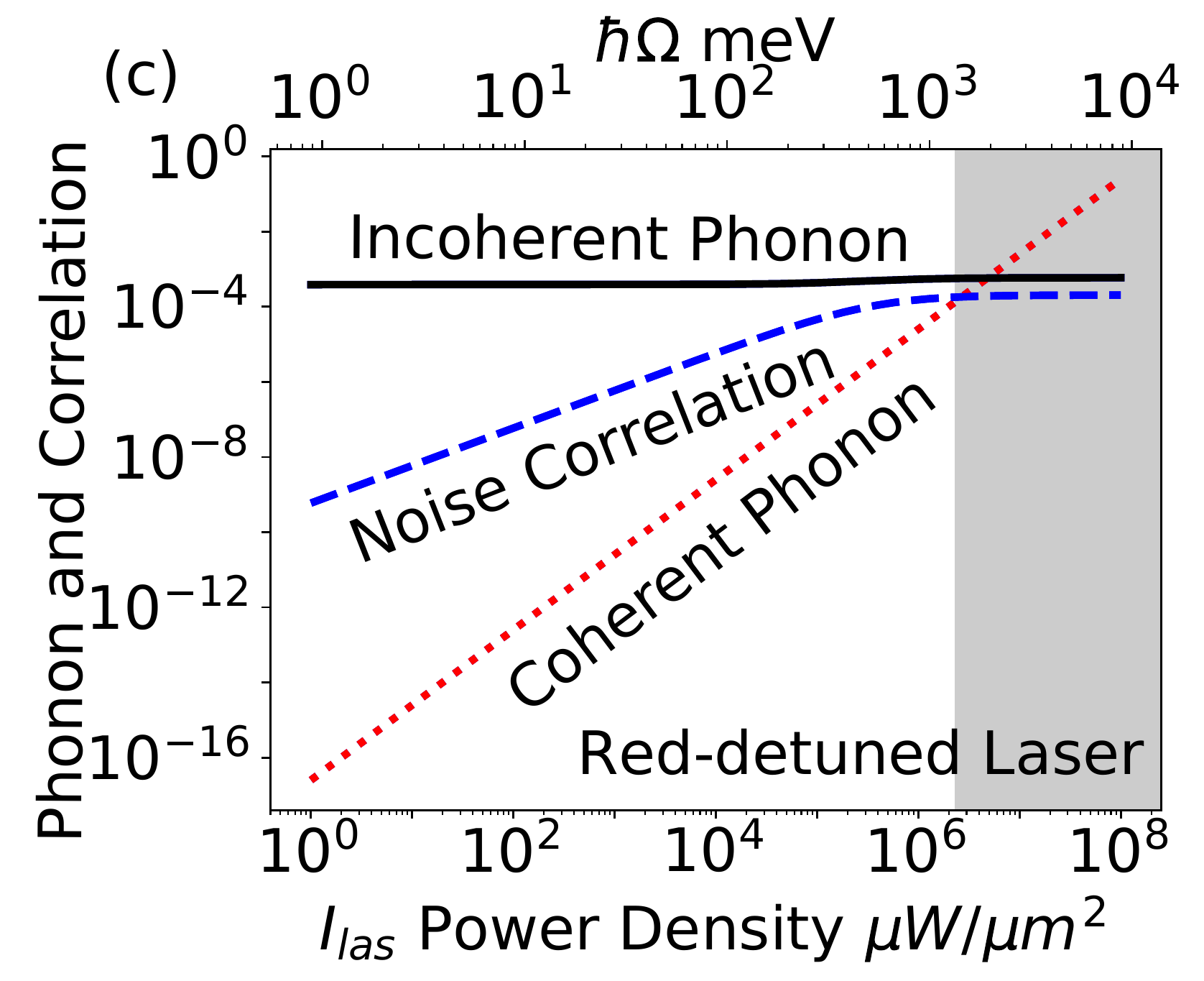}
\par\end{centering}
\caption{\label{fig:comparison}  Incoherent phonon population  $n_s = \left\langle \delta b_{s}^{\dagger}\delta b_{s}\right\rangle$ (black solid lines), noise correlation $c_{ss'} = \left\langle \delta b_{s}^{\dagger}\delta b_{s'}\right\rangle$ ($s \neq s'$, blue dashed lines) and coherent phonon $\left| \beta_s \right|^{2}$ (red dotted lines) as a function of the driving $\Omega$ (top axis) or the laser power density $I_{las}$ (bottom axis) for  a laser of frequency $\omega_l$ that is (a) blue-, (b) zero- and (c) red-detuned from the plasmonic frequency $\omega_c$ [with the frequency detunings $\hbar \Delta =\hbar(\omega_l -\omega_c) = -236,0,236$ meV], respectively. Here, the detuning is defined with respect to $\omega_c$, not the shifted value $\omega'_c$.  The gray area in (c) indicates the regimes where the coherent phonon might contribute to the Raman scattering. In the calculations we consider $N=300$ identical molecules, no homogeneous broadening $2\chi_s =0$ and the temperature $T=290 K$. Other parameters are same as used in the main text.}
\end{figure}

\subsection{Comparison of incoherent and coherent phonon population \label{sec:comparison}}

The expressions we have derived for the  the Raman signal depend on the noise properties, i.e. the incoherent phonon population $n_s = \left\langle \delta b_{s}^{\dagger}\delta b_{s}\right\rangle$ and the noise correlation $c_{ss'} = \left\langle \delta b_{s}^{\dagger}\delta b_{s'}\right\rangle$ (with $s\neq s'$), which are defined with the noise operators $\delta b_s = b_s - \beta_s$.
However, the coherent amplitudes could in principle also play a role e.g. in the chemical reactivity \citep{SI-FlemingCF} of the molecule or in the  Raman signal when going beyond the linearized Hamiltonian. To verify that these effects are negligible in our studies, we show that the incoherent phonon population is generally much larger than the coherent phonon population.

We compare in Figure \ref{fig:comparison}  the incoherent phonon population $n_s = \left\langle \delta b_{s}^{\dagger}\delta b_{s}\right\rangle$ (black solid lines), the noise correlation $c_{ss'} = \left\langle \delta b_{s}^{\dagger}\delta b_{s'}\right\rangle$  (blue dashed lines) with the coherent phonon population $|\beta_s|^2$ (red dotted lines) for systems with $300$ molecules illuminated by a (a) blue-, (b) zero- and (c) red-detuned laser of increasing intensity $I_{las}$. In the  two former cases, the incoherent phonon population and the noise correlation dominate for all the laser intensity. In particular, under the zero-detuned laser illumination these quantities actually diverge for strong laser intensity because the plasmon frequency is shifted by the optomechanical coupling (see Section S5.5 for more information). Only for the red-detuned laser with very strong intensity  $I_{las}>2\times 10^6 \mu W/\mu m^2$ does the coherent phonon population become larger than the incoherent phonon and the noise correlation. We also note that for such strong illumination the laser-plasmon coupling $\Omega$ (upper axis)  becomes comparable with the plasmon frequency $\omega_c$. In this case, the rotating wave approximation as used in our theory might  fail and the rotating term in the laser-plasmon coupling might start contributing to the response. 

\subsection{Shift, narrowing and broadening of SERS lines \label{sec:shiftsnarrowingbroadening}}
\begin{figure}
\begin{centering}
\includegraphics[scale=0.4]{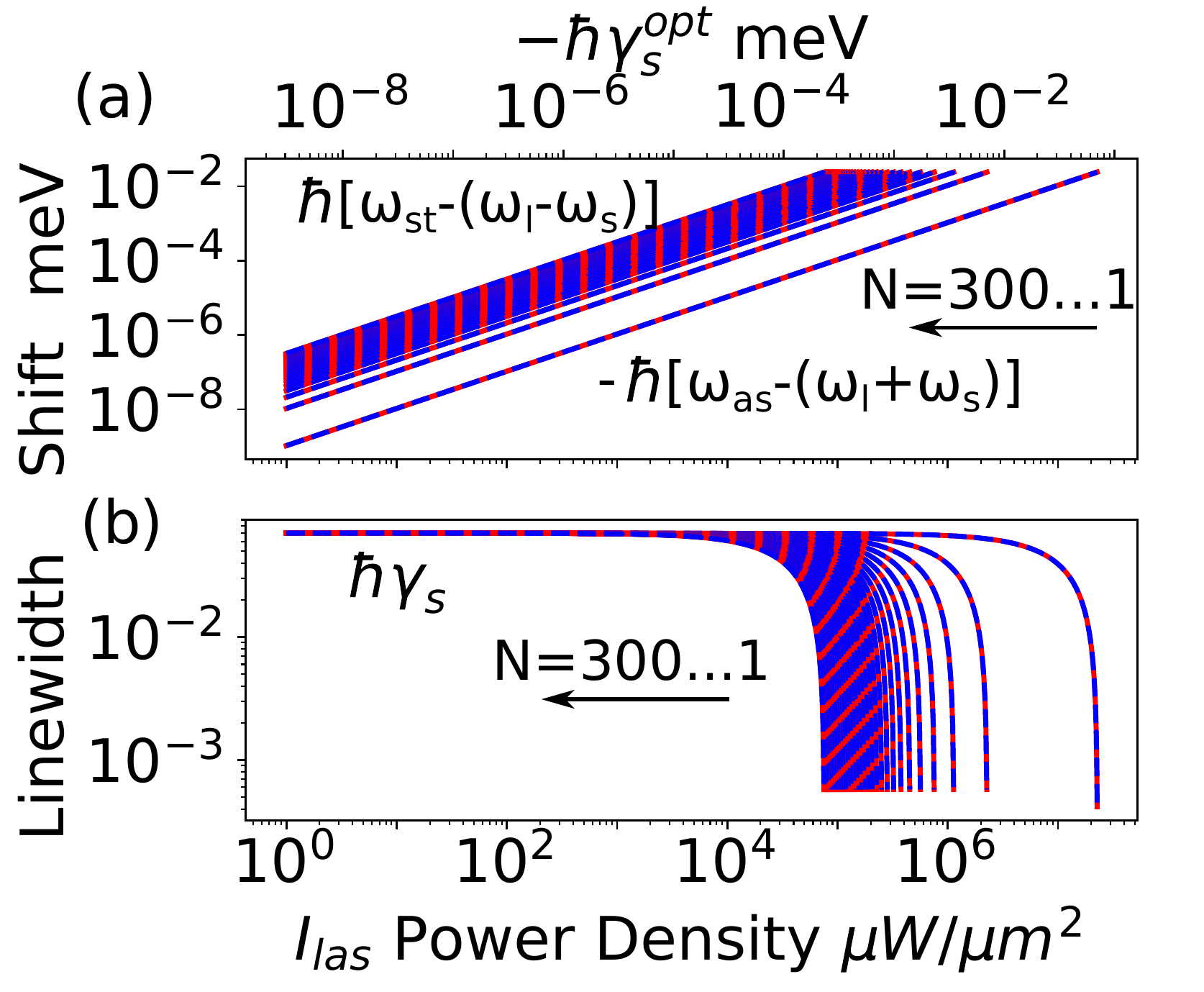}
\includegraphics[scale=0.4]{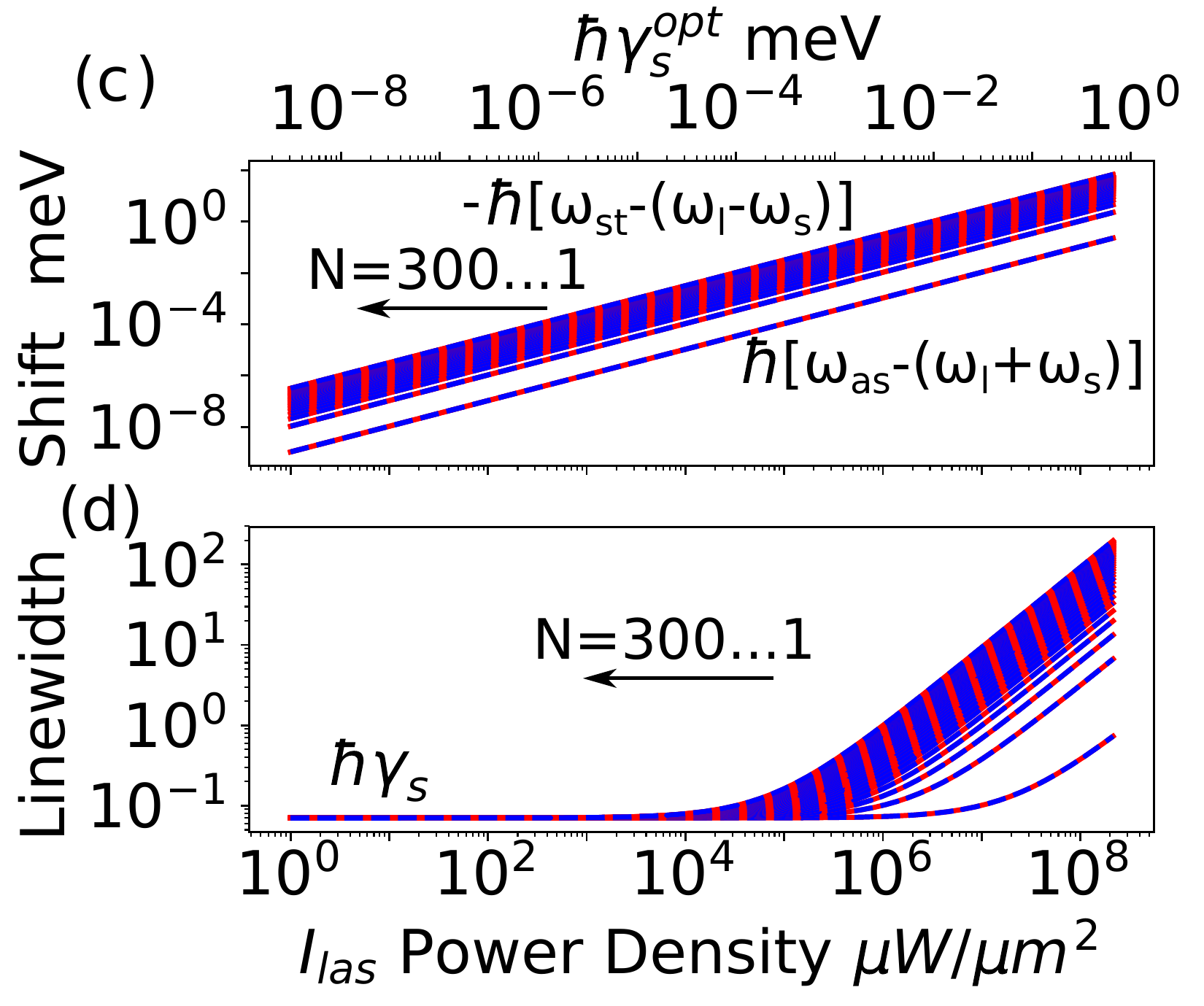}
\par\end{centering}
\caption{ \label{fig:shiftlinewidth} Spectral properties of the SERS lines for $N=1,10,20,...300$ molecules illuminated by a blue-detuned (a,b) and red-detuned (c,d) laser. We show the frequency shift $\omega_{st} - (\omega_l-\omega_s)$, $-\left[\omega_{as} - (\omega_l+\omega_s)\right]$ (note the different sign) 
and the linewidth $\Gamma_{st}$, $\Gamma_{as}$ of the Stokes (blue lines) and anti-Stokes (red lines) lines versus laser illumination $I_{las}$ (bottom axis) and the optomechanical damping rate $\Gamma_{s}^{opt}$ (top axis). For the blue- and red-detuned laser, the frequency detunings are $\hbar \Delta =\hbar(\omega_l -\omega'_c) = -236,236$ meV, respectively. We consider the temperature $T=290 K$ and no homogeneous broadening $2\chi_{s}=0$. Other parameters are the same as used in the main text. } 
\end{figure}

In the main text, we focused on analyzing the intensity integrated over the SERS lines. On the other hand, the analysis in Section \ref{sec:formulas-identical} shows that the central frequency $\omega_{st}$ ($\omega_{as}$) and the linewidth $\Gamma_{st}$ ($\Gamma_{as}$) of the  Stokes (anti-Stokes) lines also depend on the number of molecules $N$, as given by $\omega_{st}=\omega_l - \omega_s + N(\Omega_{ss}^{+} +\Omega_{ss}^{-})/2$ [$\omega_{as}=\omega_l + \omega_s - N(\Omega_{ss}^{+} +\Omega_{ss}^{-})/2$]  and $\Gamma_{st}= \gamma_s +2\chi_s + N \Gamma_s^{opt}$ ($\Gamma_{as}=\Gamma_{st}$).  We plot in Figure \ref{fig:shiftlinewidth} how these parameters vary  with laser intensity $I_{las}$ for $N=1,10,20,...300$ molecules. Figure \ref{fig:shiftlinewidth}a shows that under blue-detuned laser illumination the Stokes (blue lines) and anti-Stokes (red lines) Raman lines become blue- and red-shifted, respectively [with a positive shift $\omega_{st} - (\omega_l-\omega_s)$  and a negative shift $\omega_{as} - (\omega_l+\omega_s)$, respectively] and the shift increases linearly with the laser intensity $I_{las}$ and $N$.  For the anti-Stokes signal we plot $-\left[\omega_{as} - (\omega_l-\omega_s)\right]$, i.e. the shift multiplied by minus one, to  plot the data in a logarithmic scale.
 In addition, the linewidth of these Raman lines (Figure \ref{fig:shiftlinewidth}b) decreases linearly from $\hbar \gamma_s = 0.07$ meV to zero with increasing $I_{las}$ and increasing $N$. The spectrum thus narrows for increasing laser intensity and number of molecules. In contrast, for red-detuned laser illumination we obtain the opposite trends: as $N$ and $I_{las}$ increases the Stokes and anti-Stokes lines become red- and blue-shifted, respectively, (Figure \ref{fig:shiftlinewidth}c) and the linewidth increases, i.e. broader spectrum, (Figure \ref{fig:shiftlinewidth}d). We don't show the results for zero-detuned laser illumination because in this case both the frequency and the width of the Raman lines remain constant.

\begin{figure}
\begin{centering}
\includegraphics[scale=0.6]{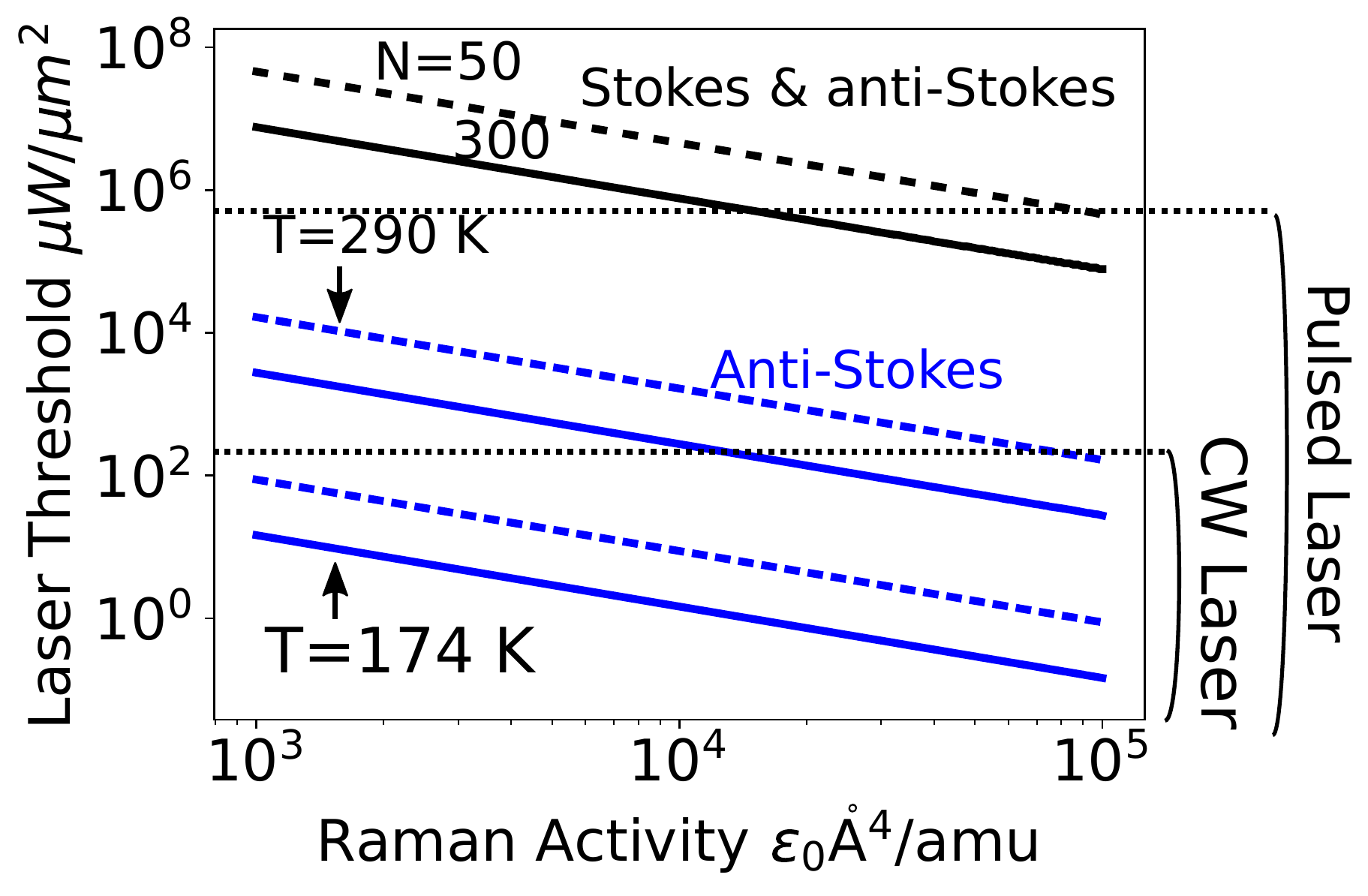}
\par\end{centering}
\caption{\label{fig:laserthreshold} Laser threshold power density to achieve the divergence of the SERS signal (i.e. parametric instability, black lines) and the quadratic scaling of the anti-Stokes SERS signal (blue lines) at $T=290$ K and $T=174$ K (i.e. thermal energy $k_B T= 25,15$ meV) versus the Raman activity of the molecular vibrational mode in systems with $N=50$ (dashed lines) and $300$ (solid lines) molecules. We also indicate the power density achievable in the experiments with a pulsed  \citep{SI-ALombardi} and continuous-wave (CW) \citep{SI-FBenz} laser.
We consider the blue-detuned laser with the frequency detuning $\hbar \Delta =\hbar(\omega_l -\omega'_c) = 236$ meV and no homogeneous broadening $\chi_{s}=0$. Other parameters are the same as used in the main text.}
\end{figure}

\subsection{Laser threshold  for molecules with different Raman activity \label{sec:laserthreshold}}
We examine in this section how the emergence of the collective effects depend on the temperature and properties of the molecules. In the main text, we have used a large Raman activity $10^5 \epsilon_{0}\mathring{\mathrm{A}}^{4}/\mathrm{amu}$  that would account for both the chemical enhancement and a possible  conformational change of the biphenyl-4-thiol molecule that we have chosen as reference \cite{SI-ALombardi,SI-YFang}. However, the Raman activity can change dramatically depending on the particular molecule considered and its exact interaction with the gold atoms. 

We first focus on the threshold laser intensity $I_{thr}$ to achieve the parametric instability, i.e. the divergence of the phonon population and the SERS signal. $I_{thr}$ can be computed  from the condition $N|\Gamma^{opt}_s| = \gamma_s$ (see the main text). The black lines in Figure \ref{fig:laserthreshold} indicate $I_{thr}$ for the Raman activity increasing from $10^3$ to $10^5$ $\epsilon_{0}\mathring{\mathrm{A}}^{4}/\mathrm{amu}$, and systems with $N=50$ (dashed lines) and $N=300$ molecules (solid lines). Other parameters take the same values as in the main text. We see that $I_{thr}$ is always very large but can be reduced by increasing the number of molecules $N$ and the Raman activity. For reference, we also indicate in the figure the typical intensity ranges that are accessible with pulsed \citep{SI-ALombardi} and continuous-wave (CW) \citep{SI-FBenz} laser. 

Next, we consider the laser intensity $I_{sr}$ necessary to observe  the superradiant $N^2$ scaling of the anti-Stokes signal. As indicated in the main text, $I_{sr}$ can be computed from the condition $N\Gamma^{st}_s =n^{th}_s \gamma_s$.  The blue lines in Figure \ref{fig:laserthreshold} show the evolution of $I_{sr}$ for the same Raman activity and number of molecules as considered for the threshold intensity $I_{thr}$ and two different temperatures $T=290$ K  and $T=174$ K (corresponding to thermal energy $k_B T= 25$ meV and $15$ meV), which reduces the  thermal phonon population $n^{th}_s$ from $3.9\times 10^{-4}$ to $2.0\times 10^{-6}$. We see that the superradiance of the anti-Stokes signal is significantly easier to be achieved than the parametric instability, and can be even accessed with continuous laser illumination, particularly for sufficiently low temperature.

\subsection{Collective effects landscape under blue-, zero- and red-detuned laser illumination \label{sec:scalingSERS}}

\begin{figure}
\includegraphics[scale=0.32]{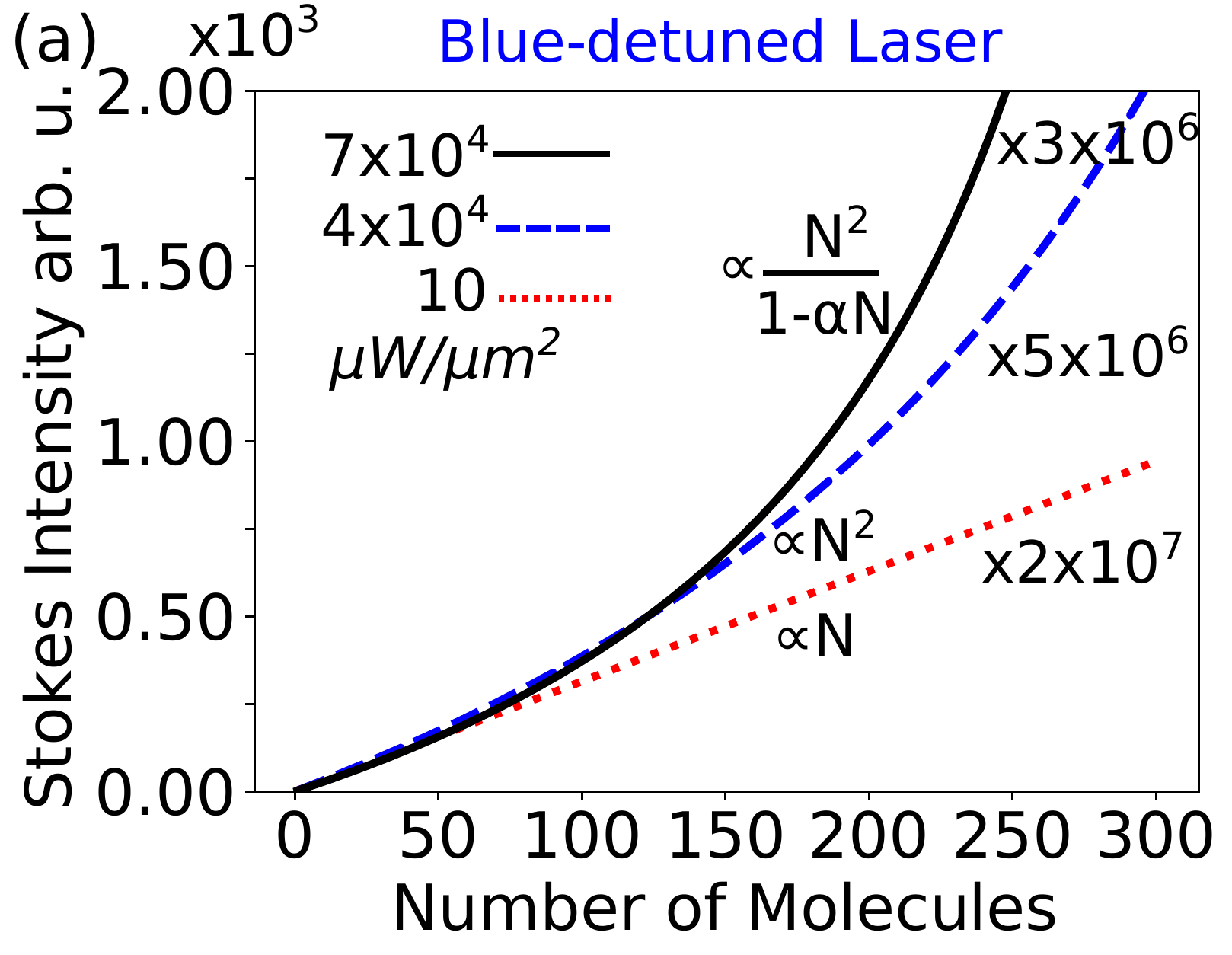}
\includegraphics[scale=0.32]{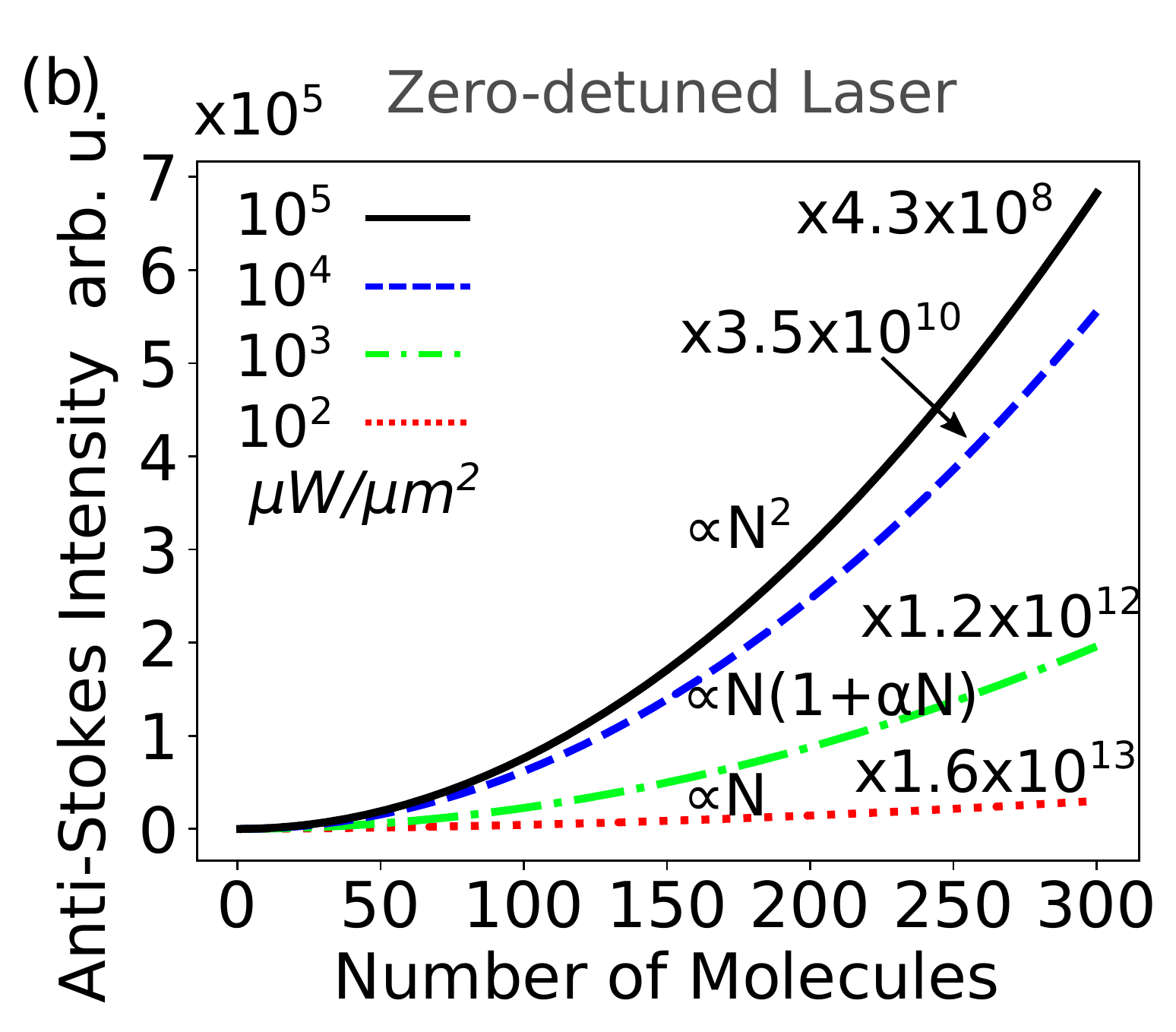}
\includegraphics[scale=0.32]{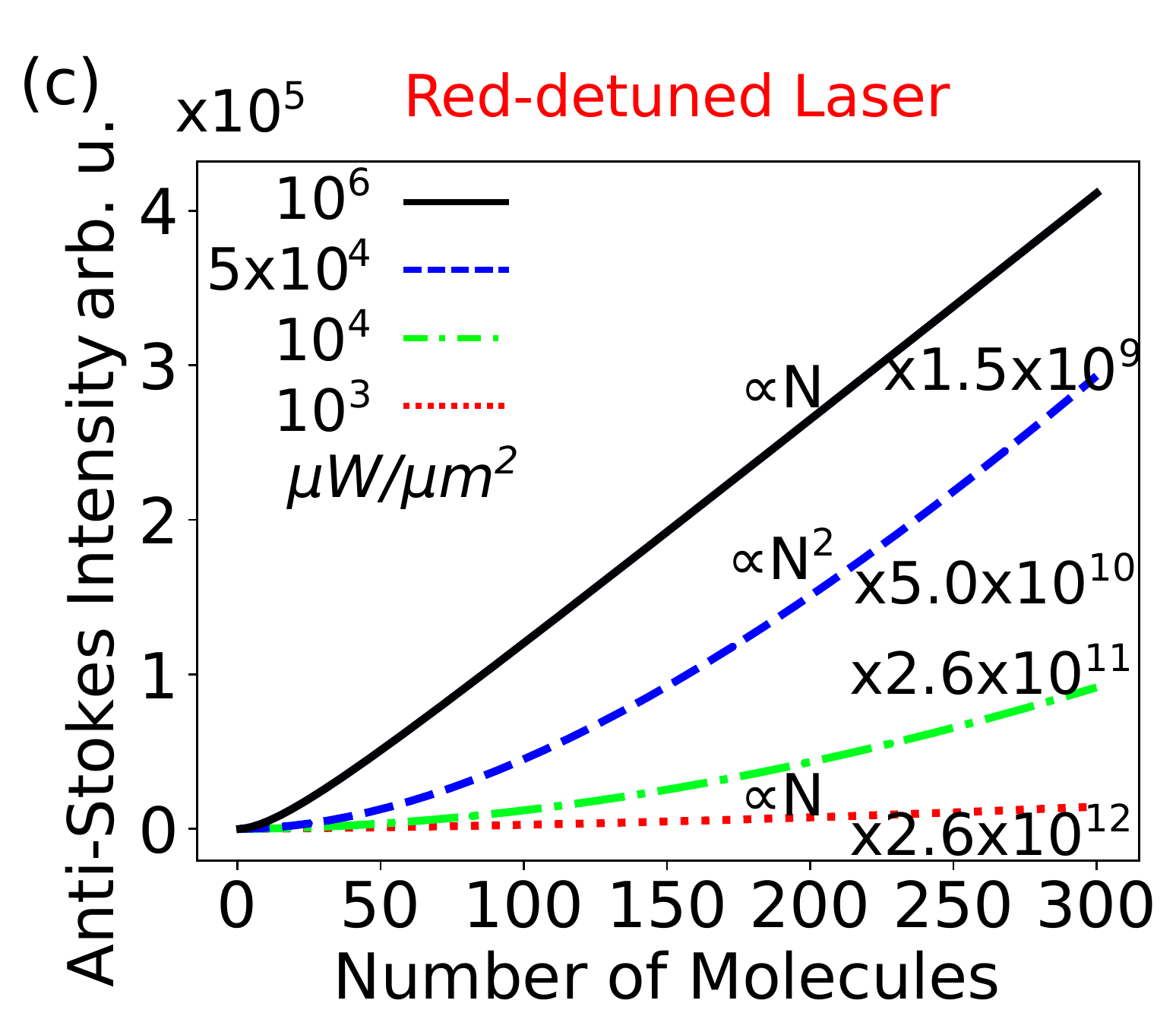}
\caption{
\label{fig:molecular-number-Section}Stokes intensity (a) and anti-Stokes intensity (b,c), scaled as indicated in the panels, as a function of the number of molecules $N$ (from $1$ to $300$) for system illuminated by a blue- (a), zero- (b) and red-detuned  (c) laser of intensities $I_{las}$ as indicated in the legends. These results correspond to vertical cuttings of the curves shown in Figure 2d-f in the main text. For the blue-, zero- and red-detuned laser, the frequency detunings are $\hbar \Delta =\hbar(\omega_l -\omega'_c) = 236,0,-236$ meV, respectively.  We consider the temperature $T=290 K$ and no homogeneous broadening $2\chi_{s}=0$. Other parameters are same as used in the main text.}
\end{figure}

In Figure 3 in the main text, we summarized the different scaling of the anti-Stokes signal with the number of molecules $N$ for a blue-detuned laser of various intensities $I_{las}$. In Figure \ref{fig:molecular-number-Section} we provide additional results  for (a) the Stokes intensity under blue-detuned illumination, and (b,c) the anti-Stokes intensity  for various illumination with a (b) zero- and (c) red-detuned laser. Comparing Figure \ref{fig:molecular-number-Section}a with Figure 3 in the main text, we see that larger intensities are required to observe the quadratic $N^2$ scaling for the Stokes than for the anti-Stokes signal. For such large intensities, it becomes difficult to distinguish between this $N^2$ scaling of the Stokes signal and the $\propto N^2/(1-\alpha N)$ scaling that is typical of the parametric instability.
Figure \ref{fig:molecular-number-Section}b shows that for the zero-detuned laser  the scaling of the anti-Stokes signal progresses from the linear $\sim N$ to super-linear $\sim N(1+N\alpha)$ [with $\alpha= \Gamma^{+}_{ss}/(\gamma_s n_s^{th})$] and finally to quadratic scaling $\sim N^2$ with increasing laser intensity. Last, we show in Figure \ref{fig:molecular-number-Section}c  that for the red-detuned laser  the scaling changes from the linear $\sim N$ to quadratic $\sim N^2$ and finally to linear scaling $\sim N$ again with increasing laser intensity. 

\begin{figure}[!ht]
\begin{centering}
\includegraphics[scale=0.9]{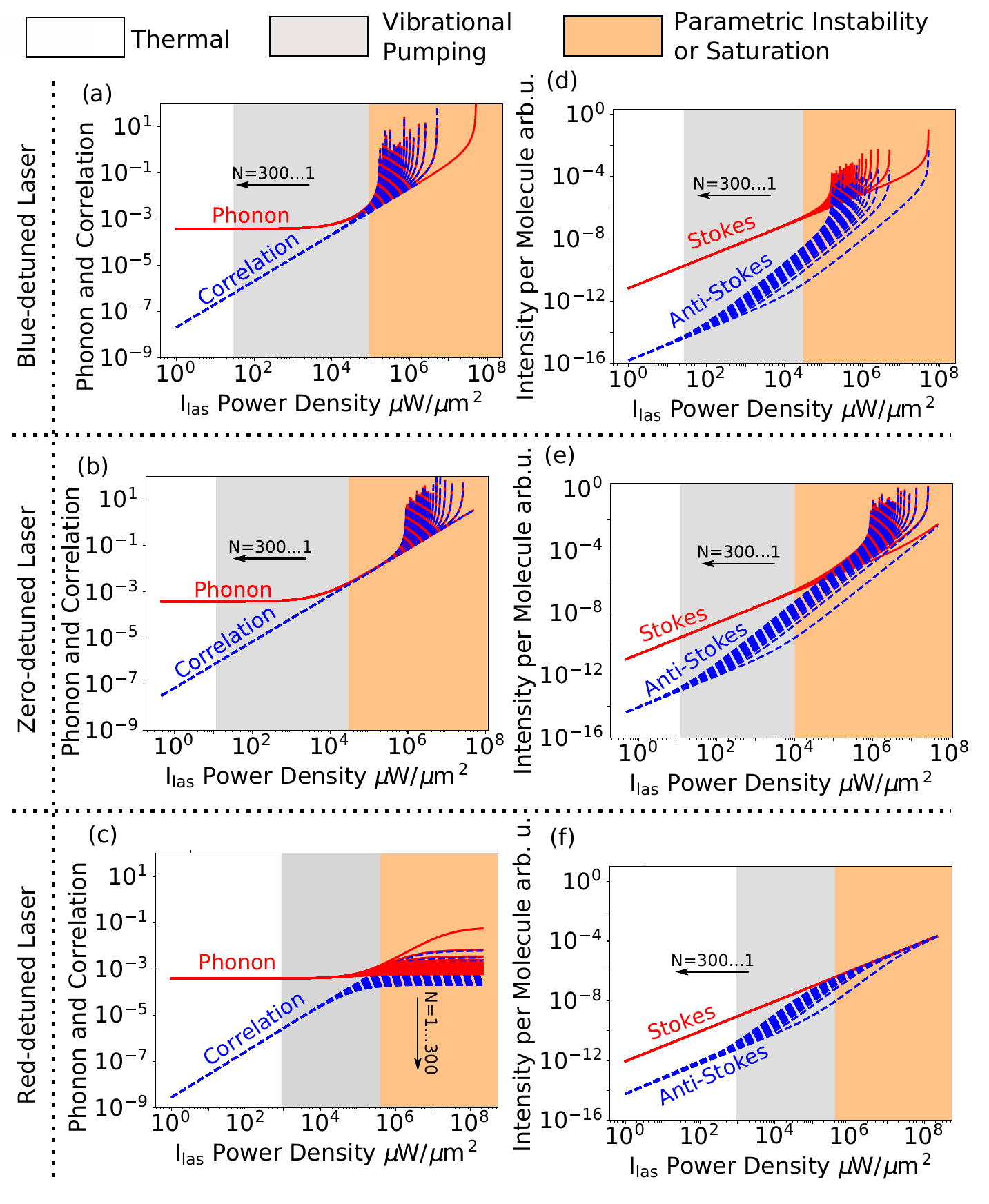}
\end{centering}
\caption{\label{fig:plasmon-shift} Same as Figure 2 in the main text except that the detuning $\Delta \omega = \omega_{l} - \omega_c$ is defined here as the difference between the laser frequency $\omega_l$ and the plasmon frequency $\omega_c$ (instead of the difference between the laser frequency and the shifted plasmon frequency $\omega_c'$).} 
\end{figure}

\subsection{Influence of phonon-induced plasmon shift on Stokes and anti-Stokes intensity \label{sec:plasmon-shift}}

We have discussed in Section \ref{sec:effectivemeq} that the molecular optomechanical coupling can shift the plasmonic resonant frequency  $\omega_c'=\omega_c - 2 \sum_s g_s {\rm Re} [\beta_s]$ by  $2 \sum_s g_s {\rm Re} [\beta_s]$  (which is $2 N g_s {\rm Re} [\beta_s] $ for identical molecules). Since the coherent phonon population $|\beta_s|^2$ scales linearly with the laser intensity $I_{las}$ (via the coherent plasmon amplitude $\alpha_l$, see Section \ref{sec:effectivemeq}), the shift scales also with the laser intensity and this complicates the analysis of the Stokes and anti-Stokes spectrum. Thus, to simplify the analysis we have considered the detuning of the laser frequency relative to the shifted plasmonic frequency $\omega_c'$ in the main text. In an experiment, this can actually be done by adequately tuning the laser frequency as $I_{las}$ increases.

To understand the effect of this phonon-induced plasmonic shift, we plot  in Figure \ref{fig:plasmon-shift} the dependence of the results on $I_{las}$ and $N$ for different laser frequencies. These results are calculated for the same conditions as those in Figure 2 of the main text, except that here  we tune  the laser frequency $\omega_l$ relative to the original plasmon frequency $\omega_c$, instead of  to the shifted one $\omega_c'$.  We see that the only significant difference compared to Figure 2 occurs for zero-detuned illumination ($\omega_l-\omega_c=0$) and very large laser intensity (Figure \ref{fig:plasmon-shift}b,e). Here, we see a divergence of the results (i.e. the incoherent phonon, the noise correlation and the Stokes and anti-Stokes intensity) while such a divergence does not occur for $\omega_l-\omega'_c=0$ (Figure 2b,e).  We notice that such a divergence is similar to what found under  blue-detuned laser illumination, i.e. parametric instability.  Indeed, this divergence can be understood as the consequence of the blue-detuned illumination with respect to the shifted frequency $\omega_c'$, which reduces due to the increased phonon-induced shift $2 N g_s {\rm Re} [\beta_s]$ as the laser intensity grows. On the other hand, this  shift is much smaller in our calculations than the detuning $|\omega_c - \omega_l|=236$ meV we choose for the  blue- and red-detuned illumination, and thus it does not affect significantly the results for these detunings ( Figure \ref{fig:plasmon-shift}a,b,d,e).


\begin{thebibliography}{99}

\bibitem{MMoskovits} Moskovits, M. Surface-enhanced Spectroscopy \emph{Rev. Mod. Phys.} \textbf{1985}, \emph{57}, 3.

\bibitem{EricLR}Ru, E. C. L.; Etchegoin, P. G.  \emph{Principles of Surface Enhanced Raman Spectroscopy and Related Plasmonic Effects}; Elsevier, Amsterdam, 2009.

\bibitem{BhaP} Bharadwaj, P.; Deutsch, B.; Novotny, L.  Optical Antennas, \emph{Adv. Opt. Photon.} \textbf{2009}, \emph{1}, 438-483.  

\bibitem{MuhP} M\"{u}hlschlegel, P.; Eisler, H.-J.; Martin, O. J. F.; Hecht, B.; Pohl D. W. Resonant Optical Antennas, \emph{Science} \textbf{2005}, \emph{308}, 1607-1609. 

\bibitem{STSivapalan} Sivapalan, S. T.; DeVetter, B. M.; Yang, T. K.; et. al. Off-Resonance Surface-Enhanced Raman Spectroscopy from Gold Nano-rod Suspensions as a Function of Aspect Ratio: Not What We Thought. \emph{ACS Nano} \textbf{2013}, \emph{7}, 2099-2105.

\bibitem{TamTH} Taminiau, T. H.; Stefani, F. D.; van Hulst, N. F.  Single Emitters Coupled to Plasmonic Nano-antennas: Angular Emission and Collection Efficiency  \emph{New J. Phys.} \textbf{2008}, \emph{10}, 105005.  

\bibitem{MuskOL} Muskens, O. L.; Giannini, V.; S\'{a}nchez-Gil, J. A.; G\'{o}mez Rivas, J. Strong Enhancement of the Radiative Decay Rate of Emitters by Single Plasmonic Nanoantennas \emph{Nano Lett.} \textbf{2007}, \emph{7}, 2871-2875.

\bibitem{RogL} Rogobete, L.; Kaminski,  F.;  Agio, M.; Sandoghdar, V.  Design of Plasmonic Nanoantennae for Enhancing Spontaneous Emission \emph{Opt. Lett.} \textbf{2007}, \emph{32}, 1623-1625. 

\bibitem{WNiu}Niu, W.; Chua, Y. A. A.; Zhang, W.; Huang, H.; Lu, X. Highly Symmetric Gold Nanostars: Crystallographic Control and Surface-Enhanced Raman Scattering Property. \emph{J. Am. Chem. Soc.} \textbf{2015}, \emph{137}, 10460-10463.

\bibitem{HaoF} Hao, F.; Nehl C. L.; Hafner, J. H.; Nordlander, P. Plasmon Resonances of a Gold Nanostar \emph{Nano Lett.} \textbf{2007}, \emph{3}, 729-732. 
 
\bibitem{KumPS} Kumar, P. S.;  Pastoriza-Santos, I.; Rodr\'{i}guez-Gonz\'{a}lez B.; et. al. High-yield Synthesis and Optical Response of Gold Nanostars. \emph{Nanotechnology} \textbf{2007}, \emph{19}, 015606. 

\bibitem{WZhu} Zhu, W.; Crozier, K. B. Quantum Mechanical Limit to Plasmonic Enhancement as Observed by Surface-Enhanced Raman Scattering. \emph{Nat. Commun.} \textbf{2014}, \emph{5}, 5228.

\bibitem{AizpJ} Aizpurua J.; Bryant, G. W. ; Richter, L. J.; Garc\'{i}a de Abajo, F. J. Optical Properties of Coupled Metallic Nanorods for Field-enhanced Spectroscopy, \emph{Phys. Rev. B} \textbf{2005} \emph{71}, 235420. 

\bibitem{RomI} Romero, I.; Aizpurua, J.; Bryant, G. W.; Garc\'{i}a de Abajo, F. J. Plasmons in Nearly Touching Metallic Nanoparticles: Singular Response in the Limit of Touching Dimers  \emph{Opt. Express} \textbf{2006}, \emph{14}, 9988-9999.  

\bibitem{EsteR} Esteban, R.; Borisov, A. G.; Nordlander, P.; Aizpurua, J. Bridging Quantum and Classical Plasmonics with a Quantum-corrected Model, \emph{Nat. Comm.} \textbf{2012}, \emph{3}, 825.  

\bibitem{NordP} Nordlander, P.; Oubre, C.; Prodan, E.; Li, K.; Stockman, M. I. et. al. Plasmon Hybridization in Nanoparticle Dimers. \emph{Nano Lett.}  \textbf{2004} \emph{4}, 899-903.  

\bibitem{ALombardi-0} Lombardi, A.; Demetriadou, A.; Weller, L.; Andrae, P.; Benz, F.; Chikkaraddy, R.; Aizpurua, J.; Baumberg, J. J. Anomalous Spectral Shift of Near- and Far-Field Plasmonic Resonances in Nanogaps. \emph{ACS Photonics} \textbf{2016}, \emph{3}, 471-477. 

\bibitem{BaumbergJJ}  Baumberg, J. J.; Aizpurua, J.; Mikkelsen, M. H.; Smith D. R.  Extreme Nanophotonics from Ultrathin Metallic Gaps. \emph{Nat. Mater.} \textbf{2019}, \emph{18}, 668-678. 

\bibitem{RZhang} Zhang, R.; Zhang, Y.; Dong, Z.; et. al. Chemical Mapping of a Single Molecule by Plasmon-Enhanced Raman Scattering. \emph{Nature} \textbf{2013}, \emph{498}, 82-86.

\bibitem{LiuS}  Liu, S.; M\"{u}ller, M.; Sun, Y.; et. al. Resolving the Correlation between Tip-enhanced Resonance Raman Scattering and Local Electronic States with 1 nm Resolution, \emph{Nano Lett.} \textbf{2019}, \emph{19}, 5725-5731. 

\bibitem{QiuXH} Qiu, X. H.; Nazin, G. V.; Ho, W. Vibrationally Resolved Fluorescence Excited with Submolecular Precision. \emph{Science} \textbf{2003}, \emph{299}, 542-54. 

\bibitem{ImadaH} Imada, H.; Miwa, K.; Imai-Imada, M.; et al. Real-space Investigation of Energy Transfer in Heterogeneous Molecular Dimers. \emph{Nature} \textbf{2016},  \emph{538}, 364-367. 

\bibitem{PettB} Pettinger, B.; Schambach, P.; J. Villag\'{o}mez, C.;  Scott N. Tip-enhanced Raman Spectroscopy: Near-fields Acting on a Few Molecules. \emph{Annu. Rev. Phys. Chem.} \textbf{2012}, \emph{63}, 379-399. 

\bibitem{StoRM} St\"{o}ckle, R. M.; Suh, Y. D.; Deckert, V.; Zenobi, R. Nanoscale Chemical Analysis by Tip-enhanced Raman Spectroscopy. \emph{Chem. Phys. Lett.} \textbf{2000}, \emph{318}, 131-136. 

\bibitem{KazE} Kazuma, E.; Jung, J.; Ueba, H.; Trenary, M.; Kim, Y. Real-space and Real-time Observation of a Plasmon-induced Chemical Reaction of a Single Molecule. \emph{Science} \textbf{2018}, \emph{360}, 521-526. 

\bibitem{DopB} Doppagne, B.; Chong, M. C.; Bulou, H.; et. al.,  Electrofluorochromism at the Single-molecule Level. \emph{Science} \textbf{361}, 251-255.

\bibitem{JRLombardi} Lombardi, J. R.; Birke, R. L.; Lu, T.; Xu, J. Charge-transfer Theory of Surface Enhanced Raman Spectroscopy: Herzberg-Teller contributions. \emph{J. Chem. Phys} \textbf{1986}, \emph{84}, 4174.

\bibitem{WKa} Willets, K.  A.; Van Duyne, R. P. Localized Surface Plasmon Resonance Spectroscopy and Sensing. \emph{Annu. Rev. Phys. Chem.} \textbf{2007}, \emph{58}, 267-297. 

\bibitem{LKq} Lin, K.-Q.; Yi J.; Zhong J.-H.; et al. Plasmonic Photoluminescence for Recovering Native Chemical Information from Surface-enhanced Raman Scattering. \emph{Nat. Comm.} \textbf{2017}, \emph{8}, 14891. 

\bibitem{QinL} Qin, L.; Zou, S.; Xue, C.; Atkinson, A.; Schatz, G. C.; Mirkin, C. A. Designing, Fabricating, and Imaging Raman Hot Spots. \emph{PNAS}  \textbf{2006}, \emph{103}, 13300-13303. 

\bibitem{LalS} Lal, S.; Grady, N. K.; Kundu, J.;  et al. Tailoring Plasmonic Substrates for Surface Enhanced Spectroscopies. \emph{Chem. Soc. Rev.} \textbf{2008}, \emph{37}, 898-911.

\bibitem{SMNie} Nie, S. M.; Emory, S. R. Probing Single Molecules and Single Nanoparticles by Surface Enhanced Raman Scattering. \emph{Science} \textbf{1997}, \emph{275}, 1102-1106.

\bibitem{LRu} Le Ru, E. C.; Etchegoin, P. G. Single-molecule Surface-enhanced Raman Spectroscopy. \emph{Annu. Rev. Phys. Chem.}, \textbf{2012}, \emph{63}, 65-87. 

\bibitem{KKneipp} Kneipp, K.; Wang, Y.; Kneipp, H.; Perelman, L. T.; Itzkan, I.; Dasari, R. R.; Feld, M. S. Single Molecule Detection Using Surface- Enhanced Raman Scattering (SERS). \emph{Phys. Rev. Lett.} \textbf{1997}, \emph{78}, 1667-1670.

\bibitem{CiallaMayD}  Cialla-May D.; Zheng  X.-S.; Weberabc K. and Popp  J. Recent Progress in Surface-enhanced Raman Spectroscopy for Biological and Biomedical Applications: from Cells to Clinics,  \emph{Chem. Soc. Rev.} \textbf{2017}, \emph{46}, 3945. 

\bibitem{KKneipp-1} Kneipp, K.; Wang, Y.; Kneipp, H.; Itzkan, I.; Dasari, R. R.; Feld, M. S. Population Pumping of Excited Vibrational States by Spontaneous Surface-Enhanced Raman Scattering.  \emph{ Phys. Rev. Lett.} \textbf{1996}, \emph{76}, 2444.

\bibitem{MaherR} Maher, R.; Etchegoin, P.; Le Ru, E.; Cohen, L. A Conclusive Demonstration of Vibrational Pumping Under Surface Enhanced Raman Scattering Conditions.  \emph{J. Phys. Chem. B}, \textbf{2006}, \emph{110}, 11757.

\bibitem{ECLRu} Le Ru E.; Etchegoin P. G. Vibrational Pumping and Heating under SERS Conditions: Fact or Myth? \emph{ Faraday Discuss.}, \textbf{2006}, \emph{132}, 63.


\bibitem{RoelliP} Roelli, P.; Galland, C.; Piro, N.; Kippenberg, T. J. Molecular Cavity Optomechanics: a Theory of Plasmon-Enhanced Raman Scattering. \emph{Nat. Nanotechnol.} \textbf{2015}, \emph{11}, 164-169.

\bibitem{MKSchmidt} Schmidt, M. K.; Esteban, R.; Gonzalez-Tudela, A.; Giedke, G.; Aizpurua, J. Quantum Mechanical Description of Raman Scattering from Molecules in Plasmonic Cavities. \emph{ACS Nano} \textbf{2016}, \emph{10}, 6291-6298.

\bibitem{MKSchmidt-1} Schmidt, M. K.; Esteban, R.; Benz, F.;
Baumberg, J. J.;  Aizpurua, J Linking Classical and Molecular Optomechanics Descriptions of SERS, \emph{Faraday Discuss.}  \textbf{2017}, \emph{205}, 31-65.

\bibitem{MKDezfouli-1}M. K. Dezfouli, R. Gordon, S. Hughes,Molecular Optomechanics in the Anharmonic Cavity-QED Regime Using Hybrid Metal-Dielectric Cavity Modes, \emph{ACS Photonics} \textbf{2019}, \emph{66}, 1400-1408.

\bibitem{MAspelmeyer} Aspelmeyer, M.; Kippenberg, T. J.; Marquardt F. Cavity Optomechanics, \emph{Rev. Mod. Phys.} \textbf{2014}, \emph{86}, 1391. 

\bibitem{SMAshrafi} Ashrafi S. M.; Malekfar, R.; Bahrampour A. R.; Feist, J. Optomechanical Heat Transfer between Molecules in a Nanoplasmonic Cavity, \emph{Phys. Rev. A} \textbf{2019}, \emph{100}, 013826

\bibitem{FBenz} Benz, F.; Schmidt, M. K.; Dreismann, A.; Chikkaraddy, R.; Zhang, Y.; Demetriadou, A.; Carnegie, C.; Ohadi, H.; de Nijs, B.; Esteban, R.; Aizpurua, J.; Baumberg, J. J. Single-molecule Optomechanics in "picocavities". \emph{Science} \textbf{2016}, \emph{354}, 726-729.

\bibitem{PRolli-1} Roelli, P.; Martin-Cano, D.; Kippenberg, T. J.and Galland C. Molecular Platform for Frequency Upconversion at the Single-photon Level,arXiv:1910.11395v1 


\bibitem{ALombardi} Lombardi, A.; Schmidt, M. K.; Weller, L.;
Deacon, W. M.; Benz, F.; de Nijs, B.; Aizpurua, J.; Baumberg, J. J. Pulsed Molecular Optomechanics in Plasmonic Nanocavities: From Nonlinear Vibrational Instabilities to Bond-Breaking \emph{Phys. Rev. X} \textbf{2018}, \emph{8}, 011016. 

\bibitem{GVrijsen} Vrijsen, G.; Hosten, O.; Lee, J.; Bernon, S.; Kasevich M. A. Raman Lasing with a Cold Atom Gain Medium in a High-Finesse Optical Cavity \emph{Phys. Rev. Lett} \textbf{2011}, \emph{107}, 063904. 

\bibitem{ASSorensen} S{\o}rensen A. S.; M{\o}lmer,  K. 
Entangling Atoms in Bad Cavities \emph{Phys. Rev. A} \textbf{2002}, \emph{66}, 022314.

\bibitem{JKlinder}  Klinder, J.; Ke{\ss}ler, H.;  Wolke, M.; Mathey, L.;  Hemmerich, A. Dynamical Phase Transition in the Open Dicke Model, \emph{PNAS} \textbf{2015}, \emph{112}, 3290-3295.  

\bibitem{JGBohnet} Bohnet, J.  G.; Chen, Z.; Weiner, J. M.; Meiser, D.; Holland, M. J.; Thompson, J. K. A Steady-state Superradiant Laser with less than One Intracavity Photon, \emph{Nature} \textbf{2012}, \emph{484}, 78-81. 

\bibitem{YZhao}Zhao, Y.; Chen,  G. H. J. Quantum Dissipative Master Equations: Some Exact Results, \emph{Chem. Phys.} \textbf{2001}, \emph{114}, 10623. 

\bibitem{MKDezfouli}Dezfouli, M. K.; Hughes, S. 
Quantum Optics Model of Surface-Enhanced Raman Spectroscopy for Arbitrarily Shaped Plasmonic Resonators, \emph{ACS Photonics} \textbf{2017}, \emph{4}, 1245-1256.

\bibitem{EstebanR} Esteban, R.; Aizpurua, J.; Bryant, G. W. Strong Coupling of Single Emitters Interacting with Phononic Infrared Antennae, \emph{New J. Phys.} \textbf{2014}, \emph{16}, 013052.

\bibitem{RChikkaraddy} Chikkaraddy, R.; de Nijs, B.; Benz, F.; Barrow S. J.; Scherman, O. A.; Rosta, E.; Demetriadou, A.; Fox, P.; Hess, O.; Baumberg, J. J. Single-molecule Strong Coupling at Room Temperature in Plasmonic Nanocavities. \emph{Nature} \textbf{2016}, \emph{535}, 127-130.

\bibitem{UrbietaM} Urbieta, M.; Barbry, B.; Zhang, Y.; Koval, P.; S\'anchez-Portal, D.; Zabala, N.; Aizpurua, J. Atomic-Scaling Lightning Rod Effect in Plasmonic Picocavities: A Classical View to a Quantum Effect, \emph{ACS Nano} \textbf{2018}, \emph{12}, 585-596.

\bibitem{FrankeS} Franke, S.; Hughes, S.; Dezfouli, M. K.; Kristensen, P. T.; Busch, K.; Knorr, A.; Richter, M. Quantization of Quasinormal Modes for Open Cavities and Plasmonic Cavity-QED, \emph{Phys. Rev. Lett.} \textbf{2019}, \emph{122}, 213901.

\bibitem{HPBreuer}Breuer, H. P.; Petruccione, F. \emph{The Theory of Open Quantum Systems}, Oxford University Press, 2002. 
 
\bibitem{PMeystre}Meystre, P.; M. Sargent \emph{Elements of Quantum Optics} Springer-Verlag Berlin and Heidelberg Gmbh \& Co. Kg, 2010

\bibitem{MOScully} Scully, M. O.; Svidzinsky, A. A. The Lamb Shift-Yesterday, Today and Tomorrow, \emph{Science} \textbf{2010}, \emph{328}, 1239-1241. 

\bibitem{ZYao} Zhang, Y.; Meng, Q.-S.; Zhang, L.;  et. al. Sub-nanometre Control of the Coherent Interaction between a Single Molecule and a Plasmonic Nanocavity \emph{Nat. Comm.}, \textbf{2017}, \emph{8}, 15225. 

\bibitem{KipfT} Kipf, T.; Agarwal, G.S. Superradiance and Collective Gain in Multimode Optomechanics. \emph{Phys. Rev. A} \textbf{2014}, \emph{90}, 053808.

\bibitem{AndreevAV} Andreev, A. V.; Emel'yanov V. I.; II'inski\u i Y. A. Collective Spontaneous Emission (Dicke Superradiance), \emph{Sov. Phys. Usp.} \textbf{1980}, \emph{23}, 493-514.

\bibitem{NPVitaliy1} Pustovit, V., N.; Shahbazyan, T. V.Cooperative Emission of Light by an Ensemble of Dipoles Near a Metal Nanoparticle: The Plasmonic Dicke Effect, \emph{Phys. Rev. Lett.} \textbf{2009}, \emph{102}, 077401.

\bibitem{MOScully1} Scully, M. O.  Collective Lamb Shift in Single Photon Dicke Superradiance, \emph{Phys. Rev. Lett.} \textbf{2009}, \emph{102}, 143601.

\bibitem{PGEtchegoin}Etchegoin, P. G.; Ru, E. C. L. Resolving Single Molecules in Surface-Enhanced Raman Scattering within the Inhomogeneous Broadening of Raman Peaks, \emph{Anal. Chem.} \textbf{2010}, \emph{82}, 2888-2892. 

\bibitem{ADelga} Delga, A.; Feist, J.; Bravo-Abad, J. and Garcia-Vidal F. J. Quantum Emitters Near a Metal Nanoparticle: Strong Coupling and Quenching, \emph{Phys. Rev. Lett.} \textbf{2014}, \emph{112}, 253601. 

\end{thebibliography}

\begin{thebibliography}{99}
\bibitem{SI-MKSchmidt-1} Schmidt, M. K.; Esteban, R.; Benz, F.; Baumberg, J. J.;  Aizpurua, J Linking Classical and Molecular Optomechanics Descriptions of SERS, \emph{Faraday Discuss.}  \textbf{2017}, \emph{205}, 31-65.

\bibitem{SI-MKSchmidt} Schmidt, M. K.; Esteban, R.; Gonzalez-Tudela, A.; Giedke, G.; Aizpurua, J. Quantum Mechanical Description of Raman Scattering from Molecules in Plasmonic Cavities. \emph{ACS Nano} \textbf{2016}, \emph{10}, 6291-6298.


\bibitem{SI-HPBreuer}Breuer, H. P. \emph{The Theory of Open Quantum Systems}, Oxford University Press, 2002. 

\bibitem{SI-TNeuman1} Neuman, T., Phd Thesis, Theory of Plasmon-enhanced Spectroscopy of Molecular Excitations: Infrared Absorption, Fluorescence, and Raman Scattering, https://addi.ehu.es/handle/10810/33005.

\bibitem{SI-PMeystre}Meystre, P.; M. Sargent \emph{Elements of Quantum Optics} Springer-Verlag Berlin and Heidelberg Gmbh \& Co. Kg, 2010

\bibitem{SI-RoelliP} Roelli, P.; Galland, C.; Piro, N.; Kippenberg, T. J. Molecular Cavity Optomechanics: a Theory of Plasmon-Enhanced Raman Scattering. \emph{Nat. Nanotechnol.} \textbf{2015}, \emph{11}, 164-169.

\bibitem{SI-TNeuman} Neuman, T.; Aizpurua, J. Origin of the Asymmetric Light Emission from Molecular Exciton-polariton, \emph{Optica} \textbf{2018}, \emph{5}, 1247-1255.

\bibitem{SI-TKipf} Kipf, T. and Agarwal, G. S. Superradiance and Collective Gain in Multimode Optomechanics. \emph{Phys. Rev. A} \textbf{2014} \emph{90}, 053808.

\bibitem{SI-FlemingCF} Crim, F. F. Bond-Selected Chemistry: Vibrational State Control of Photodissociation and Bimolecular Reaction, \emph{J. Phys. Chem.} \textbf{1996}, \emph{100} 12725-12734.

\bibitem{SI-ALombardi} Lombardi, A.; Schmidt, M. K.; Weller, L.;
Deacon, W. M.; Benz, F.; de Nijs, B.; Aizpurua, J.; Baumberg, J. J. Pulsed Molecular Optomechanics in Plasmonic Nanocavities: From Nonlinear Vibrational Instabilities to Bond-Breaking \emph{Phys. Rev. X} \textbf{2018}, \emph{8}, 11016. 

\bibitem{SI-YFang}  Fang, Y.;  Li, Y.:  Xu, H.; Sun M. Ascertaining p,p'-Dimercaptoazobenzene Produced from p-Aminothiophenol by Selective Catalytic Coupling Reaction on Silver Nanoparticles, \emph{Langmuir} \textbf{2010}, \emph{26}, 7737-7746.

\bibitem{SI-FBenz} Benz, F.; Schmidt, M. K.; Dreismann, A.; Chikkaraddy, R.; Zhang, Y.; Demetriadou, A.; Carnegie, C.; Ohadi, H.; de Nijs, B.; Esteban, R.; Aizpurua, J.; Baumberg, J. J. Single-molecule Optomechanics in "picocavities" \emph{Science} \textbf{2016}, \emph{354}, 726-729.


\end{thebibliography}
\end{document}